\documentclass[11pt]{article}

\usepackage[utf8]{inputenc}
\usepackage{latexsym}
\usepackage{amsmath}
\usepackage{amssymb}
\usepackage{amsthm}
\usepackage{hyperref}
\usepackage{cite}
\usepackage{graphicx}
\usepackage{color}
\usepackage{enumitem}
\usepackage{comment}
\usepackage{bm}
\usepackage[left=1in,right=1in,top=1in,bottom=1in]{geometry}
\usepackage{xspace}
\setlist[description]{font=\normalfont\itshape\textbullet\space}
\usepackage[all]{xy}
\usepackage{doi, url}
\usepackage{arydshln}
\usepackage{xcolor} 
\usepackage{framed}
\usepackage{physics}
\usepackage{tikz}
\usepackage{mathdots}
\usepackage{yhmath}
\usepackage{cancel}
\usepackage{siunitx}
\usepackage{array}
\usepackage{multirow}
\usepackage{gensymb}
\usepackage{tabularx}
\usepackage{extarrows}
\usepackage{booktabs}
\usetikzlibrary{fadings}
\usetikzlibrary{patterns}
\usetikzlibrary{shadows.blur}
\usetikzlibrary{shapes}

\renewcommand{\paragraph}[1]{\vspace{6pt} \noindent \textbf{#1}\xspace}

\theoremstyle{plain}
\newtheorem{theorem}{Theorem}[section]

\newtheorem{corollary}[theorem]{Corollary}
\newtheorem{lemma}[theorem]{Lemma}
\newtheorem{observation}[theorem]{Observation}
\newtheorem{proposition}[theorem]{Proposition}

\newtheorem{fact}[theorem]{Fact}

\theoremstyle{definition}

\newtheorem{remark}[theorem]{Remark}

\newtheorem{definition}[theorem]{Definition}
\newtheorem{example}[theorem]{Example}

\newcommand{\SL}{\mathrm{SL}}
\newcommand{\GL}{\mathrm{GL}}
\newcommand{\F}{\mathbb{F}}
\newcommand{\Z}{\mathbb{Z}}
\newcommand{\Q}{\mathbb{Q}}
\newcommand{\C}{\mathbb{C}}
\newcommand{\R}{\mathbb{R}}
\newcommand{\N}{\mathbb{N}}

\newcommand{\rng}{\mathrm{R}}

\newcommand{\idmat}{\mathrm{I}}
\newcommand{\IdMat}{\mathrm{I}}

\renewcommand{\Tr}{\mathrm{Tr}}

\newcommand{\rk}{\mathrm{rk}}
\newcommand{\irk}{\rho}

\newcommand{\mcrk}{\rk_{\textrm{MC}}}

\newcommand{\m}{\mathfrak{m}}

\newcommand{\sgn}{\mathrm{sgn}}

\newcommand{\poly}{\mathrm{poly}}

\newcommand{\rad}{\mathrm{rad}}

\newcommand{\M}{\mathrm{M}}
\newcommand{\T}{\mathrm{T}}
\renewcommand{\S}{\mathrm{S}}

\newcommand{\tuple}[1]{\mathbf{#1}}
\newcommand{\tens}[1]{\mathtt{#1}}
\newcommand{\spa}[1]{\mathcal{#1}}

\newcommand{\cA}{\spa{A}}
\newcommand{\cB}{\spa{B}}

\newcommand{\tA}{\tens{A}}
\newcommand{\tB}{\tens{B}}

\newcommand{\tF}{\tens{F}}
\newcommand{\tC}{\tens{C}}
\newcommand{\tD}{\tens{D}}
\newcommand{\tG}{\tens{G}}

\newcommand{\vA}{\tuple{A}}

\newcommand{\vC}{\tuple{C}}

\newcommand{\E}{\mathrm{E}}
\newcommand{\lker}{\mathrm{LKer}}

\newcommand{\stdb}{\vec{b}}

\newcommand{\ord}{\mathrm{ord}}

\newcommand{\aut}{\mathrm{Aut}}
\newcommand{\Hom}{\mathrm{Hom}}
\newcommand{\Ann}{\mathrm{Ann}}

\newcommand{\algprobm}[1]{{\sf #1}\xspace}

\newcommand{\End}{\mathrm{End}}

\newcommand{\ThreeTI}{\algprobm{3TI}}
\newcommand{\ThreeFTI}{\algprobm{3FTI}}

\newcommand{\TI}{\algprobm{TI}}
\newcommand{\TIlong}{\algprobm{Tensor Isomorphism}}
\newcommand{\pTI}{\algprobm{part-TI}}

\newcommand{\cc}[1]{\mathsf{#1}}

\newcommand{\ynote}[1]{\textcolor{blue}{\small$\bullet$\footnote{\textcolor{blue}{Youming: #1}}}}
\DeclareMathOperator{\diag}{diag}

\renewcommand{\ge}{\geqslant}

\newcommand{\too}%
{\xrightarrow{\text{\raisebox{-3pt}{$\sim$}}\,}}

\mathchardef\mhyphen="2D
\newcommand{\rowspan}{\mathtt{row}\mhyphen\mathtt{span}}

\usepackage[T1]{fontenc}
\def\DJ{{\hbox{D\kern-.8em\raise.15ex\hbox{--}\kern.35em}}}

\date{}

\title{
On the complexity of isomorphism problems for tensors, groups, and polynomials V: over commutative rings}
\author{
	Joshua A. Grochow
	\footnote{Departments of Computer Science and Mathematics, University of Colorado 
		Boulder. \tt{jgrochow@colorado.edu}}
	\and
	Youming Qiao
	\footnote{Centre for Quantum Software and Information, University of 
		Technology Sydney. \tt{youming.qiao@uts.edu.au}}
	\and
	Katherine E. Stange
	\footnote{Department of Mathematics, University of Colorado Boulder. \tt{kstange@math.colorado.edu}}
	\and
	Xiaorui Sun
	\footnote{Computer Science Department, University of Illinois at Chicago. {\tt xiaorui@uic.edu}}
}

\begin{document}
	\pagenumbering{gobble}  
	
\maketitle

\begin{abstract}
Tensors over commutative rings naturally appear in number theory, geometry, and group theory. For example, $2\times 2\times 2$ tensors over $\mathbb{Z}$ form the starting point of Bhargava's celebrated works generalising Gauss's composition law (Bhargava, \emph{Ann. Math.}, 2004). Symmetric tensors over $\mathbb{Z}$ are central to the classification of Calabi--Yau threefolds (Yau, \emph{Commun. Pure Appl. Math.}, 1978; Wall, \emph{Invent. Math.}, 1966), geometric objects of significance in string theory. Additionally, tensors over finite commutative rings closely correspond to finite nilpotent groups of class 2 (Baer, \emph{Trans. Am. Math. Soc.}, 1938).
		
In these settings, testing isomorphism of tensors is of great interest. For example, in mathematical physics, several recent works apply machine learning techniques to distinguish symmetric tensors from Calabi--Yau threefolds. For group isomorphism, a recent breakthrough of Sun (\emph{STOC}, 2023) gives the first $N^{o(\log N)}$-time algorithm for testing isomorphism of $p$-groups of class $2$ and exponent $p$ of order $N$, using tensor-based techniques. Grunewald \& Segal studied the computability of tensor isomorphism problems over $\Z$, showing that they are computable in finite time (\emph{Ann. Math.}, 1980).

In this work, we study isomorphism testing of tensors over commutative rings from a complexity-theoretic viewpoint, and its applications. Some of our main results are:
\begin{enumerate}
\item Let $\mathrm{R}$ be a commutative ring. We introduce two complexity classes: $\mathsf{3TI}_\mathrm{R}$ consisting of problems that are polynomial-time reducible to isomorphism problems of tensor products of three modules over $\mathrm{R}$, and $\mathsf{3FTI}_\mathrm{R}$ consisting of problems that are polynomial-time reducible to isomorphism problems of tensor products of three free modules over $\mathrm{R}$.
		
\item We show that some classical problems considered by Grunewald and Segal (ibid.), and the problem of classifying Calabi--Yau threefolds, are $\mathsf{3FTI}_\mathbb{Z}$-complete. We also show that many natural problems are complete for $\mathsf{3TI}_{\Z/p^e\Z}$.

\item We show that testing isomorphism of tensors in $\mathbb{Z}^2\otimes\mathbb{Z}^2\otimes\mathbb{Z}^2$ is polynomial-time equivalent to the principal ideal problem in algorithmic number theory. 
		
The key to this reduction is Bhargava's work (\emph{Ann. Math.}, 2004). Using our equivalence, a result of Hallgren (\emph{J. ACM}, 2007) then implies that $2\times 2\times 2$ tensor isomorphism over $\mathbb{Z}$ is in quantum polynomial time.
		
\item We present an $N^{O((\log N)^{8/9})}$-time algorithm for testing isomorphism of finite nilpotent groups of class $2$ and odd order $N$. 
		
This is achieved by considering tensor isomorphism over $\mathbb{Z}/p^e\mathbb{Z}$. Following the strategy of (Sun, \emph{STOC}, 2023), the algorithm is a reduction to testing the congruence of matrix tuples over $\mathbb{Z}/p^e\mathbb{Z}$, for which we present a polynomial-time solution following and generalizing (Ivanyos--Qiao, \emph{SIAM J. Comput.}, 2019), who solved the analogous problem over finite fields of odd order.
		
\end{enumerate}
\end{abstract}

\newpage

\setcounter{tocdepth}{2}

\tableofcontents

\newpage
\pagenumbering{arabic}     

\section{Introduction}


We study the problem of testing tensor isomorphism over commutative rings, significantly generalizing previous work that focused on tensors over fields. Tensors include vectors, matrices, and 3-dimensional and higher-dimensional arrays. Two 3-tensors $T_{ijk}, T'_{ijk}$ over a field are isomorphic if they become equal after a change of basis in each of their three directions, that is, if there exist invertible matrices $A,B,C$ such that
\[
T_{ijk} = \sum_{i',j',k'} A_{ii'} B_{jj'} C_{kk'} T'_{i'j'k'} \quad (\forall i,j,k).
\]
Over more general rings, we will need to refine our notion of what it means for tensors to be isomorphic, generalizing the above equation. 

Tensors over commutative rings naturally appear in number theory, computability, geometry, mathematical physics, group theory, and cryptography. In particular, the algorithmic question of testing tensor isomorphism is of great interest in these contexts. To motivate our study, we first examine the connections to these various fields.

\subsection{Motivations}\label{subsec:motivation}

\paragraph{Number theory: Bhargava's higher composition laws.} Gauss's composition law for binary quadratic forms over $\Z$ is the first historical example of studying the class group in algebraic number theory. In the celebrated works \cite{Bhargava,higher-2,higher-3,higher-4}, Bhargava discovered new composition laws for higher degree forms that go beyond class groups.
Bhargava's starting point is the examination of $2\times 2\times 2$ tensors over $\Z$, which leads to a novel and visual description of Gauss's composition law. He proves a correspondence between $\GL(2, \Z)$-orbits on $\Z^2\otimes\Z^2\otimes\Z^2$ and isomorphism classes of certain ideal class triples in quadratic rings.

\paragraph{Geometry and mathematical physics: classifying Calabi--Yau threefolds.} Calabi--Yau threefolds, that is Calabi--Yau manifolds of complex dimension 3, 
are of great interest in the study of geometry and mathematical physics \cite{Yau78,Calabi_Yau-1,Calabi_Yau-2,Calabi_Yau-3}. A key question in this context is to classify and bound the number of Calabi--Yau threefolds due to its implications to string theory \cite{Rei87,CY-count,CCFHL24}, and Yau has long conjectured that the number should be finite (see e.g. \cite{HYY24}). 

To connect the classification of Calabi--Yau threefolds with tensors over $\Z$, the classic theorem of Wall for K\"ahler manifolds is crucial \cite{Wall66}. Wall's theorem implies that the topological type of Calabi--Yau threefolds is determined by (1) Hodge numbers $h^{1,1}$, $h^{1,2}\in \Z^+$, (2) the triple intersection numbers $C_{i,j,k}\in\N$ for $i, j, k\in [h^{1,1}] = \{1,2,\dotsc,h^{1,1}\}$, which form a symmetric tensor, and (3) the second Chern class, which is a vector in $\N^{h^{1,1}}$. It is known \cite{CY-count,CCFHL24} that Calabi--Yau threefolds with the same Hodge numbers are diffeomorphic if and only if the triple intersection numbers and the second Chern class vectors are in the same orbit under $\GL(n, \Z)$. Therefore, isomorphism of the integer 3-tensors with coordinates $C_{i,j,k}$ is a crucial step in classifying such manifolds.

Recently, machine learning methods have been applied to the classification of Calabi--Yau manifolds via Wall's theorem \cite{JTT22,HYY24}. In \cite{JTT22}, it was noted that ``there is no known algorithm to check this equivalence in finite time.'' However, as we now explain, such a finite-time algorithm does exist, although its complexity has not been analyzed and it may not be efficient in practice. 


\paragraph{Computability: arithmetic group actions and the work of Grunewald \& Segal.} In a pair of articles \cite{GS80-1,GS80-2}, Grunewald and Segal showed that several problems about arithmetic group actions are decidable in finite time. A main result in \cite{GS80-1} is the decidability of the following problem. Let $\rho$ be a rational action of $\GL(n, \Z)$ on a vector space $\C^m$. Given $u, v\in \C^m$, the \emph{orbit problem} asks whether there exists $T\in \GL(n, \Z)$ such that $\rho(T)\circ u=v$.

Grunewald and Segal mentioned several applications of their solution to the orbit problem. 
\begin{itemize}
	\item (Matrix tuple conjugacy problem.) Let $(A_1, \dots, A_m), (B_1, \dots, B_m)\in \M(n, \Q)^m$. Decide if there exists $T\in\GL(n, \Z)$ such that $\forall i\in[m]$, $TA_iT^{-1}=B_i$. 
	
	Over fields, this is a classical problem in computer algebra known as the module isomorphism problem, for which polynomial-time algorithms exist~\cite{CIK97,BL08}.
	
	\item (Form tuple equivalence problem.) Let $(f_1, \dots, f_m), (g_1, \dots, g_m)\in\Q[x_1, \dots, x_n]$ be homogeneous polynomials (a.k.a. forms). Decide if there exists $T\in \GL(n, \Z)$ such that $\forall i\in[m]$, $f_i\circ T=g_i$. 
	
		For single forms, this is a classical problem (generalizing equivalence of binary quadratic forms as in Gauss' work mentioned above), which has been studied by C. Jordan, H. Poincar\'e, and C. L. Siegel; see \cite{GS-overview}.

	\item (Algebra isomorphism problem.) Let $\phi, \psi:\Z^n\times\Z^n\to\Z^n$ be two $\Z$-bilinear maps. Decide if there exists $T\in\GL(n, \Z)$, such that $\forall u, v\in \Z^n$, $\phi(T(u), T(v))=T(\psi(u, v))$. 
	\item (Multilinear form isomorphism problem.) Let $\phi, \psi:\Z^n\times\Z^n\times\dots\times \Z^n\to \Q$ be two multilinear forms. Decide if there exists $T\in\GL(n, \Z)$, such that $\phi\circ T=\psi$. 
\end{itemize}

\paragraph{Group theory: finite nilpotent groups.}
The finite group isomorphism problem is to determine whether two finite groups are isomorphic. When the groups are given by their Cayley (multiplication) tables, this problem is one of the few natural problems in $\cc{NP}$ for which no polynomial-time algorithm is known, and which is not believed to be $\cc{NP}$-complete (otherwise $\cc{PH}$ collapses to the second level).

Research on group isomorphism in theoretical computer science began in the 1970s. The first algorithm for group isomorphism had a running time of $N^{\log N + O(1)}$, where $N$ denotes the order of the groups~\cite{Mil78, FN}\footnote{Miller attributes the result to Tarjan.}. Despite extensive study, the current best algorithm for group isomorphism has a running time of $N^{(1/4) \log N + O(1)}$~\cite{Ros13} (see \cite[Sec.~2.2]{LeGallRosenbaum}).
The recent celebrated work by Babai on quasipolynomial-time algorithms for graph isomorphism~\cite{babai2016graph} has further highlighted the importance of the group isomorphism problem. That is, group isomorphism has become a bottleneck for the $n^{o(\log n)}$ time algorithm for graph isomorphism for graphs with $n$ vertices, as group isomorphism reduces to graph isomorphism \cite{KST93}.

Class-2 nilpotent groups have long been recognized as the most challenging case for the group isomorphism problem. Despite significant efforts, the best-known algorithms for this class are not asymptotically better than for general groups. 

Recently, progress has been made on special cases of class-2 nilpotent groups. In \cite{Sun23}, an algorithm with a running time of $N^{O((\log N)^{5 / 6})}$ was proposed for $p$-groups of nilpotent class 2 and exponent $p$ (meaning that every non-identity element has order $p$), for odd primes $p$. In \cite{IMQSZ24}, an algorithm with a running time of $N^{O(\sqrt{\log N})}$ was introduced for $p$-groups of Frattini class 2,  whose group elements are of exponent at most $p^2$.\footnote{A $p$-group $G$ is of \emph{Frattini class $2$}, if there exists $H\leq G$, such that $H$ is central, and both $H$ and $G/H$ are elementary abelian. 
Note, however, that $p$-groups of Frattini class $2$ do not even contain all $p$-groups of class $2$ and exponent $p^2$.} However, for the more general class of class-2 nilpotent groups---even those of odd order---no algorithm with a running time of $N^{o(\log N)}$ was known prior to our paper.

\paragraph{Cryptography: lattice isomorphism.} While we will be mostly focusing on 3-tensors, 2-tensors (or matrices) over commutative rings already pose a nice challenge with possible applications in cryptography. 

Let $A$ and $B$ be two $n\times n$ matrices over $\Q$. The \algprobm{Lattice Isomorphism Problem} asks whether there exists an $n\times n$ unimodular matrix $T$---that is, a matrix with integer entries and determinant being $\pm 1$---such that $T^tAT=B$. Haviv and Regev \cite{HR14} showed that this problem is in the complexity class Statistical Zero Knowledge ($\cc{SZK}$), and admits an algorithm in time $n^{O(n)}\cdot \poly(s)$, where $s$ is the input size. This group action has recently received considerable attention in post-quantum cryptography \cite{DW22,DG23,BBCK24}. In particular, a digital signature scheme called Hawk \cite{Hawk} based on lattice isomorphism has been implemented 
and has recently advanced to the second round of the call for post-quantum digital signature schemes by the U.S. National Institute of Standards and Technology (NIST) \cite{NIST-report}.



\subsection{Tensors over commutative rings: some intriguing phenomena}\label{subsec:intriguing}

In this section, we walk through some intriguing phenomena for tensors over commutative rings, especially when compared with tensors of fields. While some basic ring theory will be needed to state and prove our results, in this introduction, we shall only assume some familiarity with two commutative rings: the ring of integers $\Z$, and the modulo ring $\Z/m\Z$ where $m\in\N$.


\paragraph{Rank notions.} To start with, consider matrices, or $2$-tensors. 
For matrices over fields, the rank notion has several equivalent interpretations, via its row span, column span, left kernel, right kernel, non-zero minors, and decompositions into sum of rank-$1$ matrices. We also have that an $n\times n$ matrix over a field $\F$ is full-rank (invertible), if its determinant is non-zero.

For matrices over rings, these notions do not coincide in general. Consider $\rng=\Z/6\Z$ and the matrix $\begin{bmatrix}
	4 & 0 \\
	0 & 3
\end{bmatrix}$ over $\rng$. Its image is spanned by $(4, 0)$ and $(0, 3)$. Its left-kernel is spanned by $(3, 0)$ and $(0, 2)$. Furthermore, it can be decomposed as $\begin{bmatrix}
	2\\3\end{bmatrix}\begin{bmatrix}
	2 & 3
\end{bmatrix}$. This suggests that several natural ways to formulate a rank notion for matrices over commutative rings are not equivalent.

As another example, over $\Z$, the matrix $A=\begin{bmatrix}
	1 & 0 \\
	0 & 2
\end{bmatrix}$ might be considered to have full rank from several perspectives. However, as a linear map from $\Z^2$ to $\Z^2$, $A$ is not surjective, and it does not admit an inverse. Indeed, over a commutative ring $\rng$, an $n\times n$ matrix is invertible if and only if its determinant is a \emph{unit} in $\rng$, that is an element in $\rng$ admitting a multiplicative inverse. Note that $\det(A)=2$ which does not have an inverse in $\Z$.

In this work, we need to identify a suitable notion of ranks when dealing with matrices over certain $\rng$. It turns out that the notion of \emph{inner ranks} \cite{Coh85}, based on decomposition into a sum of rank-$1$, matrices works well for our purpose; see Section~\ref{subsec:rank-notion}.

\paragraph{Modules: free or not.} The ring-theoretic correspondence of vector spaces over $\F$ is modules over $\rng$. While vector spaces over $\F$ are determined up to isomorphism solely by their dimension, modules over $\rng$ can be much more abundant and complicated. 

First, the so-called free modules are in closer analogy with vector spaces, as they are isomorphic to $\rng^n$, the module of length-$n$ vectors over $\rng$. However, there exist modules for which the notion of basis in a vector space does not behave well. Consider $\rng=\Z/4\Z=\{0, 1, 2, 3\}$. Then $2\Z/4\Z=\{0, 2\}$ is a module under the natural multiplication from $\Z/4\Z$, but it does not come with a basis because $2$ in $2\Z/4\Z$ is not ``independent'', since a linear combination with non-zero coefficient results in zero, namely: $2\times 2=0\mod 4$. 

As a result, in some settings we will need to formulate isomorphism problems for elements of tensor product of modules. The $3$-dimensional array formulation is just for isomorphism problems of tensor products of \emph{free} modules. Isomorphism of tensor products of (possibly) non-free modules will be crucial for our results on finite group isomorphism.

\paragraph{Equivalence of matrices.} Let $S$ and $T$ be $m\times n$ matrices over $\rng$. We say that $S$ and $T$ are (left-right) equivalent, if there exists an $m\times m$ and $n\times n$ invertible matrices $P$ and $Q$, such that $S=PTQ^t$. When $\rng=\F$ is a field, $S$ and $T$ are equivalent if and only if they are of the same rank. When $\rng$ is a general commutative ring, equivalence of $S$ and $T$ is generally more complicated. For example, for $\rng=\Z$, the two matrices $S=\begin{bmatrix}
	1 & 0 \\
	0 & 1
\end{bmatrix}$ and $T=\begin{bmatrix}
	1 & 0 \\
	0 & 2
\end{bmatrix}$ are not equivalent, despite that their row spans are both free modules of rank $2$. Of course, over $\Z$ and more generally over principal ideal rings, equivalence classes of matrices are determined by their Smith Normal Forms. We shall make use of Smith Normal Form for some rings, but some of our results extend to rings over which Smith Normal Forms do not exist \cite{Sta16}.

\subsection{Our main results}

In this paper we study isomorphism problems of tensors over commutative rings from a  complexity-theoretic viewpoint. We first introduce a complexity class for tensor isomorphism over commutative rings. We then develop algorithms for $2\times 2\times 2$ free tensor isomorphism over $\Z$, and for testing isomorphism of class-$2$ finite nilpotent groups of odd order (with no restriction on their exponent). 

\paragraph{Tensor Isomorphism complexity class.} Over fields, the \TIlong complexity class was developed in a series of recent works \cite{TI1,TI2,TI3,TI4,GQT}. Over commutative rings, as discussed in Section~\ref{subsec:intriguing}, to define a \TIlong problem, we need to specify the (family of) modules under consideration. 

We first define the Free-Module Tensor Isomorphism complexity class. For a commutative ring $\rng$, let $\rng^n$ be the free module consisting of length-$n$ row vectors over $\rng$. That is, $\rng^n=\{(a_1, \dots, a_n)\mid a_i\in \rng\}$. Let $\GL(n, \rng)$ be the group of invertible matrices over $\rng$; recall that an $n\times n$ matrix $T$ over $\rng$ is \emph{invertible} if $\det(T)$ is a unit (i.e. invertible) in $\rng$. 

Let $U=\rng^n$, $V=\rng^m$, and $W=\rng^\ell$. The tensor product space $U\otimes V\otimes W$ consists of $\ell\times m\times n$ cubes of elements from $\rng$. That is, $\tA\in U\otimes V\otimes W$ is $\tA=(a_{i,j,k})$ where $i\in[\ell]$, $j\in[m]$, $k\in[n]$ and $a_{i,j,k}\in \rng$. Given $L=(\alpha_{i,i'}), R=(\beta_{j,j'}), T=(\gamma_{k,k'})$, $(L, R, T)$ sends $\tA$ to $\tA'=(a'_{i',j',k'})$ where 
\begin{equation}\label{eq:action}
	a'_{i',j',k'}=\sum_{i,j,k}\alpha_{i',i}\beta_{j',j}\gamma_{k',k}a_{i,j,k}.
\end{equation}
The \emph{Free Module 3-Tensor Isomorphism} problem asks, given $\tA$ and $\tB\in U\otimes V\otimes W$, whether there exists $(L, R, T)\in \GL(U)\times\GL(V)\times\GL(W)$ sending $\tA$ to $\tB$.


\begin{definition}\label{def:freeTI}
	Let $\rng$ be a commutative ring. The Free Module 3-Tensor Isomorphism complexity class $\ThreeFTI_\rng$ consists of problems that are polynomial-time reducible to the Free 3-Tensor isomorphism problem over $\rng$.
\end{definition}

More generally, for three modules $A, B, C$ over a commutative ring $\rng$, the corresponding \TIlong problem asks, given $s, t\in A\otimes B\otimes C$, whether there exist module isomorphisms $\alpha:A\to A$, $\beta:B\to B$, and $\gamma\in C\to C$, such that $(\alpha, \beta, \gamma)$ sends $s$ to $t$. 
\begin{definition}\label{def:TI}
Let $\bar A$, $\bar B$, and $\bar C$ be three families of modules over $\rng$. The $(\bar A, \bar B, \bar C)$-$\TI_\rng$ complexity class consists of these problems polynomial-time reducible to the $(A, B, C)$-$\TI_\rng$ problem for some $A\in \bar A$, $B\in \bar B$, and $C\in \bar C$.

If $\bar A=\bar B=\bar C$ is the set of all $\rng$-modules, then it is called the $\ThreeTI_\rng$ class for short. 
\end{definition}

Note that the $\ThreeFTI_\rng$ complexity class is $(\bar A, \bar B, \bar C)$-$\TI_\rng$ where $\bar A=\bar B=\bar C$ is the family of free modules. 

\paragraph{Tensor Isomorphism complete problems: five actions.} After introducing the $\ThreeTI_\rng$ and $\ThreeFTI_\rng$ complexity classes, we present complete problems for these classes. For convenience, we focus on $\ThreeFTI_{\Z}$ and $\ThreeTI_{\Z/p^e\Z}$ in the introduction.

To introduce these problems, we need the following definition. Let $\T(\ell\times m\times n, \rng)$ be the set of $3$-way arrays of size $\ell\times m\times n$ over $\rng$. Note that unlike $U\otimes V\otimes W$, $\T(\ell\times m\times n, \rng)$ does not have a group action implicitly associated with it. Indeed, some natural actions on $\T(\ell\times m\times n, \rng)$ are as follows.
\begin{definition}\label{def:5action}
	Let $U=\rng^\ell$, $V=\rng^m$, and $W=\rng^n$. 
	\begin{enumerate}
		\item Given $\tA\in\T(\ell\times m\times n, \rng)$, $(R, S, T)\in\GL(\ell, \rng)\times\GL(m,\rng)\times\GL(n,\rng)$ 
		sends $\tA=(a_{i,j,k})$ to $\tA'=(a_{i,j,k}')$ as defined in Equation~\ref{eq:action}. This corresponds to the natural action of $\GL(U)\times\GL(V)\times\GL(W)$ on $U\otimes V\otimes W$.
		\item Given $\tA\in\T(\ell\times \ell\times m, \rng)$, $(S, T)\in\GL(\ell, \rng)\times\GL(m, \rng)$ sends $\tA$ to $\tA'$ by $(S, S, T)$ on $\tA$ as defined in Equation~\ref{eq:action}. This corresponds to the natural action of $\GL(U)\times\GL(W)$ on $U\otimes U\otimes W$. 
		\item Given $\tA\in\T(\ell\times \ell\times m, \rng)$, $(S, T)\in\GL(\ell, \rng)\times\GL(m, \rng)$ sends $\tA$ to $\tA'$ by $(S, S^{-t}, T)$ on $\tA$ as defined in Equation~\ref{eq:action}. Here $S^{-t}$ denotes the transpose inverse of $S$. This corresponds to the natural action of $\GL(U)\times\GL(W)$ on $U\otimes U^*\otimes W$, where $U^* := \Hom_R(U, R)$ denotes the dual module of $U$. 
		\item Given $\tA\in\T(\ell\times \ell\times \ell, \rng)$, $S\in\GL(\ell, \rng)$ sends $\tA$ to $\tA'$ by $(S, S, S)$ on $\tA$ as defined in Equation~\ref{eq:action}. This corresponds to the natural action of $\GL(U)$ on $U\otimes U\otimes U$.
		\item Given $\tA\in\T(\ell\times \ell\times \ell, \rng)$, $S\in\GL(\ell, \rng)$ sends $\tA$ to $\tA'$ by $(S, S, S^{-t})$ on $\tA$ as defined in Equation~\ref{eq:action}. This corresponds to the natural action of $\GL(U)$ on $U\otimes U\otimes U^*$. Note that in this case $\tA$ can be viewed as recording the structure constants of some (possibly non-associative) algebra. 
	\end{enumerate}
\end{definition}

We also need the following notions for $3$-way arrays. Let $\tA\in\T(\ell\times m\times n, \F)$. The \emph{length} of $\tA$ is defined as $\ell+m+n$.
The \emph{frontal slices} of $\tA$ are $\{A_1, \dots, A_n\}\subseteq\M(\ell\times m, \F)$, where $A_k(i,j)=a_{i,j,k}$.
Let $\tA\in\T(n\times n\times n, \F)$. Then $\tA$ is \emph{symmetric} if for any $\sigma\in\S_3$, $\tA(i,j,k)=\tA(\sigma(i), \sigma(j), \sigma(k))$, and $\tA$ is \emph{anti-symmetric} if for any $\sigma\in\S_3$, $\tA(i,j,k)=\sgn(\sigma)\tA(\sigma(i), \sigma(j), \sigma(k))$. 

Our first result is about $\ThreeFTI_\rng$ where $\rng=\Z$. 
\begin{theorem}\label{thm:Z-equivalence}
	Let $\rng=\Z$. For $i, j\in[5]$, $i\neq j$, let $\tA$ and $\tB$ two $3$-way arrays of total length $L$ whose sizes admit the $i$th action defined in Definition~\ref{def:5action}. Then there exists a polynomial-time computable function $f$ that takes $\tA$ and $\tB$ and outputs $3$-way arrays $f(\tA)$ and $f(\tB)$, such that $\tA$ and $\tB$ are in the same orbit under the $i$th action if and only if $f(\tA)$ and $f(\tB)$ are in the same orbit under the $j$th action.
	
	Furthermore, the above holds even with the following additional structural restrictions:
	\begin{enumerate}
		\item For $j=2$, i.e. the action of $\GL(U)\times \GL(W)$ on $U\otimes U\otimes W$, the frontal slices of $f(\tA)$ and $f(\tB)$ are symmetric (or skew-symmetric). 
		\item For $j=4$, i.e. the action of $\GL(U)$ on $U\otimes U\otimes U$, $f(\tA)$ and $f(\tB)$ are symmetric (or anti-symmetric) $3$-way arrays. 
		\item For $j=5$, i.e. the action of $\GL(U)$ on $U\otimes U\otimes U^*$, $f(\tA)$ and $f(\tB)$ record the structure constants of associative or Lie algebras. 
	\end{enumerate}
\end{theorem}

We then have the following result on the complexity of some natural algorithmic problems over $\Z$, some of which were considered in \cite{GS-overview}. Note that in the following, we may consider tensors over $\Q$ but with actions from $\GL(n, \Z)$. The complexity class $\ThreeFTI_\Z$ can be generalised to take into this account naturally.
\begin{corollary}\label{cor:TIZ-complete}
	The following problems are $\ThreeFTI_\Z$-complete. 
	\begin{itemize}
		\item Matrix space conjugacy problem: let $(A_1, \dots, A_m), (B_1, \dots, B_m)\in \M(n, \Q)^m$. Decide if there exists $T\in\GL(n, \Z)$ and $S=(s_{i,j})\in\GL(m, \Z)$, such that $\forall i\in[m]$, $TA_iT^{-1}=\sum_{j\in[m]}s_{i,j}B_j$. 
		\item Algebra isomorphism: given $\Z$-linear forms $\phi, \psi:\Z^n\times\Z^n\to\Z^n$, decide if there exists $T\in\GL(n, \Z)$, such that $\forall u, v\in \Z^n$, $\phi(T(u), T(v))=T(\psi(u, v))$. 
		\item Trilinear form isomorphism: let $\phi, \psi:\Z^n\times\Z^n\times\Z^n\to \Q$ be two symmetric trilinear forms. Decide if there exists $T\in\GL(n, \Z)$, such that $\phi\circ T=\psi$. 
		\item Inhomogeneous trilinear form isomorphism: let $\phi, \psi:\Z^n\times\Z^n\times \Z^n\to \N$ be two trilinear forms, and $f, g:\Z^n\to \N$ be two $\Z$-linear forms. Decide if there exists $T\in\GL(n, \Z)$, such that $\phi\circ T=\psi$ and $f\circ T=g$.
	\end{itemize}
\end{corollary}
\ynote{Added here:}Note that inhomogeneous trilinear form isomorphism naturally arises from the Calabi--Yau threefold homeomorphism problem as discussed in Section~\ref{subsec:motivation}.

Our second result is about $\ThreeTI_\rng$ where $\rng=\Z/p^e\Z$ where $p$ is prime. Let $U$ be a $\Z/p^e\Z$ module. As $\Z/p^e\Z$ is a local ring (=unique maximal ideal), the rank of $U$ is well-defined as the size of a minimal generating set of $A$. As $\Z/p^e\Z$ is a principal ideal ring (all ideals are generated by single elements), $U$ is isomorphic to a direct sum of cyclic modules (=modules generated by a single element). This means that in algorithms, $U$ can be viewed as consisting of vectors where each coordinate is from $\Z/p^k\Z$ for some $k\in \{1, \dots, e\}$. The automorphism group of $U$ can be represented in a matrix format \cite{Ranum, HillarRhea}. 

These indicate that we can compute with $\Z/p^e\Z$-modules, their automorphism groups, and tensor products, in a manner not very different from the free module setting; see Section~\ref{sec:PIR} for more details. In particular, for three $\Z/p^e\Z$-modules $U$, $V$, and $W$, let $\ell=\rk(U)$, $m=\rk(V)$, and $\ell=\rk(W)$. Then $U\otimes V\otimes W$ can be recorded as a $3$-way array of size $\ell\times m\times n$, and three tuples of integers $(a_1, \dots, a_\ell)$, $(b_1, \dots, b_m)$, $(c_1, \dots, c_n)$. Then we can define five actions on such $3$-way arrays as in Definition~\ref{def:5action}.

For applications to group isomorphism, the following parameter is important. Let $U$, $V$, and $W$ be three finite modules over $\Z/p^e\Z$. We define the ``complex parameter'' of $U\otimes V\otimes W$ as $\log_p(|U|\cdot |V|\cdot |W|)$.
\begin{theorem}\label{thm:ptoe-equivalence}
	Let $\rng=\Z/p^e\Z$. Let $U, V, W$ be finite $\Z/p^e\Z$-modules. 
	For $i, j\in[5]$, $i\neq j$, let $\tA$ and $\tB$ two $3$-way arrays in $U \otimes V \otimes W$, 
	with complex parameters $L$. Then there exists a polynomial-time computable function $f$ that takes $\tA$ and $\tB$ and outputs $3$-way arrays $f(\tA)$ and $f(\tB)$, such that (1) the complex parameters of $f(\tA)$ and $f(\tB)$ are upper bounded by $L^{O(1)}$,
	and (2) $\tA$ and $\tB$ are in the same orbit under the $i$th action if and only if $f(\tA)$ and $f(\tB)$ are in the same orbit under the $j$th action.
	
	Furthermore, the above holds even with the following additional structural restrictions:
	\begin{enumerate}
		\item For $j=2$, i.e. the action of $\GL(U)\times \GL(W)$ on $U\otimes U\otimes W$, the frontal slices of $f(\tA)$ and $f(\tB)$ are symmetric (or skew-symmetric). 
		\item For $j=4$, i.e. the action of $\GL(U)$ on $U\otimes U\otimes U$, $f(\tA)$ and $f(\tB)$ are symmetric (or anti-symmetric) $3$-way arrays. 
		\item For $j=5$, i.e. the action of $\GL(U)$ on $U\otimes U\otimes U^*$, $f(\tA)$ and $f(\tB)$ record the structure constants of associative or Lie algebras. 
	\end{enumerate}
\end{theorem}

\paragraph{$2\times 2\times 2$ free tensor isomorphism over $\Z$.} 
\newcommand{\disc}{\operatorname{disc}}
\newcommand{\Cl}{\operatorname{Cl}}
We study testing isomorphism of tensors of free modules over $\Z$. By the results of Grunewald and Segal on arithmetic group actions\cite{GS80-1}, this problem is decidable. The algorithms in \cite{GS80-1} relies on Tarski's decidability of elementary theory over $\R$, so it is unlikely to be in NP \cite{BKR84,Gri88}. On the other hand, as mentioned in Section~\ref{subsec:motivation}, $\Z^2 \otimes \Z^2 \otimes \Z^2$ isomorphism was studied by Bhargava \cite{Bhargava}, so this gives hope that better complexity results can be obtained for testing isomorphism of tensors in $\Z^2 \otimes \Z^2 \otimes \Z^2$. This is indeed the case; the following result relies crucially on Bhargava's work \cite{Bhargava}.
	
	\begin{theorem}
		\label{thm:tensorZ}
		\algprobm{Tensor Isomorphism} for $\Z^2 \otimes \Z^2 \otimes \Z^2$ is polynomial-time equivalent to the \algprobm{Principality Decision Problem} for real quadratic fields. In particular, it is in $\cc{BQP}$.
	\end{theorem}

Let us briefly review the \algprobm{Principality Decision Problem}. In a number field order $\rng$, this problem is to determine, given an ideal $I$ of $\rng$ (by providing an integral basis), whether $I$ is principal.  This is the decisional portion of the \algprobm{Principal Ideal Problem} or the \algprobm{Principality Testing Problem}; see \cite[Chapter 16]{JacobsonWilliams}.  It is a subroutine of the most important algorithmic number theory problems for number fields, including class group and unit group computations.  It is also important to the security of cryptography based on ideal lattices.  In fact, there are reductions from $\gamma$-SVP (Shortest Vector Problem) to the Principal Ideal Problem for cyclotomic fields due to Cramer, Ducas, and Wesolowski \cite{CramerDucasWesolowski}.

Here we deal with the \algprobm{Principal Ideal Problem} for real quadratic number field orders.  This is the basis for the cryptographic scheme of Buchmann and Williams and others \cite{BuchmannWilliams, crypto1, crypto2}.  The best known unconditional classical methods in the maximal order are exponential \cite[Chapter 16]{JacobsonWilliams}, with some conditional subexponential algorithms \cite[Theorem 13.24]{JacobsonWilliams}.  For non-maximal orders, we additionally need to factor the discriminant and solve discrete logarithm problems in the prime fields for primes dividing the conductor; see \cite{BiasseFiekerJacobson} (which leaves us in the same position: unconditionally exponential, conditionally subexponential).  However, it can be solved by a polynomial-time quantum algorithm \cite{Hallgren}, which was later generalized to arbitrary number fields \cite{BiasseSong}.



\paragraph{Class-$2$ nilpotent group isomorphism.}
A group $G$ is class-2 nilpotent if $G / Z(G)$ is Abelian, where $Z(G)$ denotes the center of $G$, the set of elements that commute with every element of $G$. 
As our main application of the \TIlong complexity classes over $\Z/p^e\Z$, we give the first $N^{o(\log N)}$ time algorithm for testing isomorphism of two class-2 nilpotent groups of odd orders, where $N$ is the order of the two groups.

\begin{theorem}\label{thm:tensor_isomorphism_main}
There exists an algorithm that determines the isomorphism of two nilpotent class-2 groups of odd order in time $N^{O((\log N)^{8/9})}$, where $N$ is the order of the two groups.
\end{theorem}

\subsection{Comments on previous works and techniques}\label{sec:issue}

\paragraph{Equivalence between five actions on $3$-way arrays.} The equivalence of the actions in Definition~\ref{def:5action} was shown over fields in \cite{FGS19,TI1,GQT,TI4}, with two types of gadgets having been designed to achieve this. The first type of gadget was first introduced in \cite{FGS19} and further developed in \cite{TI1,GQT}. For convenience, we call this gadget the Furtorny--Grochow--Sergeichuk gadget, or FGS gadget for short. The second type of gadget was introduced in \cite{TI4}, and we call such gadget the Grochow--Qiao gadget, or GQ gadget for short. 

To use the FGS gadget in the commutative ring setting, one problem arises, namely it relies on the Krull--Schmidt theorem for quiver representations in the correctness proof \cite{FGS19}. Indeed, to use it in the orthogonal and unitary group setting, a version of Krull-Schmidt theorem for orthogonal and unitary representations of quivers was required \cite{TI3}, and such a result was shown by Sergeichuk \cite{Ser98}. However, over commutative rings, Krull--Schmidt type results are only known for certain families of commutative rings \cite{Krull-Schmidt}.

To use the GQ gadget in the commutative ring setting, there is an issue caused by the so-called non-degeneracy assumption, which is closely related to some distinctions between fields and commutative rings seen in Section~\ref{subsec:intriguing}. 
More specifically, consider the field setting. If $A \in \M(m \times n, \F)$ with $m \leq n$ is full-rank (i.e., $\rk(A) = m$), then $TA = 0$ if and only if $T =0$, for any $m \times m$ matrix $T$. Furthermore, if $A$ is not full rank, we can efficiently compute $A' \in \M(r \times n)$ such that the rowspan of $A$ and $A'$ are the same, but $A'$ is now full rank. Similarly, for equivalence of matrices under left-and-right column operations, the orbits over fields are completely determined by rank; if $A,B$ are two matrices, and we build $A',B'$ as in the preceding, then $A$ and $B$ are equivalent if and only if $A',B'$ are equivalent, but $A',B'$ are now full rank. Being full rank is a kind of ``non-degeneracy'' assumption, and over fields, reduced row echelon form generally lets us reduce from the general case to the non-degenerate case. 

In contrast, if $A, B\in \M(m\times n, \rng)$ where $\rng$ is a commutative ring, the above procedure fails due to the lack of some basic notions such as ranks and the relation between kernels and bases. For example, the inner ranks (see Section~\ref{subsec:rank-notion})of $A=\begin{bmatrix}
	2 & 0 & 0 \\
	0 & 2 & 0 
\end{bmatrix}$ and $B=\begin{bmatrix}
	3 & 0 & 0 \\
	0 & 3 & 0
\end{bmatrix}$ over $\Z/6\Z$ are both $2$, but they are not equivalent. We also note that $\ker(A)$ is non-zero, while the submodule spanned by the rows of $A$ is not cyclic (i.e. not generated by $1$ element). Several reductions in \cite{TI1,GQT} make use such non-degeneracy assumptions. 

By a new combination of the FGS and GQ gadgets, we are able to avoid the need for such non-degeneracy assumptions, to generalize the reductions to a much wider class of commutative rings.

\paragraph{From isomorphism of $\Z^2\otimes\Z^2\otimes\Z^2$ to principal decision.} For a reduction from $\Z^2\otimes\Z^2\otimes\Z^2$ isomorphism to the \algprobm{Principal Ideal Decision} problem, we make use of a well-known invariant called discriminants. We first show that if the tensors are of square discriminants (which are considered a ``degenerate'' case in the setting of number theory), isomorphism can be decided in polynomial time (Lemma~\ref{lem:tensorZ}). For the case of non-square discriminant, which is the more interesting and complicated setting from a number theoretic viewpoint, we use Bhargava's correspondence between isomorphism classes of $\Z^2\otimes\Z^2\otimes\Z^2$ and the so-called balanced triples of oriented ideals in a quadratic number field order. This correspondence can be made constructive and paves the way to the \algprobm{Principal Ideal Decision} problem. 

\paragraph{Isomorphism testing for nilpotent class-2 groups.} 
Our $N^{O((\log N)^{8/9})}$-time algorithm for testing isomorphism of nilpotent class-2 groups of odd order first applies Bear's correspondence~\cite{Bae38, khukhro,naik} to reduce the group isomorphism problem to a Lie-ring isomorphism problem. In Section~\ref{sec:group_iso_to_tensor_iso}, we then generalize reductions from \cite{TI4} to work over $\Z/p^e \Z$-Lie rings---even in the non-free-module case---and by a small trick we then reduce from Lie rings whose underlying Abelian group is arbitrary odd order, to Lie rings whose underlying Abelian group is a free $\Z/p^e \Z$-module, that is, of the form $(\Z/p^e \Z)^n$ for some $n$. 

Next, we generalize recent work on tensor isomorphism over a finite field $\mathbb{F}_p$~\cite{Sun23,IMQSZ24} to tensor isomorphism over a finite ring $\mathbb{Z} / p^e \mathbb{Z}$. One challenge in this generalization is to ensure that the algorithm's efficiency depends not only on the tensor's dimension. For instance, let $G \cong (\oplus_{i = 1}^{n-1} \Z/p\Z) \oplus \Z / p^e\Z$, $H \cong (\oplus_{i = 1}^{m-1} \Z / p \Z)\oplus \Z/p^e\Z$, $G' \cong \oplus_{i = 1}^{n} \Z/p\Z$, and $H' \cong \oplus_{i = 1}^{m} \Z/p\Z$. Although two tensors encoding bilinear maps $G \times G \rightarrow H$ and $G' \times G' \rightarrow H'$  share the same dimension, the complexity of isomorphism testing differs. This is because the bilinear map $G' \times G' \rightarrow H'$ encodes more information than $G \times G \rightarrow H$.
For the purpose of group isomorphism, it requires a running time that reflects the underlying complexity of the bilinear map encoded. This complication is the cause of the $8/9$ in the exponent in Theorem~\ref{thm:tensor_isomorphism_main}, compared to the $5/6$ \cite{Sun23} and $1/2$ \cite{IMQSZ24} of previous work.


In our solution, we handle matrix slices of the tensor based on their orders under matrix addition. 
We divide the tensor into multiple parts to decouple the correlation between matrices of high and low orders. 
For the high-order matrices, we enumerate a few matrices to establish the correspondence between high-order matrices in the two tensors. 
For the low-order matrices, we follow the strategy from the previous work over finite fields~\cite{Sun23,IMQSZ24}.
To achieve the latter, we extend the individualization and refinement technique, as well as the low-rank matrix characterization method from over finite fields~\cite{Sun23,IMQSZ24}, to finite rings. This generalization combines various measures of matrices over rings, rather than relying solely on inner rank. For example, in~\cite{Sun23}, the low-rank characterization critically depends on the fact that over a finite field, for a matrix $X$ and a row vector $x$, if $x$ is not in the row-span of $X$, then the rank of $\begin{bmatrix} X \\ x \end{bmatrix}$ is greater than the rank of $X$. However, over a ring, the rank of $\begin{bmatrix} X \\ x \end{bmatrix}$ may be the same as the rank of $X$. Therefore, in our approach, we sometimes use the number of distinct elements in the row-span of a matrix to track progress instead of inner rank.

As in \cite{Sun23, IMQSZ24}, we ultimately reduce to the \algprobm{Matrix Tuple Isometry} problem, but now over $\Z/p^e \Z$ instead of over a finite field of odd order. Solving this problem in polynomial time over odd order fields was one of the main results of Ivanyos \& Qiao \cite{IQ19}. We generalize their result to work over $\Z/p^e \Z$, and in fact over arbitrary finite local principal ideal rings of odd characteristic, which may have further applications to codes over such rings. The first part of the generalization requires using Smith Normal Form to reduce to the non-degenerate case (while this does not work over arbitrary commutative rings, it does work over $\Z/p^e \Z$ and finite PIRs). We then prove several algebraic lemmas about $\Z/p^e \Z$ that allow us to generalize their techniques to our setting. Finally, in the end the problem is reduced to a certain problem in semisimple $*$-algebras, and such algebras are algebras over a finite field, even if the algebras we started with were over $\Z/p^e \Z$ (or a finite local PIR). Thus, at that point, we can apply some of the results of \cite{IQ19} directly.

We note that for even-order nilpotent groups, not only is the group-theoretic situation itself more complicated, but it even remains open to just do the last step---solve \algprobm{Matrix Tuple Isometry}---over \emph{fields} of characteristic 2 (let alone rings) in polynomial time.

\section{Preliminaries}

\subsection{Rings and modules}

We review some basic notions of rings and modules that are used in this article. 

Let $\rng$ be a commutative unital (=has a multiplicative identity) ring. Let $0$ and $1$ denote the additive and multiplicative identities, respectively. For a non-zero $r\in \rng$, $r$ is a unit if there exists $r'$ such that $rr'=1$. It is a zero-divisor if there exists $r'$ such that $rr'=0$. It is nilpotent if there exists $n\in \N$ such that $r^n=0$. A ring is a domain if it does not contain zero-divisors.

A (left-)module $M$ over $\rng$ is an abelian group with a multiplication from $\rng$, denoted as $\cdot:\rng\times M\to M$, satisfying certain natural conditions. A submodule of $M$ is a subgroup of $M$ closed under the multiplication of $\rng$. For $S\subseteq M$, the module generated by $S$ is the smallest submodule of $M$ containing $S$, denoted by $\langle S\rangle$. We say that $S$ is a generating set of $M$ if $\langle S\rangle=M$. Note that $\rng$ is naturally an $\rng$-module.

A free module is any module isomorphic to $\rng^n$ for some $n$, where $\rng^n$ consists of length-$n$ row vectors over $\rng$, and the multiplication by $\rng$ is the usual ``scalar-vector coordinate-wise'' multiplication $r \cdot (r_1, \dotsc, r_n) = (rr_1, \dotsc, rr_n)$. 

Tensor products, symmetric powers, and exterior powers of modules over commutative rings are classical \cite{AM94}. Let $\rng$ be a commutative ring. Let $M$ and $N$ be $\rng$-modules. We use $M\otimes N$, $\S^k(M)$, and $\Lambda^k(M)$ to denote the tensor product of $M$ and $N$, the $k$th symmetric power of $M$, and the $k$th exterior power of $M$.

A subring $S$ of $\rng$ is a subset of $\rng$ closed under addition and multiplication. An ideal $I$ of $\rng$ is a subring that is also closed under multiplication from $\rng$, that is, it is a submodule of $\rng$ as an $\rng$-module. Given an ideal $I$, the quotient ring $\rng/I$ is naturally defined. We say that $\rng$ is local, if it has a unique maximal ideal. An ideal $I$ of $\rng$ is principal if it can be generated by a single element. We say that $\rng$ is a principal ideal ring (PIR) if every ideal is principal. 

The annihilator of $r \in \rng$ is $\Ann_\rng(r) = \{s \in \rng : sr = 0\}$. For a subset $T \subseteq \rng$, its annihilator is $\Ann_\rng(T) = \bigcap_{r \in T} \Ann_\rng(r) = \{s \in \rng : st = 0 \forall t \in T\}$. For $r \in \rng, T \subseteq \rng$, $\Ann_\rng(r)$ and $\Ann_\rng(T)$ are both ideals. Sometimes we may write just $\Ann(r), \Ann(T)$, if $\rng$ is clear from context.

\subsection{Matrices and tensors over commutative rings}

In this article our base ring is commutative with a multiplicative identity. 

Let $\rng$ be such a ring. The set of $m\times n$ matrices over $\rng$ is denoted by $\M(m\times n, \rng)$. When $n=m$, we may write $\M(n, \rng)$ for $\M(m\times n, \rng)$. Let $\idmat_n$ be the $n\times n$ identity matrix in $\M(n, \rng)$. 

We say that $A\in\M(n, \rng)$ is \emph{invertible}, if there exists $A'\in \M(n, \rng)$, such that $AA'=A'A=\idmat_n$. Let $\GL(n, \rng)$ be the group of $n\times n$ invertible matrices over $\rng$. 

For $A\in \M(n, \rng)$, the determinant of $A$ is denoted by $\det(A)$. It is a basic fact that $A\in \GL(n, \rng)$ if and only if $\det(A)$ is a unit in $\rng$. 

The set of length-$n$ row vectors over $\rng$ is denoted by $\rng^n$. The set of length-$n$ column vectors over $\rng$ is denoted by $^n\rng$.

For $A\in \M(m\times n, \rng)$, the row module is the submodule of $\rng^n$ consisting of $\rng$ combinations of rows of $A$. The column module is defined similarly. The left kernel of $A$ is the submodule of $\rng^m$ consisting of $v\in\rng^m$ such that $vA=0$. The right kernel of $A$ is the submodule of $^n\rng$ consisting of $w\in{}^n\rng$ such that $Aw=0$.

\paragraph{Block matrices over commutative rings.} 
For two matrices $A \in \M(m_1 \times n_1, \rng), B \in \M(m_2 \times n_2, \rng)$, we use $A \oplus B$ to denote the block-diagonal matrix
\[
A \oplus B := \begin{pmatrix} 
A & 0 \\
0 & B 
\end{pmatrix},
\]
of size $(m_1 + m_2) \times (n_1 + n_2)$.

Let $A=\begin{bmatrix}
	A_{11} & A_{12}\\
	0 & A_{22}
\end{bmatrix}\in\GL(n, \rng)$, where $A_{11}\in\M(n_1, \rng)$ and $A_{22}\in\M(n_2, \rng)$. Then $A_{11}$ and $A_{22}$ are also invertible, that is, $\det(A_{11})$ and $\det(A_{22})$ are units in $\rng$. To see this, let $\alpha=\det(A)$, $\beta=\det(A_{11})$, and $\gamma=\det(A_{22})$. So $\alpha=\beta\gamma$. As $\alpha$ is a unit, there exists $\alpha^{-1}$, such that $\alpha\cdot\alpha^{-1}=1$. So $(\beta\gamma)\alpha^{-1}=\beta(\gamma\alpha^{-1})=1$, so $\beta$ is also a unit.

\subsection{Rank notions of matrices over rings}\label{subsec:rank-notion}

Over fields, there are several equivalent definitions of the rank of a matrix, but some of these become inequivalent over more general rings. In this paper we will make use of two different notions of rank for rings over $R$. %
%
%
%
%
%
We refer the reader to \cite{Coh85} for free ideal rings over which ranks of modules could be defined. Here we only make use of the following rank notions. Let $A\in\M(m\times n, \rng)$.
\begin{enumerate}
	
	
	\item The \emph{inner rank} of $A$, $\irk(A)$, is the smallest $r\in\N$ such that $A$ can be written as $\sum_{i\in[r]}u_iv_i^t$, where $u_i\in \rng^m$ and $v_i\in\rng^n$. 
	
	Or equivalently, the smallest $r\in \N$ such that $A=BC$ where $B\in\M(m\times r, \rng)$ and $C\in\M(r\times n, \rng)$. 
	
\item For $t\in \min\{m, n\}$, let $I_t(A)$ be the ideal of $\rng$ 
generated by $t\times t$ minors of $A$. For $t=0$, let $I_0(A)=\rng$. The \emph{McCoy rank} of $A$, $\mcrk(A)$, is the largest $r\in \{0, 1, \dots,  \min\{m, n\}\}$ such that the annihilator of $I_r(A)$ is the zero ideal \cite{McCoy-Rank}. 
\end{enumerate}


To decide between inner and McCoy rank, the following example \cite[Section 4.2, Example 3]{inner-rank-comp} is useful. Consider $A=\diag(p, \dots, p)\in \M(n, \Z/p^{n+1}\Z)$. Then $\mcrk(A)=0$, as for $1\leq t\leq n$, $I_t(A)$ is generated by $p^t$, so its annihilator is generated by $p^{n+1-t}$. As seen below, $\irk(A)=n$, and this turns out to be useful for our purpose. 

However, McCoy rank is also useful in our context as it is closely related to the solvability of equations. Let $A\in \M(m\times n, \rng)$, and let $\vec{x}=(x_1, \dots, x_n)^t$. Then $Ax=\vec{0}$ has a non-zero solution over $\rng$ if and only if $\mcrk(A)<n$. In particular, when $A\in\M(n, \rng)$, $Ax=\vec{0}$ has 
non-zero solutions if and only if $\det(A)$ is a zero-divisor.

\subsection{Some basic facts of inner ranks}

Let $\rng$ be a commutative ring. Some basic properties of inner ranks are as follows \cite[pp. 248]{Coh85}. 
\begin{fact}\label{fact:inner-rank}
	Let $A \in \M(m \times n, \rng)$. Let $L\in\GL(m, \rng)$ and $R\in\GL(n, \rng)$. 
	\begin{enumerate}
		\item $0\leq \irk(A) \leq \min\{m,n\}$.
		\item $\irk(A\oplus B)\leq \irk(A)+\irk(B)$.
		\item $\irk(AB)\leq \min\{\irk(A), \irk(B)\}$. 
		\item $\irk(\begin{bmatrix}
			A & B
		\end{bmatrix})\geq \max\{\irk(A), \irk(B)\}$ and $\irk(\begin{bmatrix}
			A \\ B
		\end{bmatrix})\geq \max\{\irk(A), \irk(B)\}$.
		\item $\irk(A)=\irk(LAR)$.
	\end{enumerate}
\end{fact}

The inner rank is also equal 
and 
to the least $r$ such that row-span of $A$ (that is, the $\rng$-submodule of $\rng^n$ generated by the rows of $A$) is contained in a submodule of $\rng^n$ that is generated by $r$ elements, and similarly for the column-span of $A$ \cite[Proposition~5.4.3]{Coh85}. The latter characterization cannot, in general, be strengthened to $\rho(A)=$ the minimum number of generators of the row-span of $A$ itself; see Example~\ref{ex:nonreduced} for a counterexample.


%

\begin{example}[Inner rank can be strictly subadditive]\label{ex:Z6}
	For $\rng = \Z/6\Z$, consider the matrix $A = \begin{pmatrix} 2 & 0 \\ 0 & 3 \end{pmatrix}$. We have $A = \begin{pmatrix} 4 \\ 3 \end{pmatrix} \begin{pmatrix} 2 & 3 \end{pmatrix}$, so $\rho(A) \leq 1$ (in fact, it is equal to $1$, since only the zero matrix has inner rank zero). Thus, we do not in general have $\rho(B \oplus C) = \rho(B) + \rho(C)$. 
\end{example}

We recall the following classical definitions, see, e.\,g., \cite[pp.~6--7]{Coh85}. 
\begin{enumerate}
	\item Let $\rng$ be a ring that is not necessarily commutative. We say that $\rng$ has the \emph{unbounded generating number} property (UGN), if the following holds: for any $A\in\M(m\times n, \rng)$ and $B\in\M(n\times m, \rng)$, if $AB=\idmat_m$, then $n\geq m$. Or equivalently, 
	$\rng$ has UGN if
	$\rho(\idmat_n)=n$.\footnote{The name comes from the fact that an equivalent condition is that there is no constant upper bound on the number of generators of all $R$-modules.}
	\item Let $\rng$ be a ring that is not necessarily commutative. We say that $\rng$ is \emph{stably finite} (sometimes called ``weakly finite''), if the following holds: for any $A, B\in \M(n, \rng)$, if $AB=\idmat_n$, then $BA=\idmat_n$.
\end{enumerate}
An observation due to P. M. Cohn \cite{Coh85} is that stable finiteness implies UGN. 
Stably finite rings include all 
commutative rings, Artinian rings, and Noetherian rings.

Following \cite[pp. 248]{Coh85}, we say that a matrix $A\in\M(n, \rng)$ is \emph{full} if $\irk(A)=n$, and it was also noted there that for a ring with UGN, every $A\in \GL(n, \rng)$ is full. 

\begin{lemma}\label{lem:upper-tri}
	Let $\rng$ be a commutative ring. Let $a, b, c\in \N$. Let $A\in \M(a, \rng)$ satisfy that $\mcrk(A)=a$. Suppose $(P, Q)\in\GL(a+b, \rng)\times\GL(a+c, \rng)$ satisfies that 
	\begin{equation}\label{eq:upper-tri}
		P\begin{bmatrix}
			A & 0 \\
			0 & 0_{b\times c}
		\end{bmatrix}Q^t=\begin{bmatrix}
			A & 0 \\
			0 & 0_{b\times c}
		\end{bmatrix}.
	\end{equation} Then $P=\begin{bmatrix}
		P_{11} & P_{12}\\
		0_{a\times b} & P_{22}
	\end{bmatrix}$, $Q=\begin{bmatrix}
		Q_{11} & Q_{12}\\
		0_{a\times c} & Q_{22}
	\end{bmatrix}$.
\end{lemma}
\begin{proof}
	Let $P=\begin{bmatrix}
		P_{11} & P_{12}\\
		P_{21} & P_{22}
	\end{bmatrix}$, $Q=\begin{bmatrix}
		Q_{11} & Q_{12}\\
		Q_{21} & Q_{22}
	\end{bmatrix}$. Then Equation~\ref{eq:upper-tri} implies that 
	$$
	\begin{bmatrix}
		P_{11}AQ_{11}^t & P_{11}AQ_{21}^t\\
		P_{21}AQ_{11}^t & P_{21}AQ_{21}^t
	\end{bmatrix}
	=
	\begin{bmatrix}
		A & 0 \\
		0 & 0_{b\times c}.
	\end{bmatrix}
	$$
	This implies that
	$P_{11}AQ_{11}^t=A$. As $\mcrk(A)=a$, $\det(A)=\det(P_{11}AQ_{11}^t)$ is a non-zero-divisor. It follows that each of $\det(P_{11})$, $\det(A)$, $\det(Q_{11}^t)$ are all non-zero-divisors. Therefore, $\mcrk(AQ_{11}^t)=a$, so by $P_{21}AQ_{11}^t=0$, we have $P_{21}=0$. To show that $Q_{21}=0$ follows the same line of argument.
\end{proof}

Inner ranks may have strange behaviours. For example, non-zero scalar matrices may not have full inner ranks.

\begin{example}[Inner ranks of (non-zero) scalar matrices over commutative Noetherian rings] \label{ex:nonreduced}
	Let $R = \F[x,y,z] / \langle x^2 - z, y^2 - z, xy \rangle$, where $\F$ is a field (though $\F=\Z$ would work as well). Since $x^2 = y^2 = z$ in $R$, but $xy=0$, we have $xz = xy^2 = 0$ and similarly $yz=0$, and $z^2 = x^2 y^2 = 0$. A little Gröbner basis calculation gives us that $\{x^2-z, y^2-z, xy, yz, xz, z^2\}$ is a Gröbner basis for the defining ideal here, and thus $\{1,x,y,z\}$ is an $\F$-basis of $R$ as a vector space. But we have $\begin{pmatrix} x \\ y \end{pmatrix} \begin{pmatrix} x & y \end{pmatrix} = z I_2$, hence $\rho(z I_2) = 1$, so a non-zero scalar matrix may not have full inner rank. This is a counterexample to scalar matrices of full inner rank for many classes of rings as, in addition to being commutative Noetherian, $R$ is local Artinian, every non-unit is a zero divisor, and every zero divisor is nilpotent. (To see that it is local, note that $x^4 = y^4 = z^2 = 0$, so geometrically this is a ``fat point at the origin;'' the unique maximal ideal is $\langle x,y \rangle = Span_\F\{x,y,z\}$.) We note that in this example, it is also the case that $row(z I_2)$ is not generated by 1 element despite $\rho(z I_2)=1$. (This example can be generalized to give a ring in which $\rho(z I_n) = n-1$ but that $\rho(r I_m) = m$ for all nonzero $r$ and all $m < n$.)
\end{example}

At some places, we may require that the inner rank of $A\in \M(\ell\times m, \rng)$ is efficiently computable. The inner rank of matrices over a principal ideal domain can be computed efficiently via the Smith Normal Form \cite{SmithNF,SmithNFJournal}. In \cite{inner-rank-comp}, it is shown that the inner rank of matrices over a finite commutative principal ideal ring can be computed in polynomial time.

\subsection{Finite local principal ideal rings} \label{sec:PIR}


Let $\rng$ be a finite local PIR. We will compute with modules over finite local PIRs and their tensor products in Section~\ref{subsec:five-action-PIR} and Section~\ref{subsec:tuple-isometry}. Let us first record some examples and notations.
\begin{itemize}
	\item 
	Examples of such rings include $\Z/p^e \Z$, $\F[x] / x^n$ where $\F$ is a finite field, and Galois rings.
	
	
	\item The unique maximal ideal of $\rng$ is denoted $\m$.
	
	\item We use $p$ to denote the characteristic of $\rng/\m$.
	
	\item We use $e$ to denote the index of nilpotency of $\m$: $\m^e = 0$ but $\m^{e-1} \neq 0$. (Since $R$ is finite and local, $\m$ is necessarily nilpotent.)
	
	\item When working over a ``general'' finite local PIR, we treat basic arithmetic operations in $R$ as unit cost (or, equivalently, we work in the BSS model over $R$ \cite{BSS}). We also assume that division in $R$ is unit cost: given $a,b \in R$, decide if $a$ divides $b$, and if so, output some $c$ such that $ac = b$. We also assume that we have access to $t$, a generator of the maximal ideal. We refer to algorithms in this model as $R$-algorithms. 
	
	\item We also assume that elements $x \in R/\m^i$ are represented computationally by elements $\overline{x} \in R$ such that $\overline{x} \equiv x \pmod{\m^i}$, and the algorithm simply keeps track of when it is working modulo $\m^i$. For example, to test whether $\overline{x}, \overline{y}$ are equivalent modulo $\m^i$, an $R$-algorithm can simply check whether $\overline{x} - \overline{y}$ is divisible by $t^i$. 
		
\end{itemize}

Smith Normal Form is efficiently computable by an $\rng$-algorithm, by the standard ``Gaussian-elimination-style'' algorithm, for matrices over a finite local PIR $\rng$.\footnote{Lacking a convenient reference that does not assume either the integers or at least a domain, we outline it here. Since a local commutative PIR is a quotient of a PID \cite{hungerford}, the algorithm does the same operations as though one is over a PID, keeping track of which elements are equal or zero in the quotient PIR. In outline: standard row operations can make the (1,1) entry into the GCD of the entries in the first column, and then multiples of the first row can be added to the remaining rows to zero out the rest of the first column. This procedure can be repeated on the second column, to make the (2,2) entry the GCD of the new entries of the second column at or below the second row. And so on. The result is an upper triangular matrix. Now we repeat the analogous procedure but using column operations, resulting in a diagonal matrix.} In many cases of interest, including $\rng = \Z/p^e \Z$, 
ring operations in $\rng$ are computable in $\poly(\log|\rng|)$ time on an ordinary Turing machine. With a little more work this is probably also true over arbitrary finite local PIRs, but we will not need this in our paper.

An $\rng$-algebra $A$ is a unital ring $A$ together with an injective ring homomorphism $\iota \colon \rng \to A$, such that the image of $\iota$ (which includes the multiplicative identity $1 \in A$) is contained in the center of $A$. Because of this, the underlying additive group of $A$ is an $\rng$-module. A \emph{$*$-algebra over $\rng$} is an $\rng$-algebra $A$ together with an anti-automorphism $* \colon A \to A$ of order at most 2, that is $(a+b)^* = a^* + b^*$, $(ab)^* = b^* a^*$, and $(a^*)^* = a$ for all $a \in A$. 

Given an $R$-module $M$, its endomorphism ring $\End(M)$ consists of all $\rng$-module homomorphisms $M \to M$. In this ring, addition is pointwise---$(f+g)(x) = f(x) + g(x)$---and multiplication is given by the composition of homomorphisms. A commutative ring has all its modules direct sums of cyclic modules (this condition is sometimes called being a ``Köthe ring'') if and only if it is an Artinian PIR \cite{CohenKaplansky, Kothe}. In particular, this holds for our ring $\rng$ by assumption (since finite implies Artinian), and thus every $\rng$-module has the form $M \cong \bigoplus_i R/\m^{e_i}$ for some integers $1 \leq e_i \leq e$. If $M$ is finitely generated, there are only finitely many such summands, say $n$ of them. 

The following theorem is the natural generalization of Hillar \& Rhea's description of the endomorphism ring and automorphism group of Abelian $p$-groups \cite{HillarRhea} (originally proved by Ranum 100 years earlier \cite{Ranum}) to the endomorphism ring and automorphism group of $\rng$-modules, when $\rng$ is a commutative local PIR. 

\begin{theorem} \label{thm:HillarRheaPIR}
	Let $\rng$ be a commutative local PIR with maximal ideal $\m = \langle t \rangle$. For $M = \bigoplus_{i=1}^n \rng/\m^{e_i}$, $\End(M)$ is isomorphic to the ring of $n \times n$ matrices whose $(i,j)$ entry is an element of $\Hom(\rng/\m^{e_j}, \rng/\m^{e_i})$ for all $i,j \in \{1,\dotsc,n\}$. 
	
	Furthermore, this is isomorphic to the ring of $n \times n$ matrices whose $(i,j)$ entry is an element of $\rng/\m^{e_i}$, that is divisible by $t^{e_i - e_j}$ whenever $e_i > e_j$. Such a matrix is invertible if and only if it is invertible when every entry is taken modulo $\m$.
\end{theorem}

As the proof is nearly identical to that of Hillar and Rhea in the case of $\rng = \Z/p^e \Z$ \cite{HillarRhea}, we place it in Appendix~\ref{app:similar}.

For an $\rng$-module $M$, a \emph{basis} is a generating set $m_1, \dotsc, m_n$ with the property that $\langle m_i \rangle \cap \langle m_j : j \neq i \rangle = 0$ for all $i=1,\dotsc,n$, where we use $\langle$ angle brackets $\rangle$ to denote the generated $\rng$-module. In particular, if $\{m_1,\dotsc,m_n\}$ is a basis of $M$, we have $M = \bigoplus_{i=1}^n \langle m_i \rangle$, and $M$ is a direct sum of cyclic $\rng$-modules. Over general commutative rings, not every finitely generated module has such a basis (e.g., the ideal $\langle x, y\rangle \subseteq \F[x,y]$, as an $\F[x,y]$-module), but because modules over commutative PIRs are direct sums of cyclic modules, every module over our ring $\rng$ has a basis. 



\section{Equivalence of tensor isomorphism problems}

\subsection{A review of two gadgets for partition tensor isomorphism}

A key problem in the development of the complexity class $\cc{TI}$ is to reduce \emph{partitioned} tensor isomorphism reduces to ordinary \TI \cite{FGS19}. Two gadgets have been developed to achieve the reduction, the first one in \cite{FGS19}, and the second one in \cite{TI4}. 

To use these gadgets for commutative rings instead of fields, some issues arise, and as a result, we need to combine ideas from both gadget designs. 

In this subsection, we shall review the partitioned tensor isomorphism problem, review the two gadgets, demonstrate some issues for using them in commutative rings, and then explain on how to overcome them.

\paragraph{The partitioned tensor isomorphism problem.} Briefly speaking, the partitioned tensor isomorphism problem asks for isomorphisms between tensors whose indices in one or more directions are partitioned into blocks that must be preserved. 

\begin{definition}\label{def:pTI}
	Given $\tA, \tB\in \T(n\times m\times \ell, \F)$, and a partition of $[\ell]=D_1\cup D_2\cup \dots \cup D_L$, the \algprobm{Partitioned Tensor Isomorphism} (\pTI) problem asks whether there exists $(P, Q, R)$ that sends $\tA$ to $\tB$, where $P\in \GL(n, \F)$, $Q\in\GL(m, \F)$, and $R\in\GL(\ell_1, \F)\times\dots\times\GL(\ell_L, \F)$ viewed as a subgroup of block-diagonal matrices in $\GL(\ell, \F)$ respecting the partition.
\end{definition}
One can of course consider partitions in all three directions, but for clarity of exposition we focus just on one direction in this section. This problem can be viewed as the tensor analogue of the classical \emph{colored} graph isomorphism problem \cite{KST93}; see \cite{TI4} for a discussion.


\subsubsection{The Futorny--Grochow--Sergeichuk (FGS) gadget}\label{sec:FGS}

The first reduction from \pTI to \TI was done using a gadget designed in \cite{FGS19}, which we call the \emph{Futorny--Grochow--Sergeichuk gadget}, or FGS gadget for short. This gadget can be naturally viewed as corresponding to the star graph gadget, see e.g. the discussion in \cite[Sec.~3]{TI2}.

\begin{example}[An example of the FGS gadget]\label{ex:FGS}
To see how the FGS gadget works, consider the following example. By slicing along the third index, a $3$-way array $\tA\in\T(n\times m\times 6, \F)$ can be represented as a tuple of six matrices $\vA=(A_1, \dots, A_6)$. Suppose $\ell=6$ and $L=3$, $n\leq m$, with $[6]$ partitioned into $\{1, 2\}, \{3, 4\}, \{5, 6\}$. 

To reduce this \pTI instance to \TI following \cite{FGS19}, we construct 
\[\tilde A_1=\begin{bmatrix}
	A_1 & 0 & 0 & 0 & 0 & 0 & 0 \\
	0 & \idmat_{n+1} & 0 & 0 & 0 & 0 & 0 \\
	0 & 0 & 0 & 0 & 0 & 0 & 0 \\
	0 & 0 & 0 & 0 & 0 & 0 & 0 \\
\end{bmatrix}, 
\tilde A_2=\begin{bmatrix}
	A_ 2& 0 & 0 & 0 & 0 & 0 & 0 \\
	0 & 0 & \idmat_{n+1} & 0 & 0 & 0 & 0 \\
	0 & 0 & 0 & 0 & 0 & 0 & 0 \\
	0 & 0 & 0 & 0 & 0 & 0 & 0 \\
\end{bmatrix},\]
\[
\tilde A_3=\begin{bmatrix}
	A_3 & 0 & 0 & 0 & 0 & 0 & 0 \\
	0 & 0 & 0 & 0 & 0 & 0 & 0 \\
	0 & 0 & 0 & \idmat_{2n+2} & 0 & 0 & 0 \\
	0 & 0 & 0 & 0 & 0 & 0 & 0 \\
\end{bmatrix},
\tilde A_4=\begin{bmatrix}
	A_4 & 0 & 0 & 0 & 0 & 0 & 0 \\
	0 & 0 & 0 & 0 & 0 & 0 & 0 \\
	0 & 0 & 0 & 0 & \idmat_{2n+2} & 0 & 0 \\
	0 & 0 & 0 & 0 & 0 & 0 & 0 \\
\end{bmatrix}, \]
\[
\tilde A_5=\begin{bmatrix}
	A_5 & 0 & 0 & 0 & 0 & 0 & 0 \\
	0 & 0 & 0 & 0 & 0 & 0 & 0 \\
	0 & 0 & 0 & 0 & 0 & 0 & 0 \\
	0 & 0 & 0 & 0 & 0 & \idmat_{4n+4} & 0 \\
\end{bmatrix}, 
\tilde A_6=\begin{bmatrix}
	A_6 & 0 & 0 & 0 & 0 & 0 & 0 \\
	0 & 0 & 0 & 0 & 0 & 0 & 0 \\
	0 & 0 & 0 & 0 & 0 & 0 & 0 \\
	0 & 0 & 0 & 0 & 0 & 0 & \idmat_{4n+4} \\
\end{bmatrix}.
\]
Let $\tilde\tA=(\tilde A_1, \dots, \tilde A_6)$. By \cite[Lemma 2.2]{FGS19}, $\tA$ and $\tB$ in $\T(n\times m\times 6, \F)$ are partitioned isomorphic if and only if $\tilde\tA$ and $\tilde\tB$ are isomorphic. 

The basic argument for the FGS gadget is based on ranks. That is, we have $\rk(\tilde A_1)$ and $\rk(\tilde A_2)$ are between $n+1$ and $2n+1$, $\rk(\tilde A_3)$ and $\rk(\tilde A_4)$ are between $2n+2$ and $3n+2$, and $\rk(\tilde A_5)$ and $\rk(\tilde A_6)$ are between $4n+4$ and $5n+4$. Furthermore, the rank of any linear combination of $\tilde A_1, \dots, \tilde A_4$ is at most $4n+3$. As ranks are preserved by the left-right action, it follows that $T\in \GL(6, \F)$ appearing in an isomorphism from $\tilde \tA$ to $\tilde\tB$ has to be of the form $\begin{bmatrix}
	T_{11} & & \\
	& T_{22} & \\
	& & T_{33}
\end{bmatrix}$, that is, respecting the partitions.
\end{example}

One issue with the FGS gadget is that the sizes of $\tilde A_i$ are quadratic in the sizes of $A_i$. This is not evident in the example above, but the reader can easily see that if $[\ell]$ is partitioned into $[\ell-1]$ and $\{\ell\}$, then $I_s$ needs to be repeated $\ell-1$ times, so the sizes of $\tilde A_i$ would be of $(n+n+2n)\times (m+(\ell-1)n+2n)$. The next gadget avoids this quadratic blow up.

\subsubsection{The Grochow--Qiao (GQ) gadget}\label{sec:GQ}

The second gadget for reducing \pTI to \TI appeared in \cite{TI4}, which we call the \emph{Grochow--Qiao} gadget, or GQ gadget for short. The GQ gadget combines the rank idea in the FGS gadget with a basic linear algebra fact. Its main advantage is that the blow-up factor is linear instead of quadratic. 

\begin{example}[An example for the GQ gadget]\label{ex:GQ}
To see how the GQ gadget works, we use the same example in Example~\ref{ex:FGS}, but presented in a ``flipped'' fashion. That is, consider a $3$-way array $\tC\in\T(6\times n\times m, \F)$ represented as a tuple of $m$ matrices $\vC=(C_1, \dots, C_m)$ where $C_i\in\M(6\times n, \F)$. Suppose $\ell=6$ and $L=3$, $n\leq m$, with $[6]$ partitioned into $\{1, 2\}, \{3, 4\}, \{5, 6\}$. That is, we consider the setting where the first index set (instead of the third one) is partitioned. But this is easily seen to be the same problem as in the FGS gadget setting by ``flipping'' the indices of the tensor $\tA\in \T(n\times m\times 6, \F)$. We do this just to make the notation for the GQ gadget clearer.

Let $C_i=\begin{bmatrix}
	C_{i, 1}\\
	C_{i, 2}\\
	C_{i, 3}
\end{bmatrix}$, where $C_{i, j}\in\M(2\times n, \F)$. 

To reduce this instance of \pTI to \TI following \cite{TI4}, we construct 
$$
\tilde C_i=\left[\begin{array}{ccccccc}
	C_{i,1} & 0 & 0 & 0 & 0 & 0 & 0 \\
	C_{i,2} & 0 & 0 & 0 & 0 & 0 & 0 \\
	C_{i,3} & 0 & 0 & 0 & 0 & 0 & 0 \\  
	0 & 0 & 0 & 0 & 0 & 0 & 0 \\
	0 & 0 & 0 & 0 & 0 & 0 & 0 \\
	0 & 0 & 0 & 0 & 0 & 0 & 0\\ 
	0 & 0 & 0 & 0 & 0 & 0 & 0 \\
	0 & 0 & 0 & 0 & 0 & 0 & 0 \\
	0 & 0 & 0 & 0 & 0 & 0 & 0 	
\end{array}\right] \qquad (i=1,\dotsc,m),
\quad
\tilde C_{m+1}=\begin{bmatrix}
	0 & \idmat_2 & 0 & 0 & 0 & 0 &0 \\
0 & 0 & 0 & 0 & 0 & 0 &0\\
0 & 0 & 0 & 0 & 0 & 0 &0 \\
\idmat_n & 0 & 0 & 0 & 0 & 0 &0 \\
0 & 0 & 0 & 0 & 0 & 0 &0\\
0 & 0 & 0 & 0 & 0 & 0 &0\\
0 & 0 & 0 & 0 & \idmat_{n+6} & 0&0 \\
0 & 0 & 0 & 0 & 0 & 0 &0\\
0 & 0 & 0 & 0 & 0 & 0 &0
\end{bmatrix},
$$
\begin{equation}\label{eq:GQ}
\tilde C_{m+2}=\begin{bmatrix}
	0 & 0 & 0 & 0 & 0 & 0 &0 \\
	0 & 0 & \idmat_2 & 0 & 0 & 0 &0\\
	0 & 0 & 0 & 0 & 0 & 0 &0 \\
	0 & 0 & 0 & 0 & 0 & 0 &0 \\
	\idmat_n & 0 & 0 & 0 & 0 & 0 &0\\
	0 & 0 & 0 & 0 & 0 & 0 &0\\
	0 & 0 & 0 & 0 & 0 & 0&0 \\
	0 & 0 & 0 & 0 & 0 & \idmat_{2(n+6)} &0\\
	0 & 0 & 0 & 0 & 0 & 0 &0
\end{bmatrix},
\tilde C_{m+3}=\begin{bmatrix}
	0 & 0 & 0 & 0 & 0 &0 & 0 \\
	0 & 0 & 0 & 0 & 0 &0 & 0\\
	0 & 0 & 0 & \idmat_2 & 0  &0 \\
	0 & 0 & 0 & 0 & 0 &0 & 0 \\
	0 & 0 & 0 & 0 & 0 &0 & 0\\
	\idmat_n & 0 & 0 & 0 & 0 & 0 &0\\
	0 & 0 & 0 & 0 & 0&0 & 0 \\
	0 & 0 & 0 & 0 & 0 &0 & 0\\
	0 & 0 & 0 & 0 & 0 &0 &  \idmat_{4(n+6)}
\end{bmatrix},
\end{equation}
Let $\tilde\tC=(\tilde C_1, \dots, \tilde C_m, \tilde C_{m+1}, \tilde C_{m+2}, \tilde C_{m+2})$. By \cite{TI4}, $\tC$ and $\tD$ in $\T(6\times n\times m, \F)$ are partitioned isomorphic if and only if $\tilde\tC$ and $\tilde\tD$ are isomorphic. 

We briefly explain some key arguments. To start with, note that $\idmat_{n+6}$, $\idmat_{2(n+6)}$, and $\idmat_{4(n+6)}$ serve as a similar role as in FGS gadget, namely they prevent $\tilde C_{m+1}$, $\tilde C_{m+2}$, and $\tilde C_{m+3}$ to be added to $\tilde C_i$ with $1\leq i\leq m$, and $\tilde C_{m+1}$, $\tilde C_{m+2}$, and $\tilde C_{m+3}$ themselves mixing with each other. However, a linear combination of $\tilde C_i$, $i\in[m]$, can be added to $\tilde C_{m+1}$ (and $\tilde C_{m+2}$, $\tilde C_{m+3}$). This is where $\idmat_{n}$ in $\tilde C_{m+1}$ comes to help, namely $\idmat_{n}$ can be used to cancel what is added to the upper-left $3$ blocks of size $2\times n$. This cancellation trick is one of the innovations in \cite{TI4} that allowed them to improve the size of the gadgets. After these cancellations, any $(P, Q)\in\GL(6+3n+7(n+6), \F)\times \GL(m+6+7(n+6), \F)$ appearing in an isomorphism from $\tilde \tC$ to $\tilde \tD$ satisfies that $P\tilde C_{m+j} Q^t=\tilde D_{m+j}=\tilde C_{m+j}$ for $j=1, 2, 3$. Then the $\idmat_2$ in $\tilde C_{m+1}$, $\tilde C_{m+2}$, and $\tilde C_{m+3}$ are used to ensure that the upper-left $6\times 6$ submatrix of $P$ is of the form $\begin{bmatrix}
	P_{11} & & \\
	& P_{22} & \\
	& & P_{33}
\end{bmatrix}$, that is, respecting the partitions. However, note that here $P_{ii}$ are not necessarily invertible because of blocks of $P$ that appear outside the first $3 \times 3$ blocks. This is resolved by assuming certain non-degeneracy conditions on the original $\tA$ and $\tB$, which is made possible by a reduction from degenerate to non-degenerate cases that works over fields using algorithms for reduced row echelon form. 
\end{example}

\subsection{Rings for which diagonal matrices are full}\label{subsec:diagonal-full}

Let $\rng$ be a commutative ring. We say that $\rng$ has the \emph{diagonal-full} property, if every diagonal matrix with all diagonal entries non-zero is of full inner-rank. We see in Example~\ref{ex:Z6} that $\Z/6\Z$ does not satisfy the diagonal-full condition. We will show that domains and local rings satisfy the diagonal-full condition.

\begin{fact}\label{fact:det}
	Let $A\in \M(n, \rng)$. If $\det(A)\neq 0$, then $A$ is full.
\end{fact}
\begin{proof}
	For the sake of contradiction, suppose $A$ is not full. Then there exist matrices $B \in \M(n \times (n-1), R)$ and $C \in \M((n-1) \times n,R)$ such that $BC = A$. The determinant over a commutative ring is still multi-linear in the rows, so writing the $i$-th row of $BC$ as an $R$-linear combination of the rows of $C$, we can expand out the determinant of $BC$ as a sum of terms, each of which is some $R$-multiple of the determinant of an $n \times n$ matrix whose rows are drawn (possibly with replacement) from the rows of $C$. Since $C$ only has $n-1$ rows, every such matrix must have a repeated row, and thus its determinant is zero. This contradicts to the assumption that $\det(A)\neq 0$. 
\end{proof}

\begin{corollary}
All domains have the diagonal-full property.
\end{corollary}
\begin{proof}
Let $D=\diag(a_1, \dots, a_n)$ be a diagonal matrix over $\rng$ with $a_i\neq 0$. Then $\det(D)=\prod_{i\in[n]}a_i$. If $\rng$ is a domain, from $a_i\neq 0$ we get $\det(D)\neq 0$. The result then follows from Fact~\ref{fact:det}. 
\end{proof}

\begin{proposition}
	All local principal ideal rings have the diagonal-full property.
\end{proposition}

We note that there exist local (non-PIR) rings without the diagonal-full property (Example~\ref{ex:nonreduced}) as well as PIR rings that aren't local without the diagonal-full property (Example~\ref{ex:Z6}).

\begin{proof}
Let $\rng$ be a local PIR with the maximal ideal $\m = \langle t \rangle$. Let $D=\diag(a_1, \dots, a_n)$ be a diagonal matrix over $\rng$ with $a_i\neq 0$. By multiplying by units, we get another diagonal matrix $D'$ with the same rowspan as $D$---so $D$ is full iff $D'$ is full---and such that the diagonal entries of $D'$ are nonzero powers of $t$. 

Let $N$ be the rowspan of $D$ (equivalently, $D'$\ynote{Was: $D$}). Then $N$ is abstractly isomorphic to the module $\bigoplus_{i=1}^n R / t^{e_i}$ for some integers $e_i\geq 0$ \ynote{Was: $1 \leq e_i$}. By Nakayama's Lemma, since $\rng$ is local, and $N / t N \cong \F^n$ where $\F = R/\m$ is a field, this module is not generated by fewer than $n$ elements. 

However, suppose for the sake of contradiction that $D'$ is not full, that is, there exist $k<n$ row vectors $v_1, \dots, v_k$, such that $N$ is contained in $L$, the $\rng$-span of $v_i$'s. Let $M$ be the $k \times n$ matrix that gives the inclusion of $N$ into $L$. By converting to the Smith Normal Form (which works over PIRs), $M$ can be made diagonal and such that the diagonal entries are powers of $t$. However, this would mean that there exist a basis of $L$ and a basis of $N$ such that the basis of $N$ consists of certain $t$-power multiples of some subset of the basis of $L$. But such a basis for $N$ would give us a generating set of $N$ of size $k < n$, contradicting what we said above about the minimal number of generators of $N$ itself.
\end{proof}

\subsection{Free modules over commutative rings with the diagonal-full property}

In this subsection we turn to study tensor products of free modules over commutative rings with the diagonal-full property. 

Compared with results in Section~\ref{subsec:five-action-PIR}, we relax the conditions on the rings, i.e. not requiring the ring to be finite local PIRs. On the other hand, we do not deal with general modules, but only focus on free modules. These will allow us to obtain results for several problems over $\Z$. Note that the results in this subsection work for rings with no Smith Normal Forms, such as $\Z[x]$ \cite[Example 2.2]{Sta16}, as it is a domain so satisfying the diagonal-full property.
Also note that the reductions we get here give polynomial blow-up in the lengths, but not linear.

\begin{theorem}
		Let $\rng$ satisfy the diagonal-full property. Let $U=\rng^\ell$, $V=\rng^m$, $W=\rng^n$. The orbit problems for the following actions are polynomial-time equivalent. 
	\begin{enumerate}
		\item $\GL(U)\times\GL(V)\times\GL(W)$ on $U\otimes V\otimes W$. 
		\item $\GL(U)\times \GL(V)$ on $\S^2(U)\otimes V$ and $\Lambda^2(U)\otimes V$. 
		\item $\GL(U)\times \GL(V)$ on $U\otimes U^*\otimes V$. 
		\item $\GL(U)$ on $\S^3(U)$ and $\Lambda^3(U)$. 
		\item $\GL(U)$ on $U\otimes U\otimes U^*$.
	\end{enumerate}
\end{theorem}
\begin{proof}
	We first reduce actions 2 to 5 to $U\otimes V\otimes W$ in Section~\ref{subsec:UVW}. 
	
	We then reduce action 1 to action 2 in Section~\ref{subsec:S2UW}, and action 2 to action 4 in Section~\ref{subsec:S3U}. For these two reductions, we need to arrange two or three copies of the original tensors, so that the resulting tensors are in $\S^2(U)\otimes V$ or $\S^3(U)$. Then (original or symmetric or alternating) FGS tensors are added for the reduction correctness. 
	
	Finally we reduce action 1 to action 3 in Section~\ref{subsec:UUstarW}, and action 3 to action 5 in Section~\ref{subsec:UUUstar}. For these two reductions, we put the original tensors in an appropriate corner of a bigger all-zero tensor (so no more copies are required). The original FGS tensors are added for the reduction correctness (so no symmetric or alternating FGS tensors are required).
\end{proof}

When $\rng$ is an integral domain, let $\F$ be its fraction field. The action of $\GL(U)$ on $U\otimes U\otimes U^*$ extends naturally to an action on $\F^\ell\otimes \F^\ell\otimes (\F^\ell)^*$. Our results can be extended to this setting.

\subsubsection{Some tensors used in the gadgets}

To facilitate the discussions below, we introduce the tensor used in the FGS gadget, as well as its symmetric and alternating variants. 

\paragraph{The FGS tensor.} The FGS tensor $\tF_{s, t}$ is a tensor of size $s\times st\times t$, with frontal slices being $(F_1, \dots, F_t)$, where $F_i=\begin{bmatrix}
	0_s & \dots & 0_s & \idmat_s & 0_s & \dots & 0_s 
\end{bmatrix}$, where $\idmat_s$ is in the $i$th position. 

Let $\rng$ be a commutative ring satisfying diagonal-full. A non-zero linear combination $\sum_{i\in[t]}\alpha_i F_i$ gives $F'=\begin{bmatrix}
	\alpha_1\idmat_s & \alpha_2\idmat_s & & \dots & \alpha_t\idmat_s 
\end{bmatrix}$. Then by diagonal-full and the existence of some non-zero $\alpha_i$, we have $\irk(F')=s$.

Horizontal and vertical slices of $\tF_{s, t}$ are as follows. For $i\in [s]$, the $i$th horizontal slice is 
\begin{equation}\label{eq:FGS-horizontal}
\begin{bmatrix}
	\E_{1, i} & \E_{2, i} & \dots & \E_{s, i}
\end{bmatrix}
\end{equation}
where $\E_{j, i}\in \M(t\times s, \rng)$ is an elementary matrix. In particular, its rank is $t$. For $i\in[st]$, the $i$th vertical slice is an elementary matrix, so its rank is $1$. From Equation~\ref{eq:FGS-horizontal}, the following can be verified easily.
\begin{fact}\label{fact:FGS-tensor}
	For $T\in \GL(t, \rng)$, $\tF_{s, t}^T=\tF_{s, t} (T\otimes \idmat_s)$.
\end{fact}

\paragraph{The alternating FGS tensor.} First introduced in \cite{GQT}, the alternating FGS tensor $\tG_{s, t}$ is a tensor of size $s(t+1)\times s(t+1)\times t$, with frontal slices being $(G_1, \dots, G_t)$,
$$G_i=\begin{bmatrix}
	0 & 0 & \dots & 0 & \idmat_{s} & 0 & \dots & 0 \\
	0 & 0 &  \dots & 0 & 0 & 0 & \dots & 0 \\
	\vdots & \vdots &  \dots & \vdots & \vdots & \vdots & \dots & \vdots \\
	0 & 0 &  \dots & 0 & 0 & 0 & \dots & 0 \\
	-\idmat_{s} & 0 &  \dots & 0 & 0 & 0 & \dots & 0 \\
	0 & 0 &  \dots & 0 & 0 & 0 & \dots & 0 \\
	\vdots& \vdots & \dots & \vdots & \vdots & \vdots & \dots & \vdots \\
	0 & 0 &  \dots & 0 & 0 & 0 & \dots & 0 
\end{bmatrix},$$
where $0$ here denotes the $s\times s$ all-zero 
matrix, $\idmat_{s}$ is at the $(1, i+1)$th block position, and $-\idmat_{s}$ is at the 
$(i+1, 1)$th block position. Horizontal slices of $\tG$ are essentially first in the form of horizontal slices of $\tF_{s, t}$, and then in the form of vertical slices of $\tF_{s, t}$. Therefore, the first $s$ horizontal slices of $\tG_{s, t}$ are of rank $t$, and the last $st$ horizontal slices of $\tG_{s, t}$ are of rank $1$.
\begin{fact}\label{fact:alt-FGS-tensor}
For $T\in\GL(t, \rng)$, $\tG_{s, t}^T=\begin{bmatrix}
	\idmat_s & 0 \\
	0 & T\otimes \idmat_s
\end{bmatrix}\tG_{s, t}
\begin{bmatrix}
	\idmat_s & 0 \\
	0 & T\otimes \idmat_s
\end{bmatrix}^t$.
\end{fact}

\paragraph{The symmetric FGS tensor.} This is the same as the alternating FGS tensor, but with $-\idmat_{s}$ changed to $\idmat_s$.

\subsubsection{Reductions to $U\otimes V\otimes W$}\label{subsec:UVW}

By \cite{FGS19}, to reduce $\S^2(U)\otimes V$, $\Lambda^2(U)\otimes V$, $U\otimes U^*\otimes V$, $\S^3(U)$, $\Lambda^3(U)$, and $U\otimes U\otimes U^*$ to $U\otimes V\otimes W$, it suffices to reduce partitioned tensor isomorphism to tensor isomorphism. As explained in Section~\ref{sec:issue}, there are issues to apply either FGS or GQ gadget to the commutative ring setting. We resolve this by introducing the following gadget. This gadget is based on the FGS gadget but makes use some ideas from the GQ gadget to avoid the use of the Krull--Schmidt theorem.

\begin{example}[A hybrid of FGS and GQ gadgets]\label{ex:hybrid}
We illustrate our gadget with the example in Section~\ref{sec:FGS}. Recall that there we have 
a $3$-way array $\tA\in\T(n\times m\times 6, \rng)$ can be represented as a tuple of six matrices $\vA=(A_1, \dots, A_6)$. Suppose $\ell=6$ and $L=3$, $n\leq m$, with $[6]$ partitioned into $\{1, 2\}, \{3, 4\}, \{5, 6\}$. The FGS gadget constructs the following matrices: 
\[\tilde A_1=\begin{bmatrix}
	A_1 & 0 & 0 & 0 & 0 & 0 & 0 \\
	0 & \idmat_{n+1} & 0 & 0 & 0 & 0 & 0 \\
	0 & 0 & 0 & 0 & 0 & 0 & 0 \\
	0 & 0 & 0 & 0 & 0 & 0 & 0 \\
\end{bmatrix}, 
\tilde A_2=\begin{bmatrix}
	A_ 2& 0 & 0 & 0 & 0 & 0 & 0 \\
	0 & 0 & \idmat_{n+1} & 0 & 0 & 0 & 0 \\
	0 & 0 & 0 & 0 & 0 & 0 & 0 \\
	0 & 0 & 0 & 0 & 0 & 0 & 0 \\
\end{bmatrix},\]
\[
\tilde A_3=\begin{bmatrix}
	A_3 & 0 & 0 & 0 & 0 & 0 & 0 \\
	0 & 0 & 0 & 0 & 0 & 0 & 0 \\
	0 & 0 & 0 & \idmat_{2n+2} & 0 & 0 & 0 \\
	0 & 0 & 0 & 0 & 0 & 0 & 0 \\
\end{bmatrix},
\tilde A_4=\begin{bmatrix}
	A_4 & 0 & 0 & 0 & 0 & 0 & 0 \\
	0 & 0 & 0 & 0 & 0 & 0 & 0 \\
	0 & 0 & 0 & 0 & \idmat_{2n+2} & 0 & 0 \\
	0 & 0 & 0 & 0 & 0 & 0 & 0 \\
\end{bmatrix}, \]
\[
\tilde A_5=\begin{bmatrix}
	A_5 & 0 & 0 & 0 & 0 & 0 & 0 \\
	0 & 0 & 0 & 0 & 0 & 0 & 0 \\
	0 & 0 & 0 & 0 & 0 & 0 & 0 \\
	0 & 0 & 0 & 0 & 0 & \idmat_{4n+4} & 0 \\
\end{bmatrix}, 
\tilde A_6=\begin{bmatrix}
	A_6 & 0 & 0 & 0 & 0 & 0 & 0 \\
	0 & 0 & 0 & 0 & 0 & 0 & 0 \\
	0 & 0 & 0 & 0 & 0 & 0 & 0 \\
	0 & 0 & 0 & 0 & 0 & 0 & \idmat_{4n+4} \\
\end{bmatrix}.
\]
Let $\tilde \tA=(\tilde A_1, \dots, \tilde A_6)$. Now for $i\in[6]$, let $\tilde A_i=\begin{bmatrix}
	A_i & 0 \\
	0 & E_i
\end{bmatrix}$ where $E_i\in \M((7n+7)\times (14n+14), \rng)$. Then construct the following matrices: 
$$
\bar A_1=
\begin{bmatrix}
	A_1 & 0 & 0 \\
	0 & E_1 & 0 \\
	0 & 0 & 0 
\end{bmatrix},\dots,
\bar A_6=
\begin{bmatrix}
	A_6 & 0 & 0 \\
	0 & E_6 & 0 \\
	0 & 0 & 0 
\end{bmatrix}, 
\bar A_7=
\begin{bmatrix}
	0 & 0 & 0 \\
	0 & 0 & \idmat_{7n+7} \\
	0 & \idmat_{14n+14} & 0 
\end{bmatrix}.
$$
Let $\bar\tA$ be the 3-tensor with frontal slices being $\bar A_i$'s.
Starting from $\tB\in\T(n\times m\times 6, \rng)$ with frontal slices $(B_1, \dots, B_6)$, we construct 
$$
\bar B_1=
\begin{bmatrix}
	B_1 & 0 & 0 \\
	0 & E_1 & 0 \\
	0 & 0 & 0 
\end{bmatrix},\dots,
\bar B_6=
\begin{bmatrix}
	B_6 & 0 & 0 \\
	0 & E_6 & 0 \\
	0 & 0 & 0 
\end{bmatrix}, 
\bar B_7=
\begin{bmatrix}
	0 & 0 & 0 \\
	0 & 0 & \idmat_{7n} \\
	0 & \idmat_{14n} & 0 
\end{bmatrix}.
$$
Let $\bar\tB$ be the 3-tensor with frontal slices being $\bar B_i$'s.

Let $(P, Q, T)\in \GL(22n+7, \rng)\times \GL(m+21n+14, \rng)\times\GL(7, \rng)$ be an isomorphism from $\bar\tA$ to $\bar\tB$. We examine the structure of $T$ first. Recall that the diagonal-full property gives us that an $n\times n$ diagonal matrix with non-zero entries is of rank $n$ for any non-zero $\alpha\in\rng$. This implies matrices with identity matrices of different sizes cannot mix together. For example, any mix of $\tilde A_1$ and $\tilde A_3$, that is $\alpha\tilde A_1+\beta\tilde A_3$ where $\alpha, \beta\neq 0$, has $\begin{bmatrix}
	\alpha\idmat_{n+1} & 0 \\
	0 & \beta\idmat_{2n+2}
\end{bmatrix}$ as a submatrix, whose inner rank is $3n+3$. This would cause a mismatch of the ranks with existing slices.

Because of the above considerations, we see that $T=\begin{bmatrix}
	T_{11} & 0 & 0 & T_{14}\\
	0 & T_{22} & 0 & T_{24} \\
	0 & 0 & T_{33} & T_{34} \\
	0 & 0 & 0 & T_{44}
\end{bmatrix}$ where $T_{11}, T_{22}, T_{33}\in\GL(2, \rng)$, $T_{i}\in \M(2\times 1, \rng)$ for $i\in[3]$, and $T_{44}$ is a unit in $\rng$. Note that $T_{41}, T_{42}, T_{43}$ are possibly non-zero, because some linear combinations of $A_i$'s may result in a zero matrix, as they are not assumed to be non-degenerate. Let $\bar A_i'$, $i\in[7]$, be the matrices obtained after applying $T$ to $\bar A_i$. Then $\bar A_7'=T_{44}\cdot \begin{bmatrix}
0 & 0 & 0 \\
0 & E' & \idmat_{7n+7} \\
0 & \idmat_{14n+14} & 0 
\end{bmatrix}$ for some $E'\in \M((7n+7)\times (14n+14), \rng)$. Observe that by $T_{44}$ being a unit in $\rng$, $T_{44}\cdot \begin{bmatrix}
E' & \idmat_{7n+7} \\
\idmat_{14n+14} & 0 
\end{bmatrix}$ is invertible. By $P\bar A_7' Q=\bar B_7=\bar A_7$, we obtain that $P=\begin{bmatrix}
P_{11} & 0\\
P_{21} & P_{22}
\end{bmatrix}$ where $P_{11}\in\M(n\times n, \rng)$, and $Q=\begin{bmatrix}
Q_{11} & Q_{12} \\
0 & Q_{22}
\end{bmatrix}$ where $Q_{11}\in\M(m\times m, \rng)$. It follows that $P_{11}$ and $Q_{11}$ are invertible, so $(P_{11}, Q_{11}, \begin{bmatrix}
T_{11} & & \\
& T_{22} & \\
& & T_{33}
\end{bmatrix}
)$ forms an isomorphism from $\tA$ to $\tB$ respecting the partition.
\end{example}

\begin{remark}\label{rem:hybrid}
The main difference with the FGS gadget is the introduction of the matrix $\bar A_7$, inspired by the GQ gadget. The effect of $\bar A_7$ is to ``cover'' the newly introduced spaces for holding the original FGS gadget, and ``leave out'' the spaces holding the original tensor. This allows to avoid the use of the Krull--Schmidt theorem as in \cite{FGS19}. 

It is unclear whether the same idea can be applied to the GQ gadget. If we introduce a similar new slice as $\bar A_7$, then in the cancellation step in the GQ gadget, this new slice would be modified, which would be problematic.
\end{remark}

We now extend the idea in the above to state and prove the following.

\begin{theorem}\label{thm:pTI-TI}
	Fix a partition $[\ell]=D_1\cup \dots \cup D_L$, and suppose $n\leq m$. There exists a function $f:\T(n\times m\times\ell, \rng)\to\T(n'\times m'\times \ell', \rng)$, such that $\tA, \tB\in\T(n\times m\times \ell, \rng)$ are isomorphic as partitioned tensors if and only if $f(\tA)$ and $f(\tB)$ are isomorphic as plain tensors. In particular, $n', m', \ell'$, and the time needed to compute $f$, are upper bounded by $2^{O(L)}\cdot (n\cdot \ell+m)$.
\end{theorem}
\begin{proof}
Without loss of generality, assume the partition of $[\ell]$ is $D_1\cup D_2\cup \dots \cup D_L=\{t_0=1, \dots, t_1\}\cup \{t_1+1, \dots, t_2\}\cup \dots \cup\{t_{L-1}+1, \dots, t_L=\ell\}$. Let $s_i=t_i-t_{i-1}+1$, so $|D_i|=s_i$. Note that $\sum_{i\in[L]}s_i=\ell$.

We first introduce the original FGS gadget. Let $r:=n+1$. For the part $D_i$, let $\tF^{(i)}:=\tF_{2^{i-1}\cdot r, s_i}$ be a FGS tensor. Note that $\tF^{(i)}\in\T(2^{i-1}\cdot r\times (s_i\cdot 2^{i-1}\cdot r)\times s_i, \rng)$. We then put $\tF^{(i)}$ together diagonally, and let $\tC:=\diag(\tF^{(1)}, \tF^{(2)}, \dots, \tF^{(L)})$. Note that $\tC$ is of size $(2^L-1)\cdot r \times (\sum_{i\in[L]}s_i\cdot 2^{i-1}\cdot r)\times \ell$. Let $C_j$ be the $j$th frontal slice of $\tC$. The FGS construction then yields $\begin{bmatrix}
	A_j & 0 \\
	0 & C_j
	\end{bmatrix}$ for $j\in[\ell]$.

%

We then expand the matrix size further from $(2^L-1)\cdot r \times (\sum_{i\in[L]}s_i\cdot 2^{i-1}\cdot r)$ to $\big((2^L-1+\sum_{i\in[L]}s_i\cdot 2^{i-1})\cdot r\big)\times \big((2^L-1+\sum_{i\in[L]}s_i\cdot 2^{i-1})\cdot r\big)$. Then construct 
$\bar A_j:=\begin{bmatrix}
	A_j & 0 & 0 \\
	0 & C_j & 0 \\
	0 & 0 & 0
\end{bmatrix}$ for $j\in[\ell]$, and 
$\bar A_{\ell+1}=\begin{bmatrix}
	0 & 0 & 0 \\
	0 & 0 & \idmat_{(2^L-1)\cdot r} \\
	0 & \idmat_{\sum_{i\in[L]}s_i\cdot 2^{i-1}\cdot r} & 0
\end{bmatrix}.$
Let $\bar\tA$ be the 3-tensor whose frontal slices are $\bar A_j$ for $j\in[\ell+1]$.

From $\tB$ we construct $\bar\tB$. We claim that $\tA$ and $\tB$ are partition isomorphic if and only if $\bar\tA$ and $\bar\tB$ are isomorphic (as plain tensors). 

\paragraph{For the if direction.} This direction follows the argument for Example~\ref{ex:hybrid}. Let $(P, Q, T)$ be an isomorphism from $\bar\tA$ to $\bar\tB$. First, by the rank argument, $T=\begin{bmatrix}
	T_{11} & 0 & \dots & 0 & T_{1, L+1}\\
	0 & T_{22} & \dots & 0& T_{2, L+1}\\
	0 & 0 & \dots & 0 & T_{3, L+1} \\
	\vdots & \vdots & \ddots & \vdots&\vdots \\
	0 & 0 & \dots  & T_{LL} & T_{L, L+1}\\
	0 & 0 & \dots & 0 & T_{L+1, L+1}
\end{bmatrix}$, where $T_{ii}$ is of size $s_i\times s_i$, $T_{L+1, L+1}\in \rng$. Then $\alpha:=\T_{L+1, L+1}$ is a unit. 
Let $\bar\tA'=(\bar\tA)^T$, with frontal slices being $\bar A_1', \dots, \bar A_\ell', \bar A_{\ell+1}'$. Then $\bar A_{\ell+1}'=\begin{bmatrix}
	0 & 0 & 0 \\
	0 & D & \alpha\cdot \idmat_{(2^L-1)\cdot n} \\
	0 & \alpha\cdot \idmat_{\sum_{i\in[L]}s_i\cdot 2^{i-1}\cdot n} & 0
\end{bmatrix}$. Note that $\bar A_{\ell+1}'$ is invertible because its determinant is a power of $\alpha$. Then $P, Q$ satisfy that $P\bar A_{\ell+1}' Q=\bar B_{\ell+1}=\bar A_{\ell+1}$. By Lemma~\ref{lem:upper-tri}, we have $P=\begin{bmatrix}
P_{11} & 0\\
P_{21} & P_{22}
\end{bmatrix}$ where $P_{11}\in\M(n\times n, \rng)$, and $Q=\begin{bmatrix}
Q_{11} & Q_{12} \\
0 & Q_{22}
\end{bmatrix}$ where $Q_{11}\in\M(m\times m, \rng)$. It follows that $P_{11}$ and $Q_{11}$ are invertible, so $(P_{11}, Q_{11}, \begin{bmatrix}
T_{11} & & \\
& \ddots & \\
& & T_{LL}
\end{bmatrix}
)$ forms an isomorphism from $\tA$ to $\tB$ respecting the partition.

\paragraph{For the only if direction.} Suppose $\tA$ and $\tB$ are isomorphic as partition tensors by $(P, Q, T)\in\GL(n, \rng)\times\GL(m, \rng)\times\GL(\ell, \rng)$. Note that $T=\diag(T_1, \dots, T_L)$ where $T_i\in\GL(s_i, \rng)$. For $i\in[\ell]$, let $\tilde A_i=\begin{bmatrix}
	A_i & 0 \\
	0 & C_i
\end{bmatrix}$. Let $\tilde P=\begin{bmatrix}
P & 0 \\
0 & \idmat_{(2^L-1)r}
\end{bmatrix}$, and $\tilde Q=\begin{bmatrix}
Q & 0 \\
0 & \idmat_{\sum_{i\in[L]}s_i\cdot 2^{i-1}\cdot r}\end{bmatrix}$.

As in \cite{FGS19}, apply $(\tilde P, \tilde Q, T)$ to $\tilde \tA$ to get $\tilde \tA'$. Let $\tilde\tA'=(\tilde A_1', \dots, \tilde A_\ell')$, where $\tilde A_i'=\diag(B_i, C_i')$. For $j\in\{t_{i-1}+1, t_i\}$, $C_j'$ has the $i$th block being $C_{j,i}':=\begin{bmatrix}
	(T_i)_{j, 1} \idmat_{2^{i-1}r} & \dots & (T_i)_{j, s_i} \idmat_{2^{i-1}r}
\end{bmatrix}$. Then by Fact~\ref{fact:FGS-tensor}, right multiplying $Q':=T_i^{-1}\otimes \idmat_{2^{i-1}\cdot r}$, $C_{j,k}'Q'$ gives $B_{i,j}=\begin{bmatrix}
	0 & \dots & 0 & \idmat_{2^{i-1}\cdot r}& 0 & \dots & 0
\end{bmatrix}$ where $\idmat_{2^{i-1}\cdot r}$ is at the $j-t_{i-1}$th position. This basically completes the argument from \cite{FGS19}.

We then need to handle the newly added $\bar A_{\ell+1}$. For this we first extend $T=\diag(T_1, \dots, T_L)$ to $\bar T=\diag(T_1, \dots, T_L, 1)$. Then note that $Q'$ (and the naturally extended $\tilde Q$) from above affects $\idmat_{\sum_{i\in[L]}s_i\cdot 2^{i-1}\cdot r}$ in $\bar A_{\ell+1}$. This can be remedied by introducing $P'$ accordingly which multiplies on the left on $\idmat_{\sum_{i\in[L]}s_i\cdot 2^{i-1}\cdot r}$. This concludes the proof.
\end{proof}

\subsubsection{From $U\otimes V\otimes W$ to $\S^2(U)\otimes V$ and $\Lambda^2(U)\otimes V$}\label{subsec:S2UW}
In this section we consider reductions from $U\otimes V\otimes W$ to $\S^2(U)\otimes V$ and $\Lambda^2(U)\otimes V$. We present the one from $U\otimes V\otimes W$ to $\Lambda^2(U)\otimes V$ and the $\S^2(U)\otimes V$ case follows in the same way. 

When $\rng=\F$ is a field, two reductions from $U\otimes V\otimes W$ to $\S^2(U)\otimes V$ were presented, one using the FGS gadget \cite{TI1} and the other using the GQ gadget \cite{TI4}. The issue with the GQ gadget has been discussed in Section~\ref{sec:issue}, and we do not know how to remedy the non-degeneracy issue for the GQ gadget. For FGS gadgets, the issue of relying on Krull--Schmidt actually does not appear in this setting. However, there is still an issue with the proof in \cite{TI1} caused by non-degeneracy. This can be remidied by adding another FGS gadget, as we show now. 
\begin{theorem} \label{thm:AltMatSpIsom}
	Let $U=\rng^\ell$, $V=\rng^n$, $W=\rng^m$, with $n=\max\{\ell, m, n\}$.\footnote{Assuming that $n$ is the largest is without loss of generality as we can ``rotate'' the three vector spaces if this is not satisfied.} There is a polynomial-time algorithm $\algprobm{P}$ that takes $\tA\in U\otimes V\otimes W$ and outputs $\algprobm{P}(\tA)\in \Lambda^2(U')\otimes V'$ with $U'= \rng^{\ell'}$ and $V'=\rng^{m'}$ with $\ell', m'=O\big((\ell+m+n)^3\big)$,
	such that $\tA, \tB\in U\otimes V\otimes W$ are isomorphic if and only if $\algprobm{P}(\tA)$ and $\algprobm{P}(\tB)$ are isomorphic.
%
\end{theorem}

\begin{proof}
When $\rng=\F$ is a field, the reduction in \cite{TI1} makes use of the FGS gadget in two directions as in Figure~\ref{fig:3-tensor_isometry}. 
\begin{figure}[!htbp]
	\resizebox{18cm}{!}{
		\hspace*{6cm}
	\xymatrix@R=8pt@C=6pt{
		&&&&& *{} \ar@{-}'[rrrrr] 
		\ar@{-}'[dddd]|{\ell}'[dddddr]|{m}'[dddddrrrrrr]'[ddddrrrrr]'[rrrrr]'[rrrrrdr]'[rrrrrdrdddd]
		\ar@{}'[ddddrrrrr]|{A} &&&&& *{} & 
		\\ 
		&&&&&&&&&&& *{} \ar@{-}'[rrrr]'[rrrrdr]'[rrrrdrd]'[rrrrd]'[d]'[drd]'[drdrrrr] &&&& 
		*{} \ar@{-}'[d] &&& 
		\\ 
		&&&&&&&&&&& *{} \ar@{.}'[rdrrrr]|{I_{2n+1}} &&&& *{} & 
		*{} 
		\\ 
		&&&&&&&&&&& *{} \ar@{--}'[rd]  
		& *{} \ar@{-}'[d]'[drd]'[drdrrrr]'[drrrrr]'[rrrr]'[rrrrd]'[rrrrddr] &&&& *{}  %
		\\ 
		*{} \ar@{-}'[rrrrr]|{\ell} \ar@{-}'[dddd]|{n} \ar@{}'[ddddrrrrr]|{-A^t}
		&&&&& *{} \ar@{-}'[rrrrr]|{n} \ar@{-}'[dddd] &&&&& *{}  
		&*{} \ar@{--}'[rdrd]
		& *{} \ar@{-}'[rrrr] \ar@{.}'[rrrrrd]|{I_{2n+1}} &&&& *{}  
		& *{} 
		\\ 
		&&&&&& *{} \ar@{-}'[dddd] 
		\ar@{--}'[rdrdrd]'[rdrdrdrrrrr]'[rdrdrdrdrrrrrrd][rdrdrdrdrrrrrrdddddddddd] %
		& \ar@{--}'[rdrdrd]'[rdrdrdddddddd] 
		&&&& *{} \ar@{--}'[rdrd] 
		&& *{} \ar@{}'[drrrr]|{\ddots} &&&& *{} 
		\\ 
		&&&&&& *{} 
		\ar@{--}'[rdrdrd]'[rrrdddrrrrrrrrr] 
		&&&&&&& *{} &&&& *{} 
		\\ 
		&&&&&&&&&&&&& *{} \ar@{-}'[rrrr] 
		\ar@{-}'[u]'[urrrr]'[rrrr]'[rrrrrd]'[rrrrr]'[urrrr] 
		\ar@{-}'[rd]'[rdrrrr] 
		\ar@{.}'[rdrrrr]|{I_{2n+1}} 
		& & & & *{} & 
		*{} 
		\\ 
		*{} \ar@{-}'[rrrrr] \ar@{-}'[rd]'[rdrrrrr] 
		& & & 
		& &  *{} \ar@{-}'[rd] 
		& & & & *{} \ar@{--}'[dddd] 
		& & && &  *{} & & & &  *{} 
		\ar@{-}'[rrrrr]'[rrrrrd]'[d]'[ddr]'[ddrrrrrr]'[rrrrrd]'[rrrrr]'[rrrrrdr]'[rrrrrdrd]
		\ar@{-}'[d] 
		& & & & & *{} 
		\\ 
		& *{} \ar@{-}'[ddd]'[dddrd]'[dddrdr]'[dddr]'[r]'[rrd]'[rrdddd] 
		& *{} \ar@{.}'[rdddd] 
		& & & *{}  \ar@{--}'[rdrd] 
		& *{} 
		\ar@{--}'[rdrd] 
		&&& *{} 
		&&&&&&&&& *{} \ar@{.}'[rdrrrrr]|{I_{4n+2}} &&&&& *{}  
		& *{} 
		\\ 
		&&& *{} \ar@{}'[rrrrddd]|{\ddots} 
		&&&&&&&&&&&&&&&& *{}  \ar@{-}'[d]'[drrrrr]'[rrrrr]'[rrrrrdr]'[rrrrrdrd]'[drd]'[d] 
		&&&&&*{} 
		\\ 
		&&&&&&&*{} \ar@{-}'[ddd]'[dddrd]'[dddrdr]'[dddr]'[r]'[rrd]'[rrdddd] \ar@{-}'[r] %
		& *{} \ar@{.}'[rdddd] 
		&&&&&&&&&&& *{} \ar@{.}'[rrrrrdr]|{I_{4n+2}} 
		&&&&&*{} \ar@{-}'[rd] 
		&*{} 
		\\ 
		&*{} \ar@{-}'[r] &*{} 
		&&&&&&&*{} 
		\ar@{--}'[rdrdrd]'[rdrdrdrrrrrrrrrr] 
		&&&&&&&&&&&*{} \ar@{}'[drrrrrr]|{\ddots} 
		&&&&&*{} 
		\\ 
		&&*{}&*{} 
		&&&&&&&&&&&&&&&&&&*{} 
		\ar@{-}'[rrrrr]'[rrrrrd]'[d]'[ddr]'[ddrrrrrr]'[rrrrrd]'[rrrrr]'[rrrrrdr]'[rrrrrdrd]
		\ar@{-}'[d] 
		&&&&&*{} 
		\\ 
		&&&&&&&*{} \ar@{-}'[r] & *{} 
		&&&&&&&&&&&&&*{} \ar@{.}'[rdrrrrr]|{I_{4n+2}}&&&&&*{}
		& *{} 
		\\ 
		&&&&&&&&*{} 
		&*{}  \ar@{-}'[dddd]'[ddddrd]'[ddddrdr]'[ddddr]'[r]'[rrd]'[rrddddd] \ar@{-}'[r]  %
		&*{} \ar@{.}'[rddddd] 
		&&&&&&&&&&&&*{} &&&&&*{} 
		\\ 
		&&&&&&&&&&&*{} 
		&&&&&&&&&&&&&&&&&&&&&&&&&&&&&&&&&&&
		\\ 
		&&&&&&&&&&& \ar@{}'[rdrdrd]|{\ddots} 
		&&&&&&&&&&&&&&&&&&&&&&&&&&&&&&&&&&&&&& 
		\\ 
		&&&&&&&&&&&&&&*{} \ar@{-}'[dddd]'[ddddrd]'[ddddrdr]'[ddddr]'[r]'[rrd]'[rrddddd] 
		\ar@{-}'[r]  & *{} \ar@{.}'[rddddd] 
		&&&&&&&&&&&&&&&&&&&&&&&&&&&&&&
		\\ 
		&&&&&&&&&*{} \ar@{-}'[r] & *{} 
		&&&&&&*{} 
		&&&&&&&&&&&&&&&&&&&&&&&&&&&&& 
		\\ 
		&&&&&&&&&&*{} & *{} 
		&&&&&&&&&&&&&&&&&&&&&&&&&&&&&&&&&&&&&&
		\\ 
		&&&&&&&&&&&&&*{} &*{} 
		&&&&&&&&&&&&&&&&&&&&&&&&&&&&&&&
		\\ 
		&&&&&&&&&&&& &&*{} \ar@{-}'[r] &*{}  
		&&&&&&&&&&&&&&&&&&&&&&&&&&&&&&&
		\\ 
		&&&&&&&&&&&&&&&*{} &*{} 
		&&&&&&&&&&&&&&&&&&&&&&&&&&&&&&&
	}
}
	\caption{ \label{fig:3-tensor_isometry} Pictorial representation of the reduction 
		from $U\otimes V\otimes W$ to $\Lambda^2(U)\otimes W$ in \cite{FGS19}.}
\end{figure}
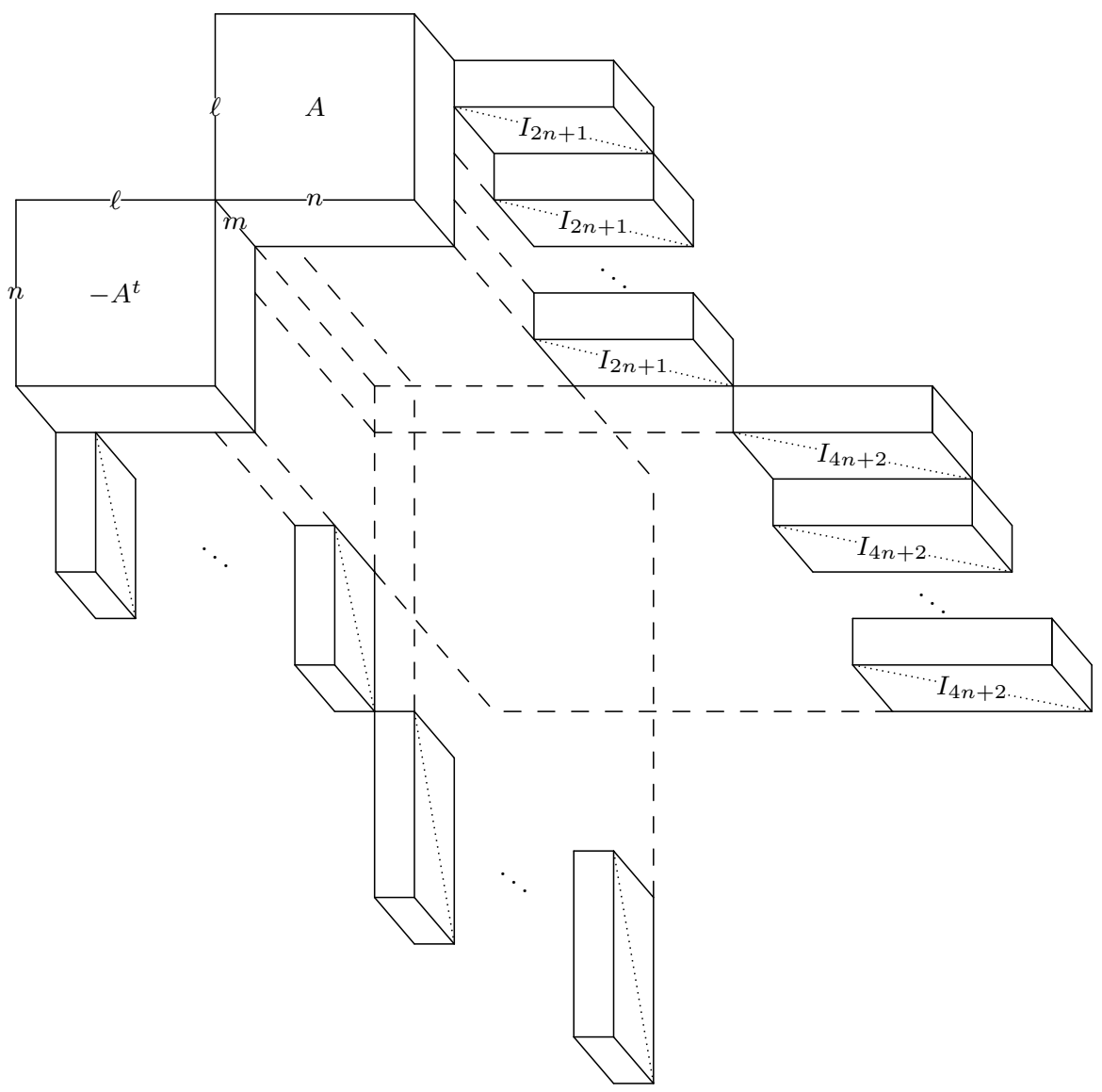
The proof in \cite{TI1} assumed that $\tA\in U\otimes V\otimes W$ is non-degenerate in the $W$ direction, i.e. the frontal slices of $\tA$ are linearly independent as in $\M(\ell\times n, \F)$. We need to get rid of this assumption. This can be done by introducing a gadget making the last $(2n+1)\ell+(4n+2)n$ slices special. 

More specifically, starting from $\tA\in\T(\ell\times n\times m, \rng)$ with frontal slices $(A_1, \dots, A_m)$, we first follow \cite{TI1} to construct $\tilde \tA\in \T(\big(\ell+n+(2n+1)+(4n+2)\big)\times\big(\ell+n+(2n+1)+(4n+2)\big)\times\big(m+(2n+1)\ell+(4n+2)n\big), \rng)$, as shown in Figure~\ref{fig:3-tensor_isometry}. Note that this is just to put two  FGS tensor $\tF_{2n+1, \ell}$ and $\tF_{4n+2, n}$ together diagonally. Then, install one copy along the first direction, and another copy multiplied with $-1$ along the second direction to maintain alternating.

We now add a new gadget. Let $N=(2n+1)\ell+(4n+2)n$, the number of the newly added frontal slices from above. First, construct the alternating FGS tensor $\tG=\tG_{2n+1, N}$. Note that $\tG$ is of size $(N+1)\cdot (2n+1)\times (N+1)\cdot (2n+1)\times N$. We then expand $\tG$ to a tensor of size $(N+1)\cdot (2n+1)\times (N+1)\cdot (2n+1)\times (m+N)$ by putting $m$ many zero frontal slices in the front.
%
We then attach $\tG$ to $\tilde\tA$ as $\bar\tA=\begin{bmatrix}
	\tilde\tA & 0 \\
	0 & \tG
\end{bmatrix}$. This is illustrated in Figure~\ref{fig:TI-AMSI}.
\begin{figure}[!htbp]
	
	\centering

	\tikzset{every picture/.style={line width=0.75pt}} 
	
	\begin{tikzpicture}[x=0.75pt,y=0.75pt,yscale=-0.7,xscale=0.7]
		
		\draw  [fill={rgb, 255:red, 245; green, 166; blue, 35 }  ,fill opacity=0.8 ] (109.71,132.33) -- (164.43,77.61) -- (164.43,12.34) -- (123.46,12.34) -- (68.74,67.06) -- (68.74,132.33) -- cycle ; \draw   (164.43,12.34) -- (109.71,67.06) -- (109.71,132.33) ; \draw   (109.71,67.06) -- (68.74,67.06) ;
		\draw  [fill={rgb, 255:red, 74; green, 144; blue, 226 }  ,fill opacity=0.5 ] (88.57,21.24) -- (111.67,-1.86) -- (139.86,-1.86) -- (139.86,39.63) -- (116.76,62.73) -- (88.57,62.73) -- cycle ; \draw   (139.86,-1.86) -- (116.76,21.24) -- (88.57,21.24) ; \draw   (116.76,21.24) -- (116.76,62.73) ;
		\draw  [fill={rgb, 255:red, 245; green, 166; blue, 35 }  ,fill opacity=0.8 ] (139.86,-1.34) -- (194.59,-56.07) -- (259.86,-56.07) -- (259.86,-15.1) -- (205.14,39.63) -- (139.86,39.63) -- cycle ; \draw   (259.86,-56.07) -- (205.14,-1.34) -- (139.86,-1.34) ; \draw   (205.14,-1.34) -- (205.14,39.63) ;
		\draw  [fill={rgb, 255:red, 74; green, 144; blue, 226 }  ,fill opacity=0.5 ] (88.94,88.43) -- (109.71,67.06) -- (109.33,41.36) -- (68.22,41.95) -- (47.46,63.32) -- (47.83,89.03) -- cycle ; \draw   (109.33,41.36) -- (88.57,62.73) -- (88.94,88.43) ; \draw   (88.57,62.73) -- (47.46,63.32) ;
		\draw  [fill={rgb, 255:red, 184; green, 233; blue, 134 }  ,fill opacity=0.5 ] (343.2,205.32) -- (416.98,131.55) -- (502.4,131.55) -- (502.4,216.22) -- (428.62,290) -- (343.2,290) -- cycle ; \draw   (502.4,131.55) -- (428.62,205.32) -- (343.2,205.32) ; \draw   (428.62,205.32) -- (428.62,290) ;
		\draw  [fill={rgb, 255:red, 189; green, 16; blue, 224 }  ,fill opacity=0.5 ] (259.86,-15.1) -- (329.52,-84.76) -- (415.93,-84.76) -- (415.93,-57.78) -- (346.27,11.87) -- (259.86,11.87) -- cycle ; \draw   (415.93,-84.76) -- (346.27,-15.1) -- (259.86,-15.1) ; \draw   (346.27,-15.1) -- (346.27,11.87) ;
		\draw  [fill={rgb, 255:red, 189; green, 16; blue, 224 }  ,fill opacity=0.5 ] (192.42,166.65) -- (264.7,94.37) -- (264.7,5.33) -- (236.71,5.33) -- (164.43,77.61) -- (164.43,166.65) -- cycle ; \draw   (264.7,5.33) -- (192.42,77.61) -- (192.42,166.65) ; \draw   (192.42,77.61) -- (164.43,77.61) ;
		\draw  [color={rgb, 255:red, 208; green, 2; blue, 27 }  ,draw opacity=1 ][dash pattern={on 4.5pt off 4.5pt}] (48.05,21.78) -- (324,21.78) -- (324,226.92) -- (48.05,226.92) -- cycle ;
		\draw  [color={rgb, 255:red, 208; green, 2; blue, 27 }  ,draw opacity=1 ][dash pattern={on 4.5pt off 4.5pt}] (68.85,-1.42) -- (344.8,-1.42) -- (344.8,203.72) -- (68.85,203.72) -- cycle ;
		\draw  [color={rgb, 255:red, 208; green, 2; blue, 27 }  ,draw opacity=1 ][dash pattern={on 4.5pt off 4.5pt}] (139.98,-84.76) -- (416.98,-84.76) -- (416.98,131.55) -- (139.98,131.55) -- cycle ;

	\end{tikzpicture}

	\caption{\label{fig:TI-AMSI}A pictorial demonstration of the newly added gadget for reducing from $U\otimes V\otimes W$ to $\Lambda^2(U)\otimes W$. The orange blocks correspond to the FGS gadgets in Figure~\ref{fig:3-tensor_isometry} with $\idmat_{2n+1}$. The purple blocks correspond to the FGS gadgets in Figure~\ref{fig:3-tensor_isometry} with $\idmat_{4n+2}$. The green block is the newly added alternating FGS gadget with $\idmat_{2n+1}$. Note that the green block is actually quite wide, as the frontal slices are square matrices of size $(2n+1)\cdot (N+1)$ where $N=O(\ell n+n^2)$. That is, in this picture its size is not shown proportionally.}
\end{figure}
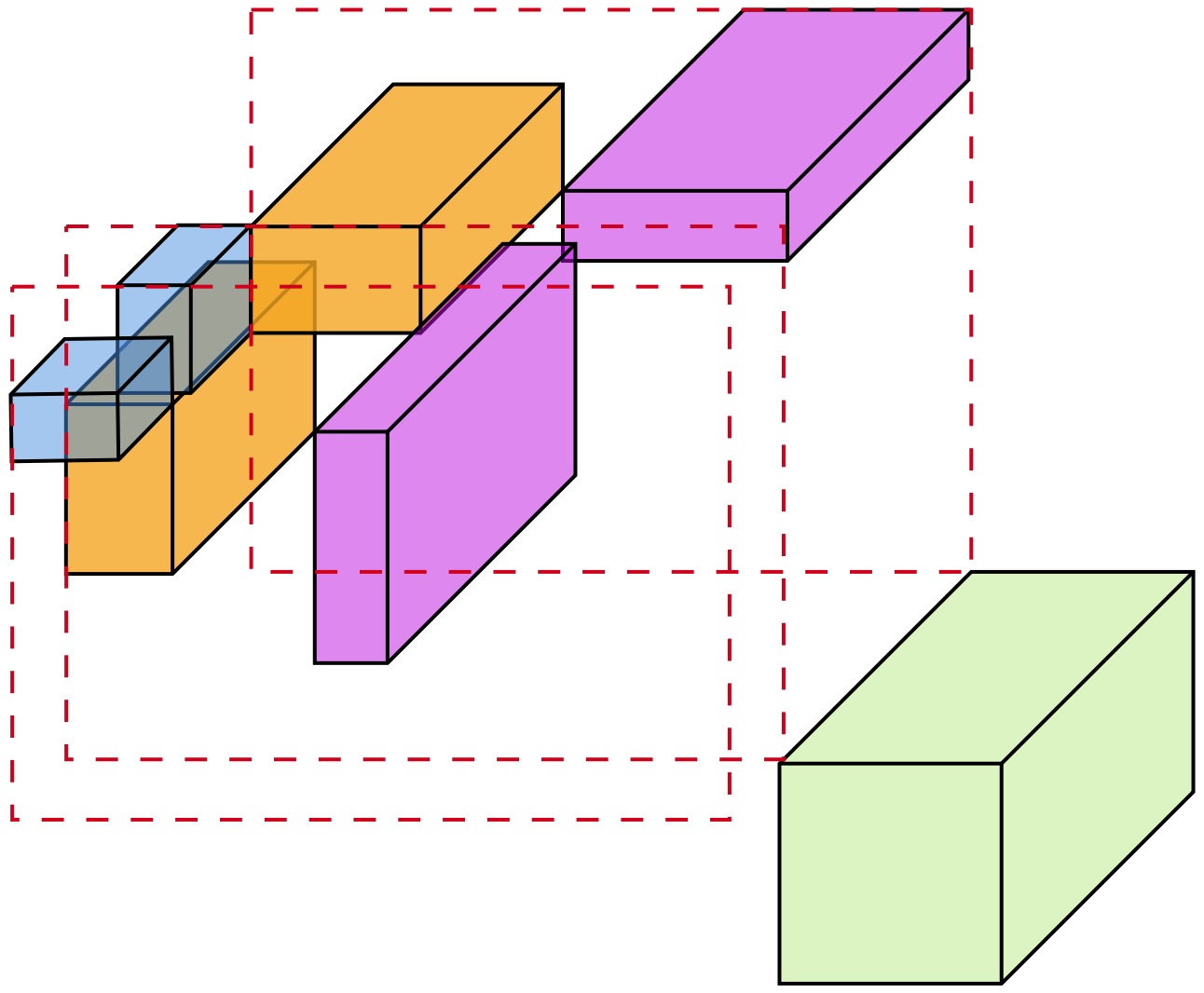

Let $U'=\rng^{\ell+n+(2n+1)+(4n+2)+(N+1)\cdot (2n+1)}$. Let $V'=\rng^{m+N}$. We claim that $\tA$ and $\tB$ in $U\otimes V\otimes W$ are isomorphic if and only if $\bar\tA$ and $\bar\tB$ are isomorphic in $\Lambda^2(U')\otimes V'$. 

For the only if direction, we start from $(P, Q, T)\in\GL(U)\times\GL(V)\times\GL(W)$ sending $\tA$ to $\tB$ and construct $(P', T')\in \GL(U')\times \GL(V')$ sending $\bar\tA$ to $\bar\tB$. We first extend $(P, Q, T)$ to an isomorphism $(\tilde P, \tilde T)$ from $\tilde\tA$ to $\tilde\tB$ as in \cite{TI1}. In particular, $\tilde P=\diag(P, Q, U)$ and $\tilde T=\diag(T, V)$ for $V\in\GL(N, \rng)$. We then see that $\tilde T$ acts on the newly added $\tG$ gadget on the third direction. This can be remedied by introducing the congruence action of $T'$ on the first two direction as in Fact~\ref{fact:alt-FGS-tensor}.


For the if direction, suppose $(P, T)\in\GL(U')\times\GL(V')$ sends $\bar \tA$ to $\bar\tB$. We first deal with the $V'$ direction. Note that the ranks of the first $m$ frontal slices of $\bar\tA$ are within $(\ell+n)\leq 2n$ as we assumed $n=\max\{n, m, \ell\}$. The ranks of the last $N$ slices are no less than $2n+1$ because of the green block in Figure~\ref{fig:TI-AMSI}.
This ensures that in any isomorphism $(P, T)$ from $\bar\tA$ to $\bar\tB$, $T$ is of the form $\begin{bmatrix}
	T_{11} & T_{12} \\
	0 & T_{22}
\end{bmatrix}$ where $T_{11}$ is of size $m\times m$. 

We then deal with the $U'$ direction, which requires a more careful examination of the alternating FGS gadget. We first write $P\in \GL(U')$ as 
\begin{equation}\label{eq:P55}
\begin{bmatrix}
	P_{11} & P_{12} & P_{13} & P_{14} & P_{15}\\
	P_{21} & P_{22} & P_{23} & P_{24} & P_{25}\\
	P_{31} & P_{32} & P_{33} & P_{34} & P_{35}\\
	P_{41} & P_{42} & P_{43} & P_{44} & P_{45}\\
	P_{51} & P_{52} & P_{53} & P_{54} & P_{55}\\
\end{bmatrix},
\end{equation}
where $P_{11}\in\M(\ell, \rng)$, $P_{22}\in\M(n, \rng)$, $P_{33}\in\M(3n+3, \rng)$, $P_{44}\in\M(2n+1, \rng)$, and $P_{55}\in\M(N\cdot (2n+1), \rng)$. We now consider the ranks of the horizontal slices of $\bar\tA$. 
\begin{enumerate}
	\item For the first $\ell$ slices, the ranks are between $2n+1$ and $3n+1$.
	\item For the next $n$ slices, the ranks are between $4n+2$ and $5n+2$. 
	\item For the next $3n+3$ slices, they are the vertical slices of the FGS gadget, so their ranks are either $\ell$ or $n$. Recall that we set $n=\max\{\ell, m, n\}$.
	\item For the next $2n+1$ slices, they are the vertical slices of the FGS gadget, so their ranks are $N$. 
	\item For the next $N\cdot (2n+1)$ slices, they are the horizontal slices of the FGS gadget, so their ranks are $1$. 
\end{enumerate}
We first note that the ranks of (4)-slices are $N=\ell(n+1)+m(2n+2)$, which is too large, so no other slices can afford to have these added to them. It follows that $P_{4i}=0$ for $i\neq $. We then note that the ranks of (5)-slices are $1$, which is too small, so they cannot have other slices added to them. It follows that $P_{i5}=0$ for $i\neq 5$. That is, $P$ is of the form 
$$\begin{bmatrix}
	P_{11} & P_{12} & P_{13} & P_{14} & 0\\
	P_{21} & P_{22} & P_{23} & P_{24} & 0\\
	P_{31} & P_{32} & P_{33} & P_{34} & 0\\
	0 & 0 & 0 & P_{44} & 0\\
	P_{51} & P_{52} & P_{53} & P_{54} & P_{55}\\
\end{bmatrix},$$
from which we infer that $P_{44}$ and $P_{55}$ are invertible. Furthermore, 
$P'=\begin{bmatrix}
	P_{11} & P_{12} & P_{13}\\
	P_{21} & P_{22} & P_{23}\\
	P_{31} & P_{32} & P_{33}
\end{bmatrix}
$ is invertible. We can then follow the rank argument as 
in \cite[Proposition 5.1]{TI1} 
to show that 
$P'=\begin{bmatrix}
	P_{11} & 0 & 0\\
	0 & P_{22} & 0\\
	P_{31} & P_{32} & P_{33}
\end{bmatrix},
$ which gives that $P_{11}$ and $P_{22}$ are invertible. Now observe that the first $m$ frontal slices of $\bar\tA$ only have non-zero entries in the blue region (the original tensors), and along the $V'$ direction, $T=\begin{bmatrix}
	T_{11} & T_{12} \\
	0 & T_{22}
\end{bmatrix}$ does not bring the last $N$ slices to the first $m$ slices. These together give that $P_{11}, P_{22}, T_{11}$, which are all invertible, form an isomorphism from $\tA$ to $\tB$.
\end{proof}
%
%

\subsubsection{From $\Lambda^2(U)\otimes V$ (resp. $\S^2(U)\otimes V$) to $\Lambda^3(U)$ (resp. $\S^3(U)$) }\label{subsec:S3U}

In this section we present reductions from $\Lambda^2(U)\otimes V$ (resp. $\S^2(U)\otimes V$) to $\Lambda^3(U)$ (resp. $\S^3(U)$). We focus on the one from $\Lambda^2(U)\otimes V$ to $\Lambda^3(U)$, as the one from $\S^2(U)\otimes V$ to $\S^3(U)$ follows the same strategy.

We first state the result formally as follows.
\begin{theorem} \label{thm:Lambda3U}
	Let $U=\rng^\ell$ and $V=\rng^m$. There is a polynomial-time algorithm $\algprobm{P}$ that takes $\tA\in \Lambda^2(U)\otimes V$ and outputs $\algprobm{P}(\tA)\in \Lambda^3(U')$ with $U'= \rng^{\ell'}$, with $\ell'=O\big((\ell+m)^3\big)$,
	such that $\tA$ and $\tB\in \Lambda^2(U)\otimes V$ are isomorphic if and only if $\algprobm{P}(\tA)$ and $\algprobm{P}(\tB)\in \Lambda^3(U')$ are isomorphic.
	%
\end{theorem}
\begin{proof}[Proof sketch of Theorem~\ref{thm:Lambda3U}] Over fields, a reduction from $\Lambda^2(U)\otimes V$ to $\Lambda^3(U)$ was presented in \cite{GQT}. The reduction over commutative rings makes use of some steps there but requires a further copy of the alternating FGS gadget. Here, we present a proof sketch of Theorem~\ref{thm:Lambda3U}, with an emphasis on the newly added FGS gadget. A detailed proof can be obtained without much effort by going over the proof of \cite[Proposition 4.1]{GQT}.

Let $\tA\in \Lambda^2(U)\otimes V$. We represent $\tA$ as a $3$-tensor as in Figure~\ref{fig:tA}.
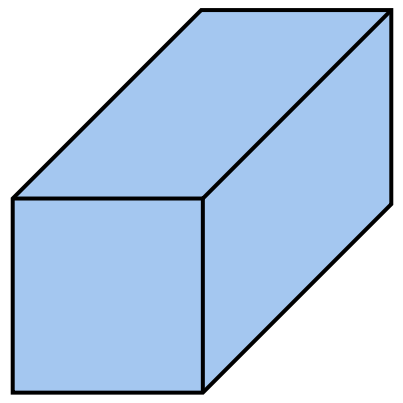
\begin{figure}[!htbp]
	
	\centering

	\tikzset{every picture/.style={line width=0.75pt}} 
	
	\begin{tikzpicture}[x=0.75pt,y=0.75pt,yscale=-1,xscale=1]
		
		\draw  [fill={rgb, 255:red, 74; green, 144; blue, 226 }  ,fill opacity=0.5 ] (69.33,120.76) -- (116.63,73.47) -- (164.33,73.47) -- (164.33,122.17) -- (117.04,169.47) -- (69.33,169.47) -- cycle ; \draw   (164.33,73.47) -- (117.04,120.76) -- (69.33,120.76) ; \draw   (117.04,120.76) -- (117.04,169.47) ;

	\end{tikzpicture}
	
	\caption{\label{fig:tA}A pictorial demonstration of $\tA\in \Lambda^2(U)\otimes V$.}
\end{figure}

The first step is to make a tensor that is alternating in three directions. This is done by setting a $2\times 2\times 2$ block tensor, with rotated $\tA$ in the $(1, 1, 2)$ and $(1, 2, 1)$ positions, and $-\tA$ in the $(2, 1, 1)$ position. Thus we get a tensor $\tilde\tA$ in Figure~\ref{fig:3tA}.

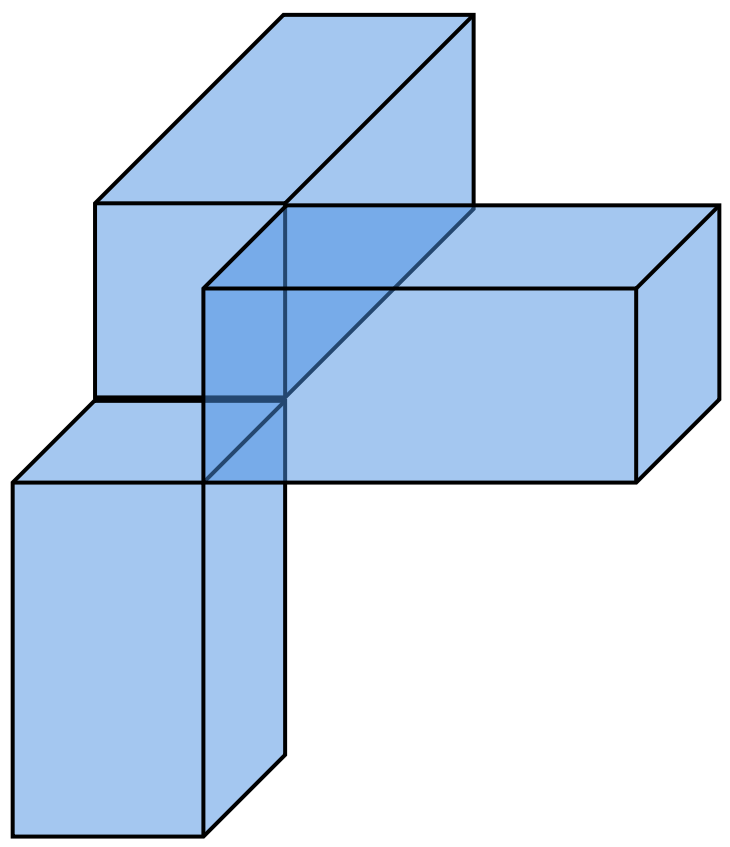
\begin{figure}[!htbp]

\centering

\tikzset{every picture/.style={line width=0.75pt}} 

\begin{tikzpicture}[x=0.75pt,y=0.75pt,yscale=-1,xscale=1]
	
	\draw  [fill={rgb, 255:red, 74; green, 144; blue, 226 }  ,fill opacity=0.5 ] (234.67,115.76) -- (281.96,68.47) -- (329.67,68.47) -- (329.67,117.17) -- (282.37,164.47) -- (234.67,164.47) -- cycle ; \draw   (329.67,68.47) -- (282.37,115.76) -- (234.67,115.76) ; \draw   (282.37,115.76) -- (282.37,164.47) ;
	\draw  [fill={rgb, 255:red, 74; green, 144; blue, 226 }  ,fill opacity=0.5 ] (214,185.84) -- (234.51,165.33) -- (282.37,165.33) -- (282.37,254.16) -- (261.86,274.67) -- (214,274.67) -- cycle ; \draw   (282.37,165.33) -- (261.86,185.84) -- (214,185.84) ; \draw   (261.86,185.84) -- (261.86,274.67) ;
	\draw  [fill={rgb, 255:red, 74; green, 144; blue, 226 }  ,fill opacity=0.5 ] (261.86,137.14) -- (282.73,116.27) -- (391.33,116.27) -- (391.33,164.97) -- (370.46,185.84) -- (261.86,185.84) -- cycle ; \draw   (391.33,116.27) -- (370.46,137.14) -- (261.86,137.14) ; \draw   (370.46,137.14) -- (370.46,185.84) ;

\end{tikzpicture}

\caption{\label{fig:3tA}A pictorial demonstration of an alternating tensor $\tilde\tA$ from $\tA\in \Lambda^2(U)\otimes V$.}
\end{figure}

The second step is to install the alternating FGS tensor $\tG=\tG_{\ell+m+1, \ell}$ to $\tilde\tA$, with three copies appropriately rotated and signed, as shown in Figure~\ref{fig:3tA_FGS}.

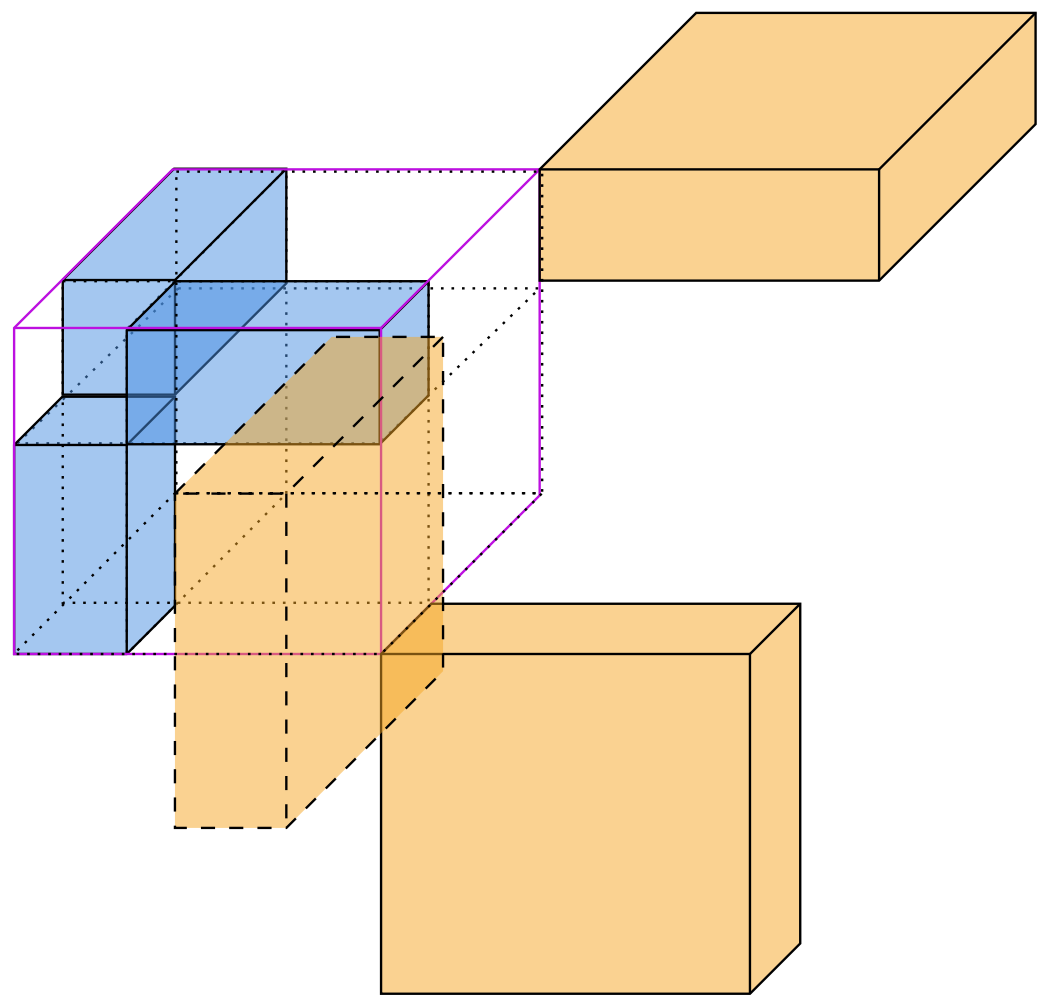
\begin{figure}[!htbp]
	
	\centering

\tikzset{every picture/.style={line width=0.75pt}} 
\resizebox{7cm}{!}{

	\tikzset{every picture/.style={line width=0.75pt}} 
	
	\begin{tikzpicture}[x=0.75pt,y=0.75pt,yscale=-1,xscale=1]
		
		\draw  [dash pattern={on 0.84pt off 2.51pt}] (158.24,-9.72) -- (311.84,-9.72) -- (243.56,56.11) -- (89.96,56.11) -- cycle ;
		\draw  [dash pattern={on 0.84pt off 2.51pt}] (204.2,77.48) -- (204.2,-60.56) -- (136.39,7.6) -- (136.39,145.64) -- cycle ;
		\draw  [fill={rgb, 255:red, 74; green, 144; blue, 226 }  ,fill opacity=0.5 ] (109.2,-13.26) -- (156.5,-60.56) -- (204.2,-60.56) -- (204.2,-11.85) -- (156.9,35.44) -- (109.2,35.44) -- cycle ; \draw   (204.2,-60.56) -- (156.9,-13.26) -- (109.2,-13.26) ; \draw   (156.9,-13.26) -- (156.9,35.44) ;
		\draw  [fill={rgb, 255:red, 74; green, 144; blue, 226 }  ,fill opacity=0.5 ] (88.53,56.82) -- (109.04,36.31) -- (156.9,36.31) -- (156.9,125.13) -- (136.39,145.64) -- (88.53,145.64) -- cycle ; \draw   (156.9,36.31) -- (136.39,56.82) -- (88.53,56.82) ; \draw   (136.39,56.82) -- (136.39,145.64) ;
		\draw  [fill={rgb, 255:red, 74; green, 144; blue, 226 }  ,fill opacity=0.5 ] (136.39,8.03) -- (157.18,-12.76) -- (264.64,-12.76) -- (264.64,35.74) -- (243.86,56.52) -- (136.39,56.52) -- cycle ; \draw   (264.64,-12.76) -- (243.86,8.03) -- (136.39,8.03) ; \draw   (243.86,8.03) -- (243.86,56.52) ;
		\draw  [color={rgb, 255:red, 189; green, 16; blue, 224 }  ,draw opacity=1 ] (88.53,7.11) -- (155.92,-60.28) -- (311.84,-60.28) -- (311.84,78.26) -- (244.45,145.64) -- (88.53,145.64) -- cycle ; \draw  [color={rgb, 255:red, 189; green, 16; blue, 224 }  ,draw opacity=1 ] (311.84,-60.28) -- (244.45,7.11) -- (88.53,7.11) ; \draw  [color={rgb, 255:red, 189; green, 16; blue, 224 }  ,draw opacity=1 ] (244.45,7.11) -- (244.45,145.64) ;
		\draw  [dash pattern={on 0.84pt off 2.51pt}] (109.2,-12.76) -- (264.64,-12.76) -- (264.64,123.88) -- (109.2,123.88) -- cycle ;
		\draw  [fill={rgb, 255:red, 245; green, 166; blue, 35 }  ,fill opacity=0.5 ] (244.45,145.64) -- (265.81,124.29) -- (422.56,124.29) -- (422.56,268.64) -- (401.2,290) -- (244.45,290) -- cycle ; \draw   (422.56,124.29) -- (401.2,145.64) -- (244.45,145.64) ; \draw   (401.2,145.64) -- (401.2,290) ;
		\draw  [fill={rgb, 255:red, 245; green, 166; blue, 35 }  ,fill opacity=0.5 ] (311.84,-60.28) -- (378.33,-126.77) -- (522.56,-126.77) -- (522.56,-79.5) -- (456.07,-13.01) -- (311.84,-13.01) -- cycle ; \draw   (522.56,-126.77) -- (456.07,-60.28) -- (311.84,-60.28) ; \draw   (456.07,-60.28) -- (456.07,-13.01) ;
		\draw  [fill={rgb, 255:red, 245; green, 166; blue, 35 }  ,fill opacity=0.5 ][dash pattern={on 4.5pt off 4.5pt}] (156.87,77.48) -- (223.45,10.9) -- (270.79,10.9) -- (270.79,152.95) -- (204.2,219.54) -- (156.87,219.54) -- cycle ; \draw  [dash pattern={on 4.5pt off 4.5pt}] (270.79,10.9) -- (204.2,77.48) -- (156.87,77.48) ; \draw  [dash pattern={on 4.5pt off 4.5pt}] (204.2,77.48) -- (204.2,219.54) ;
		\draw  [dash pattern={on 0.84pt off 2.51pt}] (156.88,77.32) -- (312.91,77.32) -- (244.45,145.64) -- (88.41,145.64) -- cycle ;
		\draw  [dash pattern={on 0.84pt off 2.51pt}] (157.47,-59.32) -- (312.91,-59.32) -- (312.91,77.32) -- (157.47,77.32) -- cycle ;

	\end{tikzpicture}
	
}

\caption{\label{fig:3tA_FGS}A pictorial demonstration of installing three alternating FGS tensors to  $\tilde\tA$ as in \cite{GQT}. Here, the FGS tensors are in orange colors, and unfilled entries are all $0$. They are appopriately rotated to match the sizes. The front tensor is also multiplied with $-1$.}
\end{figure}

The construction as in Figure~\ref{fig:3tA_FGS} works over the field setting. It makes the assumption that the frontal slices of $\tA$ are linearly independent. To remove this assumption, we need to first add three copies of $\tG_{2(\ell+m+1), m}$, appropriately rotated and signed, as in Figure~\ref{fig:3tA_2FGS}.

\begin{figure}[!htbp]
	
	\centering

	\tikzset{every picture/.style={line width=0.75pt}} 
	\resizebox{10cm}{!}{
		

	\begin{tikzpicture}[x=0.75pt,y=0.75pt,yscale=-1,xscale=1]
		
		\draw  [color={rgb, 255:red, 208; green, 2; blue, 27 }  ,draw opacity=1 ][dash pattern={on 4.5pt off 4.5pt}] (240.16,-128.08) -- (240.16,36.06) -- (139.09,139.1) -- (139.09,-25.05) -- cycle ;
		\draw  [fill={rgb, 255:red, 74; green, 144; blue, 226 }  ,fill opacity=0.5 ] (121.7,-39.83) -- (151.94,-70.07) -- (182.44,-70.07) -- (182.44,-38.93) -- (152.2,-8.69) -- (121.7,-8.69) -- cycle ; \draw   (182.44,-70.07) -- (152.2,-39.83) -- (121.7,-39.83) ; \draw   (152.2,-39.83) -- (152.2,-8.69) ;
		\draw  [fill={rgb, 255:red, 74; green, 144; blue, 226 }  ,fill opacity=0.5 ] (108.49,4.97) -- (121.6,-8.14) -- (152.2,-8.14) -- (152.2,48.65) -- (139.09,61.76) -- (108.49,61.76) -- cycle ; \draw   (152.2,-8.14) -- (139.09,4.97) -- (108.49,4.97) ; \draw   (139.09,4.97) -- (139.09,61.76) ;
		\draw  [fill={rgb, 255:red, 74; green, 144; blue, 226 }  ,fill opacity=0.5 ] (139.09,-26.22) -- (152.37,-39.51) -- (221.08,-39.51) -- (221.08,-8.51) -- (207.79,4.78) -- (139.09,4.78) -- cycle ; \draw   (221.08,-39.51) -- (207.79,-26.22) -- (139.09,-26.22) ; \draw   (207.79,-26.22) -- (207.79,4.78) ;
		\draw  [color={rgb, 255:red, 189; green, 16; blue, 224 }  ,draw opacity=1 ] (108.49,-26.81) -- (151.57,-69.89) -- (251.25,-69.89) -- (251.25,18.68) -- (208.17,61.76) -- (108.49,61.76) -- cycle ; \draw  [color={rgb, 255:red, 189; green, 16; blue, 224 }  ,draw opacity=1 ] (251.25,-69.89) -- (208.17,-26.81) -- (108.49,-26.81) ; \draw  [color={rgb, 255:red, 189; green, 16; blue, 224 }  ,draw opacity=1 ] (208.17,-26.81) -- (208.17,61.76) ;
		\draw  [fill={rgb, 255:red, 245; green, 166; blue, 35 }  ,fill opacity=0.8 ] (251.25,-70) -- (309.65,-128.4) -- (414.56,-128.4) -- (414.56,-92.93) -- (356.16,-34.53) -- (251.25,-34.53) -- cycle ; \draw   (414.56,-128.4) -- (356.16,-70) -- (251.25,-70) ; \draw   (356.16,-70) -- (356.16,-34.53) ;
		\draw  [fill={rgb, 255:red, 245; green, 166; blue, 35 }  ,fill opacity=0.8 ][dash pattern={on 4.5pt off 4.5pt}] (152.18,18.18) -- (194.75,-24.39) -- (225.01,-24.39) -- (225.01,73.03) -- (182.44,115.6) -- (152.18,115.6) -- cycle ; \draw  [dash pattern={on 4.5pt off 4.5pt}] (225.01,-24.39) -- (182.44,18.18) -- (152.18,18.18) ; \draw  [dash pattern={on 4.5pt off 4.5pt}] (182.44,18.18) -- (182.44,115.6) ;
		\draw  [color={rgb, 255:red, 208; green, 2; blue, 27 }  ,draw opacity=1 ][dash pattern={on 4.5pt off 4.5pt}] (122.45,-38.5) -- (322.04,-38.5) -- (322.04,140.21) -- (122.45,140.21) -- cycle ;
		\draw  [color={rgb, 255:red, 208; green, 2; blue, 27 }  ,draw opacity=1 ][dash pattern={on 4.5pt off 4.5pt}] (151.57,-69.89) -- (343.36,-69.89) -- (343.36,115.6) -- (151.57,115.6) -- cycle ;
		\draw  [fill={rgb, 255:red, 189; green, 16; blue, 224 }  ,fill opacity=0.5 ] (319.36,141.58) -- (345.34,115.6) -- (508.96,115.6) -- (508.96,264.02) -- (482.98,290) -- (319.36,290) -- cycle ; \draw   (508.96,115.6) -- (482.98,141.58) -- (319.36,141.58) ; \draw   (482.98,141.58) -- (482.98,290) ;
		\draw  [fill={rgb, 255:red, 189; green, 16; blue, 224 }  ,fill opacity=0.5 ] (414.34,-90.91) -- (485.67,-162.25) -- (620.96,-162.25) -- (620.96,-108.99) -- (549.63,-37.65) -- (414.34,-37.65) -- cycle ; \draw   (620.96,-162.25) -- (549.63,-90.91) -- (414.34,-90.91) ; \draw   (549.63,-90.91) -- (549.63,-37.65) ;
		\draw  [color={rgb, 255:red, 208; green, 2; blue, 27 }  ,draw opacity=1 ][dash pattern={on 4.5pt off 4.5pt}] (309.65,-128.4) -- (309.65,36.99) -- (207.79,140.83) -- (207.79,-24.56) -- cycle ;
		\draw  [color={rgb, 255:red, 208; green, 2; blue, 27 }  ,draw opacity=1 ][dash pattern={on 4.5pt off 4.5pt}] (414.34,-90.91) -- (202.89,-90.91) -- (108.49,4.97) -- (319.94,4.97) -- cycle ;
		\draw  [color={rgb, 255:red, 208; green, 2; blue, 27 }  ,draw opacity=1 ][dash pattern={on 4.5pt off 4.5pt}] (414.34,-34.12) -- (202.89,-34.12) -- (108.49,61.76) -- (319.94,61.76) -- cycle ;
		\draw  [fill={rgb, 255:red, 189; green, 16; blue, 224 }  ,fill opacity=0.5 ][dash pattern={on 3.75pt off 3pt on 7.5pt off 1.5pt}] (240.16,36.06) -- (329.88,-53.65) -- (396.86,-53.65) -- (396.86,102.49) -- (307.14,192.21) -- (240.16,192.21) -- cycle ; \draw  [dash pattern={on 3.75pt off 3pt on 7.5pt off 1.5pt}] (396.86,-53.65) -- (307.14,36.06) -- (240.16,36.06) ; \draw  [dash pattern={on 3.75pt off 3pt on 7.5pt off 1.5pt}] (307.14,36.06) -- (307.14,192.21) ;
		\draw  [fill={rgb, 255:red, 245; green, 166; blue, 35 }  ,fill opacity=0.8 ] (208.17,61.76) -- (221.91,48.02) -- (322.04,48.02) -- (322.04,140.02) -- (308.31,153.76) -- (208.17,153.76) -- cycle ; \draw   (322.04,48.02) -- (308.31,61.76) -- (208.17,61.76) ; \draw   (308.31,61.76) -- (308.31,153.76) ;

	\end{tikzpicture}
		
}

\caption{\label{fig:3tA_2FGS}A pictorial demonstration of installing three more alternating FGS tensors to  $\tilde\tA$. Here, the newly added FGS tensors are in purple colors, and unfilled entries are all $0$. They are appopriately rotated to match the sizes. The front tensor is also multiplied with $-1$.}
\end{figure}
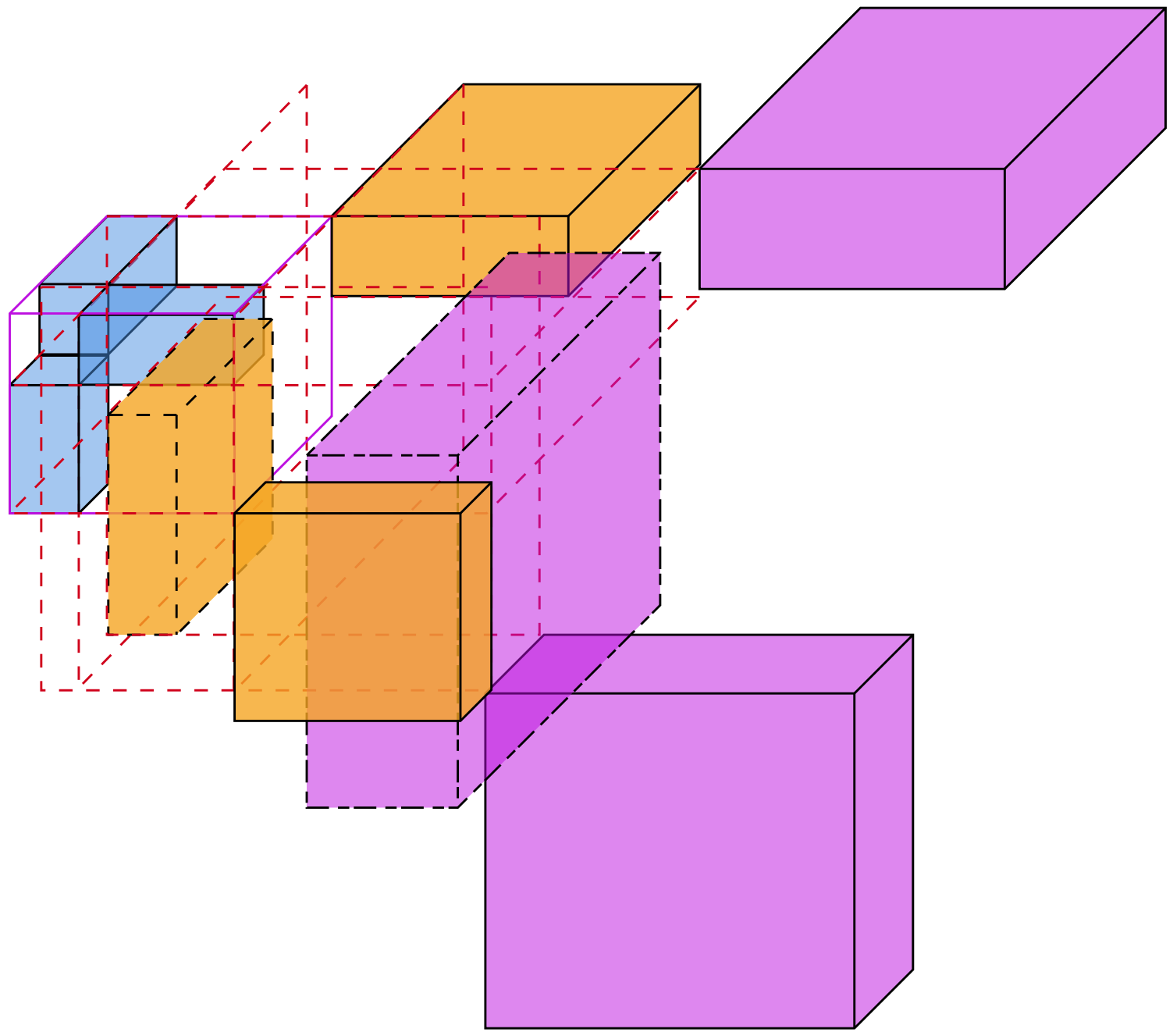

After this step, let $N=
(\ell+1)\cdot(\ell+m+1)+(m+1)\cdot(2\ell+2m+2)$ be the sum of the sizes of the frontal slices of the two added alternating FGS gadgets. We then add further alternating FGS gadgets $\tG_{4(\ell+m+1), N}$ to the three directions as shown in Figure~\ref{fig:3tA_3FGS}.

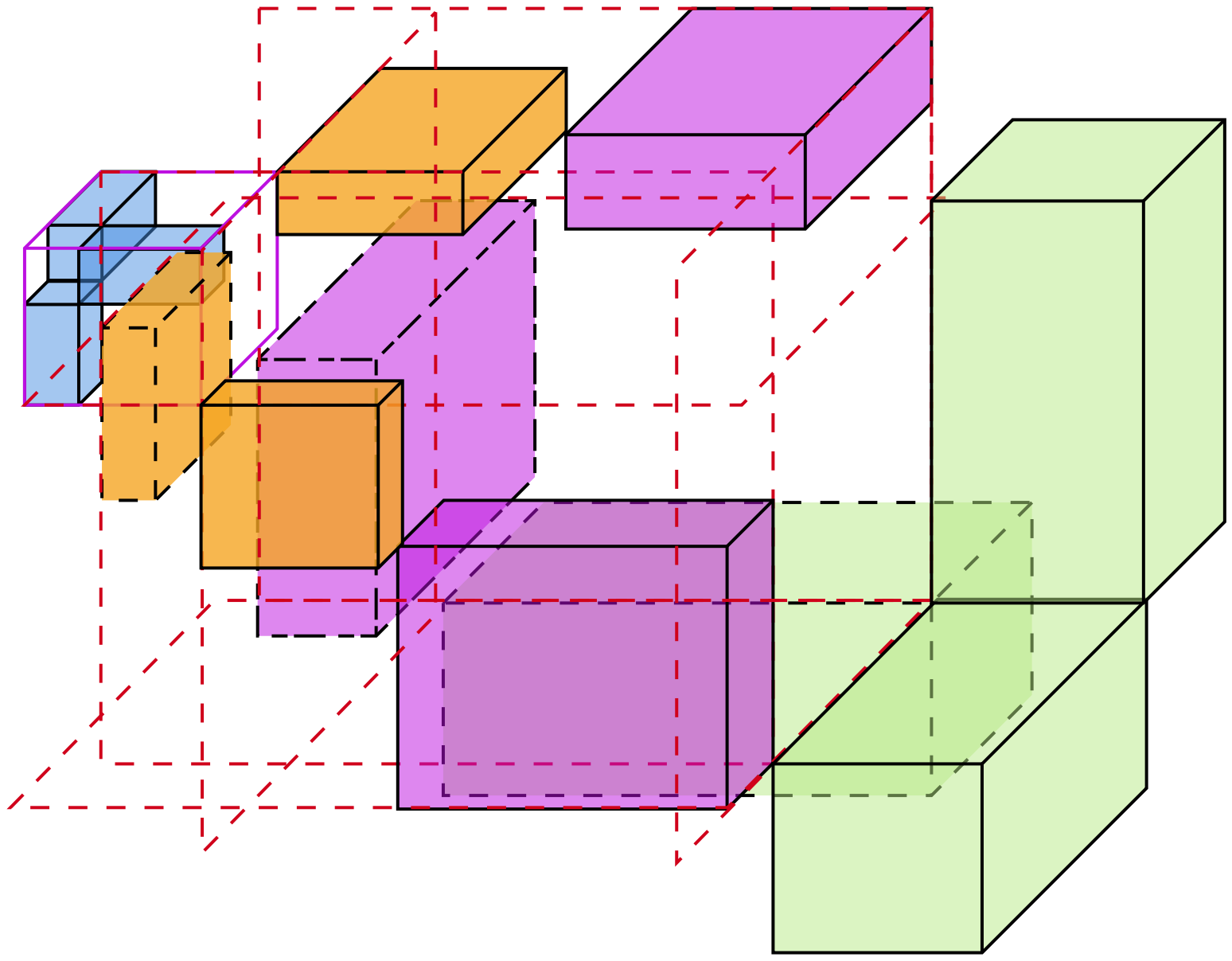
\begin{figure}[!htbp]
	
	\centering

	\tikzset{every picture/.style={line width=0.75pt}} 
	
	\begin{tikzpicture}[x=0.75pt,y=0.75pt,yscale=-1,xscale=1]
		
		\draw  [fill={rgb, 255:red, 184; green, 233; blue, 134 }  ,fill opacity=0.5 ][dash pattern={on 4.5pt off 4.5pt}] (197.28,179.79) -- (228.97,148.1) -- (382.9,148.1) -- (382.9,208.76) -- (351.22,240.45) -- (197.28,240.45) -- cycle ; \draw  [dash pattern={on 4.5pt off 4.5pt}] (382.9,148.1) -- (351.22,179.79) -- (197.28,179.79) ; \draw  [dash pattern={on 4.5pt off 4.5pt}] (351.22,179.79) -- (351.22,240.45) ;
		\draw  [fill={rgb, 255:red, 189; green, 16; blue, 224 }  ,fill opacity=0.5 ][dash pattern={on 3.75pt off 3pt on 7.5pt off 1.5pt}] (138.75,103.05) -- (188.81,52.99) -- (226.18,52.99) -- (226.18,140.11) -- (176.13,190.17) -- (138.75,190.17) -- cycle ; \draw  [dash pattern={on 3.75pt off 3pt on 7.5pt off 1.5pt}] (226.18,52.99) -- (176.13,103.05) -- (138.75,103.05) ; \draw  [dash pattern={on 3.75pt off 3pt on 7.5pt off 1.5pt}] (176.13,103.05) -- (176.13,190.17) ;
		\draw  [fill={rgb, 255:red, 74; green, 144; blue, 226 }  ,fill opacity=0.5 ] (72.66,60.71) -- (89.53,43.83) -- (106.55,43.83) -- (106.55,61.21) -- (89.68,78.08) -- (72.66,78.08) -- cycle ; \draw   (106.55,43.83) -- (89.68,60.71) -- (72.66,60.71) ; \draw   (89.68,60.71) -- (89.68,78.08) ;
		\draw  [fill={rgb, 255:red, 74; green, 144; blue, 226 }  ,fill opacity=0.5 ] (65.29,85.7) -- (72.61,78.39) -- (89.68,78.39) -- (89.68,110.07) -- (82.36,117.39) -- (65.29,117.39) -- cycle ; \draw   (89.68,78.39) -- (82.36,85.7) -- (65.29,85.7) ; \draw   (82.36,85.7) -- (82.36,117.39) ;
		\draw  [fill={rgb, 255:red, 74; green, 144; blue, 226 }  ,fill opacity=0.5 ] (82.36,68.3) -- (89.78,60.89) -- (128.11,60.89) -- (128.11,78.18) -- (120.69,85.6) -- (82.36,85.6) -- cycle ; \draw   (128.11,60.89) -- (120.69,68.3) -- (82.36,68.3) ; \draw   (120.69,68.3) -- (120.69,85.6) ;
		\draw  [color={rgb, 255:red, 189; green, 16; blue, 224 }  ,draw opacity=1 ] (65.29,67.97) -- (89.33,43.93) -- (144.94,43.93) -- (144.94,93.35) -- (120.91,117.39) -- (65.29,117.39) -- cycle ; \draw  [color={rgb, 255:red, 189; green, 16; blue, 224 }  ,draw opacity=1 ] (144.94,43.93) -- (120.91,67.97) -- (65.29,67.97) ; \draw  [color={rgb, 255:red, 189; green, 16; blue, 224 }  ,draw opacity=1 ] (120.91,67.97) -- (120.91,117.39) ;
		\draw  [fill={rgb, 255:red, 245; green, 166; blue, 35 }  ,fill opacity=0.8 ] (144.94,43.87) -- (177.53,11.29) -- (236.06,11.29) -- (236.06,31.08) -- (203.47,63.66) -- (144.94,63.66) -- cycle ; \draw   (236.06,11.29) -- (203.47,43.87) -- (144.94,43.87) ; \draw   (203.47,43.87) -- (203.47,63.66) ;
		\draw  [fill={rgb, 255:red, 245; green, 166; blue, 35 }  ,fill opacity=0.8 ][dash pattern={on 4.5pt off 4.5pt}] (89.66,93.07) -- (113.42,69.32) -- (130.3,69.32) -- (130.3,123.68) -- (106.55,147.43) -- (89.66,147.43) -- cycle ; \draw  [dash pattern={on 4.5pt off 4.5pt}] (130.3,69.32) -- (106.55,93.07) -- (89.66,93.07) ; \draw  [dash pattern={on 4.5pt off 4.5pt}] (106.55,93.07) -- (106.55,147.43) ;
		\draw  [color={rgb, 255:red, 208; green, 2; blue, 27 }  ,draw opacity=1 ][dash pattern={on 4.5pt off 4.5pt}] (89.33,43.93) -- (301.28,43.93) -- (301.28,230.49) -- (89.33,230.49) -- cycle ;
		\draw  [fill={rgb, 255:red, 189; green, 16; blue, 224 }  ,fill opacity=0.5 ] (182.94,161.92) -- (197.44,147.43) -- (301.28,147.43) -- (301.28,230.23) -- (286.78,244.73) -- (182.94,244.73) -- cycle ; \draw   (301.28,147.43) -- (286.78,161.92) -- (182.94,161.92) ; \draw   (286.78,161.92) -- (286.78,244.73) ;
		\draw  [fill={rgb, 255:red, 189; green, 16; blue, 224 }  ,fill opacity=0.5 ] (235.94,32.21) -- (275.74,-7.59) -- (351.22,-7.59) -- (351.22,22.12) -- (311.42,61.92) -- (235.94,61.92) -- cycle ; \draw   (351.22,-7.59) -- (311.42,32.21) -- (235.94,32.21) ; \draw   (311.42,32.21) -- (311.42,61.92) ;
		\draw  [color={rgb, 255:red, 208; green, 2; blue, 27 }  ,draw opacity=1 ][dash pattern={on 4.5pt off 4.5pt}] (194.88,-6.4) -- (194.88,183.46) -- (121.28,258.49) -- (121.28,68.63) -- cycle ;
		\draw  [color={rgb, 255:red, 208; green, 2; blue, 27 }  ,draw opacity=1 ][dash pattern={on 4.5pt off 4.5pt}] (355.68,52.09) -- (129.58,52.09) -- (65.3,117.39) -- (291.4,117.39) -- cycle ;
		\draw  [fill={rgb, 255:red, 245; green, 166; blue, 35 }  ,fill opacity=0.8 ] (120.91,117.43) -- (128.57,109.77) -- (184.44,109.77) -- (184.44,161.1) -- (176.78,168.76) -- (120.91,168.76) -- cycle ; \draw   (184.44,109.77) -- (176.78,117.43) -- (120.91,117.43) ; \draw   (176.78,117.43) -- (176.78,168.76) ;
		\draw  [color={rgb, 255:red, 208; green, 2; blue, 27 }  ,draw opacity=1 ][dash pattern={on 4.5pt off 4.5pt}] (139.27,-7.59) -- (351.22,-7.59) -- (351.22,178.96) -- (139.27,178.96) -- cycle ;
		\draw  [color={rgb, 255:red, 208; green, 2; blue, 27 }  ,draw opacity=1 ][dash pattern={on 4.5pt off 4.5pt}] (351.22,178.96) -- (125.12,178.96) -- (60.83,244.26) -- (286.93,244.26) -- cycle ;
		\draw  [color={rgb, 255:red, 208; green, 2; blue, 27 }  ,draw opacity=1 ][dash pattern={on 4.5pt off 4.5pt}] (351.22,-7.59) -- (351.22,179.79) -- (270.88,261.69) -- (270.88,74.3) -- cycle ;
		\draw  [fill={rgb, 255:red, 184; green, 233; blue, 134 }  ,fill opacity=0.5 ] (301.28,230.49) -- (353.13,178.63) -- (419.03,178.63) -- (419.03,238.15) -- (367.17,290) -- (301.28,290) -- cycle ; \draw   (419.03,178.63) -- (367.17,230.49) -- (301.28,230.49) ; \draw   (367.17,230.49) -- (367.17,290) ;
		\draw  [fill={rgb, 255:red, 184; green, 233; blue, 134 }  ,fill opacity=0.5 ] (351.22,53.08) -- (376.82,27.48) -- (443.68,27.48) -- (443.68,154.19) -- (418.08,179.79) -- (351.22,179.79) -- cycle ; \draw   (443.68,27.48) -- (418.08,53.08) -- (351.22,53.08) ; \draw   (418.08,53.08) -- (418.08,179.79) ;

	\end{tikzpicture}

	\caption{\label{fig:3tA_3FGS}A pictorial demonstration of installing further three more alternating FGS tensors to  $\tilde\tA$. Here, the newly added FGS tensors are in green colors, and unfilled entries are all $0$. They are appopriately rotated to match the sizes. The front tensor is also multiplied with $-1$. Note that the frontal slices of the newly added FGS tensors are of size $(N+1)\cdot 4(\ell+m+1)=O((\ell+m)^3)$, so the green blocks are actually quite wide, which is not shown proportionally in the picture.}
\end{figure}

Let $U'=\rng^{\ell+m+N+(N+1)\cdot 4(\ell+m+1)}$. Let the resulting alternating tensor from the above process be $\bar\tA$. We need to show that $\tA$ and $\tB$ are isomorphic as in $\Lambda^2(U)\otimes V$ if and only if $\bar\tA$ and $\bar\tB$ are isomorphic as in $\Lambda^3(U')$.

\paragraph{The if direction.} Recall that $U'=\rng^{\ell+m+N+(N+1)\cdot 4(\ell+m+1)}$. For the if direction, suppose $\bar\tA$ and $\bar\tB$ are isomorphic as in $\Lambda^3(U')$ via 
$$P=\begin{bmatrix}
	P_{11} & P_{12} & P_{13} & P_{14} & P_{15}\\
	P_{21} & P_{22} & P_{23} & P_{24} & P_{25}\\
	P_{31} & P_{32} & P_{33} & P_{34} & P_{35}\\
	P_{41} & P_{42} & P_{43} & P_{44} & P_{45}\\
	P_{51} & P_{52} & P_{53} & P_{54} & P_{55}\\
\end{bmatrix},$$
where $P_{11}$ is of size $\ell\times \ell$, $P_{22}$ is of size $m\times m$, $P_{33}$ is of size $N\times N$, $P_{44}$ is of size $4(\ell+m+1)\times4(\ell+m+1)$, and $P_{55}$ is of size $4N(\ell+m+1)\times 4N(\ell+m+1)$. We then examine the ranks of the frontal slices.
\begin{enumerate}
	\item For the first $\ell$ slices, their ranks are between $2(\ell+m+1)$ and $2(\ell+m+1)+\ell+m$.
	\item For the next $m$ slices, their ranks are between $4(\ell+m+1)$ and $4(\ell+m+1)+\ell+m$. 
	
	\item For the next $N$ slices, their ranks are between $8(\ell+m+1)$ and $8(\ell+m+1)+2(\ell+m)$.
	
	\item For the next $4(\ell+m+1)$ slices, their ranks are $2N=\Theta((\ell+m)^2)$. 
	
	\item For the last few slices, their ranks are $2$.
\end{enumerate}
Because of these rank statistics, by an analysis similar to that after Equation~\ref{eq:P55}, we obtain that $P$ is of the form 
\begin{equation}\label{eq:P-form}
	\begin{bmatrix}
	P_{11} & 0 & 0 & 0 & 0\\
	0 & P_{22} & 0 & 0 & 0\\
	0 & 0 & P_{33} & 0 & 0\\
	0 & 0 & 0 & P_{44} & 0\\
	P_{51} & P_{52} & P_{53} & P_{54} & P_{55}\\
\end{bmatrix}.
\end{equation}

We claim that the $(1, 1, 2)$ block of $P^t\bar\tA^P P$ is $P_{11}^t\tA^{P_{22}}P_{11}$. If this claim holds, then by $P$ sending $\bar\tA$ to $\bar\tB$, we have $P_{11}^t\tA^{P_{22}}P_{11}=\tB$, showing that $\tA$ and $\tB$ are isomorphic in $\Lambda^2(U)\otimes V$.

We first examine how a matrix $P$ in the form of Equation~\ref{eq:P-form} acts on the frontal slices with indices in $[\ell+1, \ell+m]$. Note that the frontal slices in this range are of the form. 
$S=\begin{bmatrix}
	S_{11} & & & & \\
	& & & & \\
	& & S_{33} & & \\
	& & & & \\
	& & & & 
\end{bmatrix}$. Therefore, $\tilde S:=P^tSP=\begin{bmatrix}
P_{11}^tS_{11}P_{11} & & & & \\
& & & & \\
& & P_{33}^tS_{33}P_{33} & & \\
& & & & \\
& & & & 
\end{bmatrix}$. 

Then consider the action of $P$ on the third direction in the $(1, 1)$ direction, which consists of $(1, 1, i)$ blocks where $i\in[5]$. By the form of $P$ in Equation~\ref{eq:FGS-horizontal}, the resulting $(1, 1, 2)$ block is a linear combination of the $(1, 1, 2)$ block and the $(1, 1, 5)$ block. Consider the $(1, 1, 5)$ block of $P^t\bar\tA P$. As the last $4N(\ell+m+1)$ frontal slices are of the form $T=\begin{bmatrix}
	& & & & \\
	& & & & \\
	& & & T_{34} & T_{35}\\
	& & -T_{34}& & \\
	& & -T_{35}& & 
\end{bmatrix}$. So the congruence of $P$ gives a matrix with the $(1, 1)$ block still $0$. Therefore, the $(1, 1, 5)$ block of $P^t\bar\tA P$ is all zero. 

%
Based on the above, we conclude that the $(1, 1, 2)$ block of $P^t\bar\tA^P P$ is $P_{11}^t\tA^{P_{22}}P_{11}$. 

\paragraph{The only if direction.} Suppose $\tA$ and $\tB$ are isomorphic in $\Lambda^2(U)\otimes V$ by $(P, Q)\in \GL(U)\times\GL(V)$. To transform that to an isomorphism $P'$ from $\bar\tA$ to $\bar\tB$, we use Fact~\ref{fact:alt-FGS-tensor} three times, to look after the three FGS gadgets, that is the orange, purple, and green regions in Figure~\ref{fig:3tA_3FGS}, in order: The orange region depends on $P$, the green region on $Q$, and the green regieon depends on the outcomes from the previous two applications of Fact~\ref{fact:alt-FGS-tensor}. 

This concludes the proof.
\end{proof}

\subsubsection{From $U\otimes V\otimes W$ to $U\otimes U^*\otimes W$}\label{subsec:UUstarW}

The reduction from $U\otimes V\otimes W$ to $U\otimes U^*\otimes W$ in the field case was presented in \cite{TI1}. The construction there is simple but its proof assumes non-degeneracy. Here, we use the GQ gadget, in the same line for reducing $U\otimes V\otimes W$ to $\Lambda^2(U')\otimes W'$ as in \cite{TI4}. Interestingly, the GQ gadget is applicable to reducing $U\otimes V\otimes W$ to $U\otimes U^*\otimes W$ (but not $\Lambda^2(U')\otimes W'$), because in the conjugaction action, the restriction on the transformation matrices caused by the newly added slice are more strigent.

\begin{theorem} \label{thm:3-tensor_conjugacy}
	Let $U=\rng^\ell$, $V=\rng^m$, $W=\rng^n$. There is a polynomial-time algorithm $\algprobm{P}$ that takes $\tA\in U\otimes V\otimes W$ and outputs $\algprobm{P}(\tA)\in U'\otimes U'^*\otimes V'$ with $U'= \rng^{\ell'}$ and $V'=\rng^{m'}$ with $\ell', m'=O(\ell+m+n)$, such that $\tA, \tB\in U\otimes V\otimes W$ are isomorphic if and only if $\algprobm{P}(\tA)$ and $\algprobm{P}(\tB)$ are isomorphic.
	%
\end{theorem}
\begin{proof}
	\paragraph{The algorithm procedure $\algprobm{P}$.}
	For $\tA$ in $U\otimes V\otimes W$, let the frontal slices of $\tA$ be $A_1, \dotsc, A_n$, where $A_i\in\M(\ell\times m, \rng)$. 
	
	Fix an integer $r$ such that $r>\max\{\ell,m\}$. The algorithm $\algprobm{P}$ constructs the following matrices. 
	\begin{itemize}
		\item For $i=1,\dotsc,n$,
		\[
		\tilde A_{i} =
		\begin{bmatrix}
			0_{\ell} & A_i & 0_{\ell \times \ell} & 0_{\ell \times r}  \\
			0 & 0_{m}  \\
			& & 0_\ell \\
			& & & 0_{r} \\
		\end{bmatrix}.
		\]
		
		\item A standard alternating slice of rank $\ell+r$, connected to the first $n$ rows by an $\idmat_n$ in the appropriate place:
		\[
		\tilde A_{n+1} = 
		\begin{bmatrix}
			0_\ell& 0_{\ell \times m}  & \idmat_\ell  & 0_{\ell \times r} \\
			0_{m \times \ell} & 0_m & \\
			0 &  & 0_\ell \\
			& & & \idmat_r \\
		\end{bmatrix}.
		\]
	\end{itemize}
	Let $\ell'=\ell+m+\ell+r$ and $m'=n+1$. Let $U'=\rng^{\ell'}$ and $V'=\rng^{m'}$. Then $\algprobm{P}(\tA)=(\tilde A_1, \dots, \tilde A_n, \tilde A_{n+1})\in U'\otimes U'^*\otimes V'$.
	Let $\tA, \tB\in U\otimes V\otimes W$. For convenience, let $\cA=\algprobm{P}(\tA)$ and $\cB=\algprobm{P}(\tB)$. We now show that $\tA\cong \tB$ if and only if $\cA\cong \cB$.
	
	\paragraph{$\tA\cong \tB\Rightarrow\cA\cong \cB$.} Suppose $\tA \cong \tB$, via $(P,Q,R)\in\GL(U)\times\GL(V)\times\GL(W)$, that is, $(P \tA Q^t)^R = \tB$, or in coordinates
	\[
	\tB(i,j,k) = \sum_{i'j'k'} P_{ii'} Q_{jj'} R_{kk'} \tA(i',j',k').
	\]
	
	Recall that $U'=\rng^{\ell+m+\ell+2r}$ and $V'=\rng^{n+1}$. 
	Let $\tilde P = \diag(P, Q^{-t}, P, \idmat_{r})\in \GL(U')$, $\tilde R = \begin{bmatrix} R & 0 \\ 0 & 1 \end{bmatrix}\in\GL(V')$.  It can be verified that $(\tilde P, \tilde R)$ sends $\cA$ to $\cB$. 
	
	\paragraph{$\tA\cong \tB\Leftarrow\cA\cong \cB$.} Suppose that $\cA$ and $\cB$ are isomorphic via $(P, R)\in\GL(U')\times\GL(V')$, that is, $\sum_{i'} R_{ii'} P \tilde A_{i'} P^{-1} = B_i$ for $i = 1,\dotsc,\ell+1$. Let 
	\[
	R = \begin{bmatrix}
		R_{11} & R_{12} \\
		R_{21} & R_{22}
	\end{bmatrix}
	\]
	where $R_{11}$ is $n \times n$ and $R_{22}$ is $1 \times 1$. By a rank argument, $R_{12} = 0$. Therefore, $R=\begin{bmatrix}
		R_{11} & 0 \\
		R_{21} & R_{22}
	\end{bmatrix}$ is a block matrix. By the discussion on block matrices in $\GL(n, \rng)$, we have $R_{22}=\sigma$ which is a unit in $\rng$. 
	Let $(\tilde A_1',\dotsc, \tilde A_{n+1}') = (\tilde A_1, \dotsc, \tilde A_{n+1})^R$. Since the actions of $P$ and $R$ commute with one another, we have $P \tilde A_i' P^{-1} = \tilde B_i$ for all $i=1,\dotsc,\ell+1$. 
	
	We have 
	\[
	\tilde A_{n+1}' = \begin{bmatrix}
		0 & X & \sigma  \IdMat_\ell & 0  \\
		0 & 0 & 0 & 0 \\
		0 & 0 & 0 & 0 \\
		0 & 0 & 0 & \sigma I_r
	\end{bmatrix},
	\]
	where $X = \sum_{i=1}^\ell (R_{21})_{1i} A_i$, $\sigma\in\rng$ is a unit, and the block sizes are $\ell, m, \ell, r$ (in order).  
	Let
	$
	P_0 := \begin{bmatrix}
		\IdMat_\ell & \\
		& \IdMat_m &   \\
		& X/\sigma & \IdMat_\ell  \\
		& & & \IdMat_{r}
	\end{bmatrix},
	$
	so 
	$P_0^{-1}=\begin{bmatrix}
		\IdMat_\ell & \\
		& \IdMat_m &   \\
		& -X/\sigma & \IdMat_\ell  \\
		& & & \IdMat_{r}
	\end{bmatrix}.
	$
	Note that dividing $\sigma$ is possible as $\sigma$ is a unit. For $i=1,\dotsc,n$, because of the zeros in the third block-column and block-row of $\tilde A_i'$, $P_0A_i'P_0^t=A_i'$.
	Then $\tilde A_{n+1}'$ is transformed as follows:
	\begin{eqnarray*}
		\begin{bmatrix}
			\IdMat_\ell & \\
			& \IdMat_m &   \\
			& X/\sigma & \IdMat_\ell  \\
			& & & \IdMat_{r}
		\end{bmatrix}
		\begin{bmatrix}
			0 & X & \sigma I \\
			0 & 0 & 0 \\
			0 & 0 & 0 \\
			& & & \sigma I_r
		\end{bmatrix}
		\begin{bmatrix}
			\IdMat_\ell & 0 & 0 \\
			0 &  \IdMat_m &  \\
			0 & -X/\sigma &  \IdMat_\ell \\
			& & & \IdMat_{r}
		\end{bmatrix} \\
		=
		\begin{bmatrix}
			0 & 0 & \sigma \IdMat_\ell \\
			0 & 0 & 0 \\
			0 & 0 & 0 \\
			& & & \sigma I_{2r}
		\end{bmatrix}
		= \sigma \tilde A_{n+1}.
	\end{eqnarray*}
	To summarise, we have $P_0(\tilde A'_1, \dotsc, \tilde A'_{n+1})P_0^{-1} = (\tilde A'_1, \dotsc, \tilde A'_{n}, \sigma \tilde A_{n+1})$. Let $\cA'' = ( \tilde A'_1, \dotsc, \tilde A'_{n}, \sigma \tilde A_{n+1} )$. Since we already have $P \cA' P^{-1} = \cB$ and $\cA'' = P_0 \cA' P_0^{-1}$, we get 
	\[
	(P P_0^{-1}) \cA'' (P_0 P^{-1}) = \cB.
	\]
	Let $P' = P P_0^{-1}$. Then we have $P' \sigma \tilde A_{n+1} (P')^{-1} = \tilde B_{n+1}$. The following lemma indicates certain restrictions on the structure of $P'$.
	
	\begin{lemma} \label{lem:block2}
		Let $\ell,m,r \geq 0$. Suppose a matrix
		\[
		P' = \begin{bmatrix} 
			P_{11} & \dotsb & P_{14} \\
			\vdots & \ddots  & \vdots \\
			P_{41} & \dotsb & P_{44}
		\end{bmatrix},
		\]
		with diagonal block sizes $\ell,m,\ell,r,r$, 
		is such that
		\begin{equation}\label{eq:conj}
			\sigma P'
			\begin{bmatrix}
				0_\ell & 0 & \IdMat_\ell  \\
				0 & 0_m & 0  \\
				0 & 0 & 0_\ell \\
				&  &  &  I_r
			\end{bmatrix}
			(P')^{-1}
			=
			\begin{bmatrix}
				0_\ell & 0 & \IdMat_\ell  \\
				0 & 0_m & 0  \\
				0 & 0 & 0_\ell \\
				&  &  &  I_r
			\end{bmatrix},
		\end{equation}
		for some nonzero scalar $\sigma$. Then 
		$P'=
		\begin{bmatrix}
			P_{11} & P_{12} & P_{13} & P_{14}\\
			0 & P_{22} & P_{23} & 0\\
			0 & 0 & P_{33} & 0 \\
			0 & 0 & P_{43} & P_{44}
		\end{bmatrix}.
		$ In particular, if $P'$ is invertible, $P_{11}$, $P_{22}$, $P_{33}$, $P_{44}$ are all invertible.
	\end{lemma}
	\begin{proof}
		We transform Equation~\ref{eq:conj} to 
		$
		\sigma P'
		\begin{bmatrix}
			0_\ell & 0 & \IdMat_\ell  \\
			0 & 0_m & 0  \\
			0 & 0 & 0_\ell \\
			&  &  &  I_r
		\end{bmatrix}
		=
		\begin{bmatrix}
			0_\ell & 0 & \IdMat_\ell  \\
			0 & 0_m & 0  \\
			0 & 0 & 0_\ell \\
			&  &  &  I_r
		\end{bmatrix}
		P'.
		$
		The left-hand-side gives 
		$\sigma \begin{bmatrix}
			0 & 0 & P_{11} & P_{14}\\
			0 & 0 & P_{21} & P_{24}\\
			0 & 0 & P_{31} & P_{34}\\
			0 & 0 & P_{41} & P_{44}
		\end{bmatrix}, 
		$
		and the right-hand-side gives 
		$
		\begin{bmatrix}
			P_{31} & P_{32} & P_{33} & P_{34}\\
			0 & 0 & 0 & 0 \\
			0 & 0 & 0 & 0 \\
			P_{41} & P_{42} & P_{43} & P_{44}\\
		\end{bmatrix}.
		$
		As $\sigma$ is a unit, we get that 
		$P'=
		\begin{bmatrix}
			P_{11} & P_{12} & P_{13} & P_{14}\\
			0 & P_{22} & P_{23} & 0\\
			0 & 0 & P_{33} & 0 \\
			0 & 0 & P_{43} & P_{44}
		\end{bmatrix}.
		$
	\end{proof}
	
	Now let us consider what happens within the first $\ell$ frontal slices. From the action of $R$ on the original tuple $(\tilde A_1, \dotsc, \tilde A_{\ell+1})$, and the fact that $P_0$ did not affect the first $\ell$ matrices, we have $\tilde A_i' = \sum_{i'} R_{ii'} \tilde A_i$ for $i=1,\dotsc,\ell$. Such a slice has the form
	\[
	\tilde A_i' = \begin{bmatrix}
		0 & A_i' \\
		0 & 0 \\
		& & 0_{\ell+2r}
	\end{bmatrix}
	\]
	where all empty blocks are zero, and $A_i' = \sum_{i'} R_{ii'} A_i$. Let us see how $P'$ affects such a slice. Since $(P',R)$ sends $(\tilde A_1', \dotsc, \tilde A_{\ell}', \sigma \tilde A_{\ell+1})$ to $(\tilde B_1, \dotsc, \tilde B_{\ell+1})$, it must be the case that $P' \tilde A_i' (P')^{-1} = \tilde B_i$ for $i=1,\dotsc,\ell$. We will now focus only on the upper $2 \times 2$ blocks, grouping the remaining blocks all together:
	\begin{align*}
		\begin{bmatrix} 
			P_{11} & P_{12} & * \\
			0 & P_{22} & * \\
			* & * & *
		\end{bmatrix}
		\begin{bmatrix}
			0 & A_i' \\
			0 & 0 \\
			& & 0_{\ell+r}
		\end{bmatrix}
		& =
		\begin{bmatrix}
			0 & B_i' \\
			0 & 0 \\
			& & 0_{\ell+r}
		\end{bmatrix}
		\begin{bmatrix} 
			P_{11} & P_{12}& * \\
			0 & P_{22} & *  \\
			* & * & *
		\end{bmatrix}
	\end{align*}
	(Because of the zeros in $\tilde A_i'$, the $*$ blocks play no role.) The (1,2) entry here gives us the equation
	\[
	P_{11}A_i'=BP_{22}.
	\]
	As $P'$ is invertible, we have $P_{11}$ and $P_{22}$ are invertible, so we have
	\[
	P_{11} A_i' P_{22}^{-1} = B_i.
	\]
	Recalling that $A_i' = \sum_{i'} R_{ii'} A_i$, the preceding equation, when applied for all $i=1,\dotsc,n$, gives us
	\begin{equation}\label{eq:coset}
		(P_{11} \tA P_{22}^t)^{R_{11}} = \tB,
	\end{equation}
	where $P_{11}$, $P_{22}$, $R_{11}$ are all invertible. Therefore, $\tA \cong \tB$ as tensors, as desired.
\end{proof}

\subsubsection{From $U\otimes U\otimes V$ to $U\otimes U\otimes U^*$}\label{subsec:UUUstar}

In this section we present a reduction from $U\otimes U\otimes V$ to $U\otimes U\otimes U^*$. A reduction in the field case was shown in \cite[Lemma 6.4]{TI1}. The proof there relies on linear independence along the $V$ direction. Here we use the FGS gadget to avoid using the linear independence assumption.

\begin{theorem} \label{thm:congruence-algebra}
	Let $U=\rng^\ell$ and $V=\rng^m$. There is a polynomial-time algorithm $\algprobm{P}$ that takes $\tA\in U\otimes U\otimes V$ and outputs $\algprobm{P}(\tA)\in U'\otimes U'\otimes U'^*$ with $U'= \rng^{\ell'}$, $\ell'=O((\ell+m)^2)$, such that $\tA$ and $\tB\in U\otimes U\otimes V$ are isomorphic if and only if $\algprobm{P}(\tA)$ and $\algprobm{P}(\tB)\in U'\otimes U'\otimes U'^*$ are isomorphic.
	
	In particular, for $\tA\in \S^2(U)\otimes V$, $\algprobm{P}$ outputs $\algprobm{P}(\tA)\in U'\otimes U'\otimes U^*$ that reprensents the structure constants of a commutative associative algebra. For $\tA\in \Lambda^2(U)\otimes V$, $\algprobm{P}$ outputs $\algprobm{P}(\tA)\in U'\otimes U'\otimes U^*$ that reprensents the structure constants of a Lie algebra. 
	%
\end{theorem}
\begin{proof}
We first present the reduction procedure $\algprobm{P}$. For convenience we shall reduce from $U\otimes U\otimes V^*$ to $U'\otimes U'\otimes U'$, as the original $U\otimes U\otimes V$ is easily seen to be equivalent to this case. Furthermore, we also focus on $\tA\in\Lambda^2(U)\otimes V$, as the other cases can be proved by mimicking this case in a straightforward way.

To start with, we represent $\tA\in \Lambda^2(U)\otimes V^*$ as a $3$-tensor as in Figure~\ref{fig:tA2}.
\begin{figure}[!htbp]
	
	\centering

	\tikzset{every picture/.style={line width=0.75pt}} 
	
	\begin{tikzpicture}[x=0.75pt,y=0.75pt,yscale=-1,xscale=1]
		
		\draw  [fill={rgb, 255:red, 74; green, 144; blue, 226 }  ,fill opacity=0.5 ] (69.33,120.76) -- (116.63,73.47) -- (164.33,73.47) -- (164.33,122.17) -- (117.04,169.47) -- (69.33,169.47) -- cycle ; \draw   (164.33,73.47) -- (117.04,120.76) -- (69.33,120.76) ; \draw   (117.04,120.76) -- (117.04,169.47) ;

	\end{tikzpicture}
	
	\caption{\label{fig:tA2}A pictorial demonstration of $\tA\in \Lambda^2(U)\otimes V$.}
\end{figure}
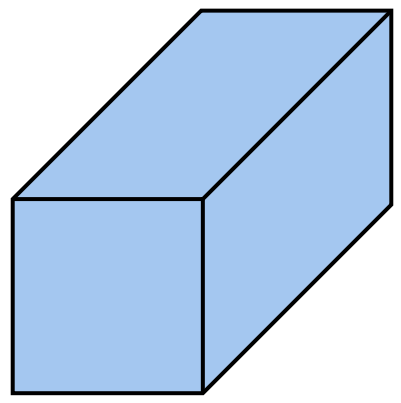

First, following \cite{TI1}, we construct $\tilde\tA\in (U\oplus V)\otimes (U\oplus V)\otimes (U\oplus V)^*$, by putting $\tA$ in the appropriate block, shown in Figure~\ref{fig:tA2_alg}.

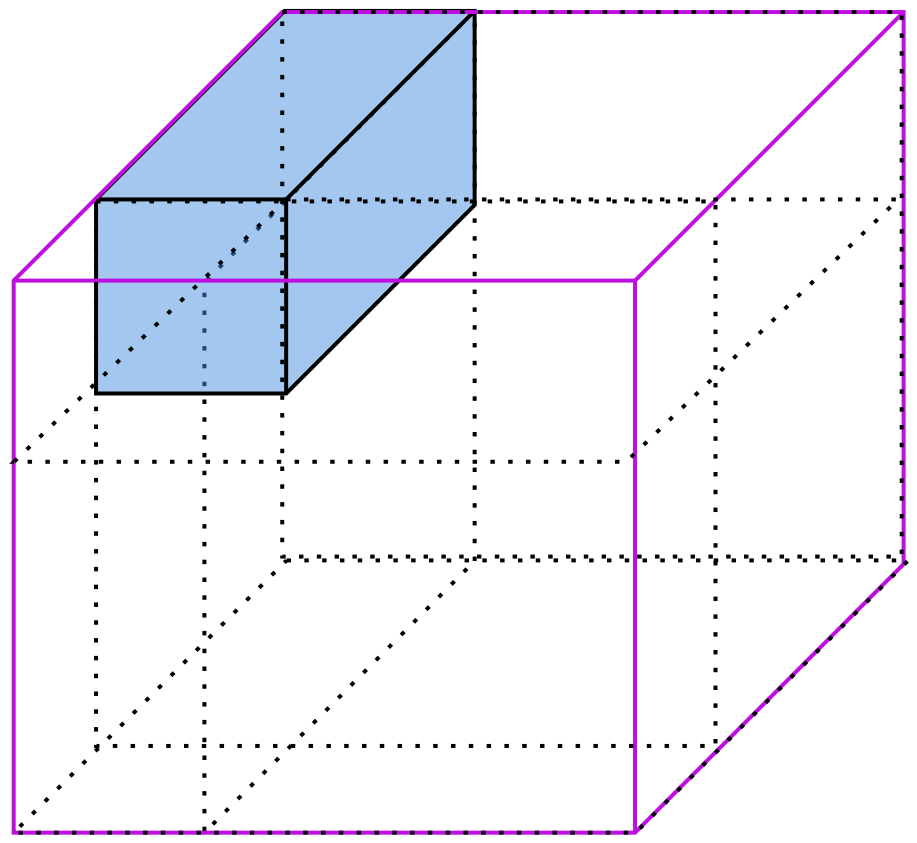
\begin{figure}[!htbp]
	
	\centering

	\tikzset{every picture/.style={line width=0.75pt}} 
	
	\begin{tikzpicture}[x=0.75pt,y=0.75pt,yscale=-1,xscale=1]
		
		\draw  [dash pattern={on 0.84pt off 2.51pt}] (365,177.2) -- (365,39.16) -- (297.19,107.32) -- (297.19,245.36) -- cycle ;
		\draw  [fill={rgb, 255:red, 74; green, 144; blue, 226 }  ,fill opacity=0.5 ] (270,86.46) -- (317.3,39.16) -- (365,39.16) -- (365,87.87) -- (317.7,135.16) -- (270,135.16) -- cycle ; \draw   (365,39.16) -- (317.7,86.46) -- (270,86.46) ; \draw   (317.7,86.46) -- (317.7,135.16) ;
		\draw  [color={rgb, 255:red, 189; green, 16; blue, 224 }  ,draw opacity=1 ] (249.33,106.83) -- (316.72,39.44) -- (472.64,39.44) -- (472.64,177.98) -- (405.25,245.36) -- (249.33,245.36) -- cycle ; \draw  [color={rgb, 255:red, 189; green, 16; blue, 224 }  ,draw opacity=1 ] (472.64,39.44) -- (405.25,106.83) -- (249.33,106.83) ; \draw  [color={rgb, 255:red, 189; green, 16; blue, 224 }  ,draw opacity=1 ] (405.25,106.83) -- (405.25,245.36) ;
		\draw  [dash pattern={on 0.84pt off 2.51pt}] (270,86.96) -- (425.44,86.96) -- (425.44,223.6) -- (270,223.6) -- cycle ;
		\draw  [dash pattern={on 0.84pt off 2.51pt}] (317.7,86.46) -- (471.3,86.46) -- (403.03,152.29) -- (249.43,152.29) -- cycle ;
		\draw  [dash pattern={on 0.84pt off 2.51pt}] (316.72,39.44) -- (472.16,39.44) -- (472.16,176.08) -- (316.72,176.08) -- cycle ;
		\draw  [dash pattern={on 0.84pt off 2.51pt}] (317.8,177.04) -- (473.83,177.04) -- (405.37,245.36) -- (249.33,245.36) -- cycle ;

	\end{tikzpicture}
	
\caption{\label{fig:tA2_alg}A pictorial demonstration of $\tilde\tA\in (U\oplus V)\otimes (U\oplus V)\otimes (U\oplus V)^*$.}
\end{figure}

To show that $\tA\cong \tB$ if and only if $\tilde\tA\cong\tilde\tB$ as in \cite{TI1}, non-degeneracy along $V$ is required. To avoid this, we add two FGS gadgets with certain twists. First, we construct FGS gadgets $\tG_{2(\ell+m+1), \ell}$ and $\tG_{4(\ell+m+1), m}$. We then construct $\tG'_{2(\ell+m+1), \ell}$, a $2(\ell+m+1)\cdot (\ell+1)\times 2(\ell+m+1)\cdot (\ell+1)\times \ell$ tensor, by putting  $\tG_{2(\ell+m+1), \ell}$ to the left-lower subtensor, and zero elsewhere. That is, $\tG'_{2(\ell+m+1), \ell}=\begin{bmatrix}
	0 & 0 \\
	\tG_{2(\ell+m+1), \ell} & 0 
\end{bmatrix}$. Similarly construct $\tG'_{4(\ell+m+1), m}=\begin{bmatrix}
0 & 0 \\
\tG_{4(\ell+m+1), m} & 0 
\end{bmatrix}$. Now install $\tG'_{2(\ell+m+1), \ell}$ and $\tG'_{4(\ell+m+1), m}$ to $\tilde\tA$ as in Figure~\ref{fig:tA2_alg-new}.

\begin{figure}[!htbp]
	
	\centering

	\tikzset{every picture/.style={line width=0.75pt}} 
	
	\begin{tikzpicture}[x=0.75pt,y=0.75pt,yscale=-0.7,xscale=0.7]
		
		\draw  [fill={rgb, 255:red, 74; green, 144; blue, 226 }  ,fill opacity=0.5 ] (66.11,281.92) -- (84.47,263.56) -- (102.98,263.56) -- (102.98,282.46) -- (84.63,300.82) -- (66.11,300.82) -- cycle ; \draw   (102.98,263.56) -- (84.63,281.92) -- (66.11,281.92) ; \draw   (84.63,281.92) -- (84.63,300.82) ;
		\draw  [color={rgb, 255:red, 189; green, 16; blue, 224 }  ,draw opacity=1 ] (58.09,289.82) -- (84.25,263.67) -- (144.76,263.67) -- (144.76,317.44) -- (118.61,343.59) -- (58.09,343.59) -- cycle ; \draw  [color={rgb, 255:red, 189; green, 16; blue, 224 }  ,draw opacity=1 ] (144.76,263.67) -- (118.61,289.82) -- (58.09,289.82) ; \draw  [color={rgb, 255:red, 189; green, 16; blue, 224 }  ,draw opacity=1 ] (118.61,289.82) -- (118.61,343.59) ;
		\draw  [fill={rgb, 255:red, 245; green, 166; blue, 35 }  ,fill opacity=0.8 ] (180.08,229.37) -- (215.53,193.92) -- (279.22,193.92) -- (279.22,215.45) -- (243.77,250.91) -- (180.08,250.91) -- cycle ; \draw   (279.22,193.92) -- (243.77,229.37) -- (180.08,229.37) ; \draw   (243.77,229.37) -- (243.77,250.91) ;
		\draw  [fill={rgb, 255:red, 245; green, 166; blue, 35 }  ,fill opacity=0.8 ] (84.61,422.88) -- (110.46,397.04) -- (128.83,397.04) -- (128.83,456.18) -- (102.98,482.02) -- (84.61,482.02) -- cycle ; \draw   (128.83,397.04) -- (102.98,422.88) -- (84.61,422.88) ; \draw   (102.98,422.88) -- (102.98,482.02) ;
		\draw  [fill={rgb, 255:red, 189; green, 16; blue, 224 }  ,fill opacity=0.5 ] (422.88,166.87) -- (466.19,123.56) -- (548.32,123.56) -- (548.32,155.89) -- (505.01,199.2) -- (422.88,199.2) -- cycle ; \draw   (548.32,123.56) -- (505.01,166.87) -- (422.88,166.87) ; \draw   (505.01,166.87) -- (505.01,199.2) ;
		\draw  [color={rgb, 255:red, 208; green, 2; blue, 27 }  ,draw opacity=1 ][dash pattern={on 4.5pt off 4.5pt}] (201.28,163.36) -- (201.28,381.82) -- (85.44,499.9) -- (85.44,281.44) -- cycle ;
		\draw  [color={rgb, 255:red, 208; green, 2; blue, 27 }  ,draw opacity=1 ][dash pattern={on 4.5pt off 4.5pt}] (376.08,214.72) -- (142.33,214.72) -- (57.28,301.12) -- (291.03,301.12) -- cycle ;
		\draw  [fill={rgb, 255:red, 189; green, 16; blue, 224 }  ,fill opacity=0.5 ] (163.67,554.11) -- (212.77,505.01) -- (248.26,505.01) -- (248.26,584.21) -- (199.16,633.31) -- (163.67,633.31) -- cycle ; \draw   (248.26,505.01) -- (199.16,554.11) -- (163.67,554.11) ; \draw   (199.16,554.11) -- (199.16,633.31) ;
		\draw  [color={rgb, 255:red, 126; green, 211; blue, 33 }  ,draw opacity=1 ] (375.12,214.72) -- (215.36,214.72) -- (144.48,286.72) -- (304.24,286.72) -- cycle ;
		\draw  [color={rgb, 255:red, 65; green, 117; blue, 5 }  ,draw opacity=1 ] (679.68,122.4) -- (466,122.4) -- (375.12,214.72) -- (588.8,214.72) -- cycle ;
		\draw  [color={rgb, 255:red, 126; green, 211; blue, 33 }  ,draw opacity=1 ] (163.84,283.36) -- (163.84,422.11) -- (102.98,482.02) -- (102.98,343.27) -- cycle ;
		\draw  [color={rgb, 255:red, 65; green, 117; blue, 5 }  ,draw opacity=1 ] (255.99,328.75) -- (255.99,542.43) -- (163.67,633.31) -- (163.67,419.63) -- cycle ;

	\end{tikzpicture}

\caption{\label{fig:tA2_alg-new}A pictorial demonstration of $\bar\tA\in U'\otimes U'\otimes U'^*$. The orange and purple regions indicate the positions of $\tG_{2(\ell+m+1), \ell}$ and $\tG_{4(\ell+m+1), m}$, respectively.}
\end{figure}
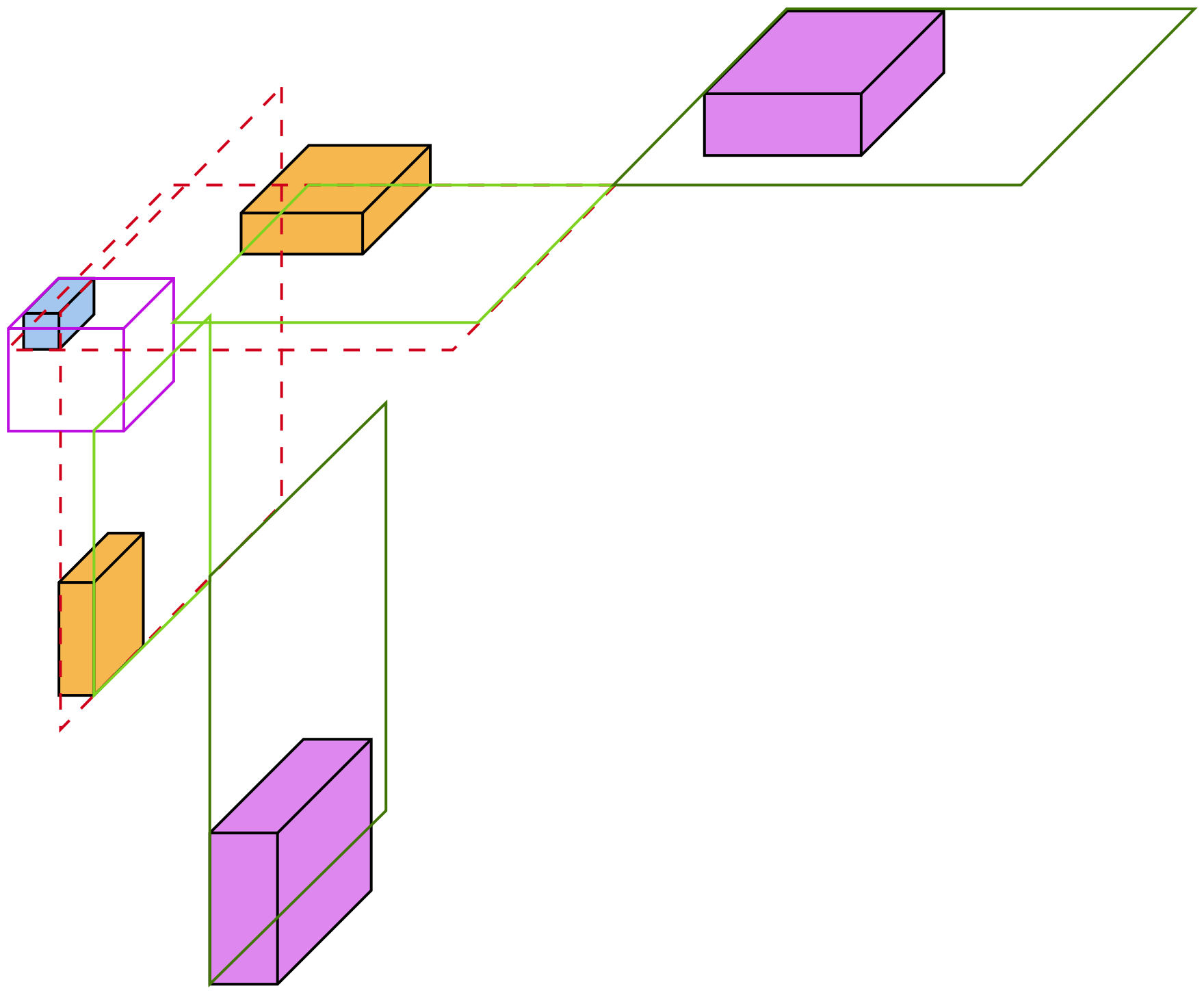

Let the resulting tensor be $\bar\tA$. Let $U'=\rng^{\ell+m+2(\ell+m+1)\cdot(\ell+1)+4(\ell+m+1)\cdot(m+1)}$. We claim that $\tA, \tB\in \Lambda^2(U)\otimes V^*$ are isomorphic if and only if $\bar\tA$ and $\bar\tB$ are isomorphic. The only if direction is straightforward by Fact~\ref{fact:FGS-tensor}. For the if direction, let $P\in \GL(U')$ be an isomorphism from $\bar\tA$ to $\bar\tB$. Because of rank considerations, we see that $P=\begin{bmatrix}
	P_{11} & 0 & 0 \\
	0 & P_{22} & 0 \\
	P_{31} & P_{32} & P_{33}
\end{bmatrix}$ where $P_{11}$ is of size $\ell\times \ell$, and $P_{22}$ is of size $m\times m$. So $P_{11}$ and $P_{22}$ are invertible. By examining the $(1, 1, 2)$-position blocks of $P^t\bar\tA P$ and $\bar\tB^P$, we see that $(P_{11}, P_{22})$ form an isomorphism from $\tA$ to $\tB$. 

Finally, we note that a Lie algebra with structural constants in the form of Figure~\ref{fig:tA2_alg-new} satisfies that for any $u, v, w$, $[u, [v, w]]=0$. That is, this is a nilpotent class-$2$ Lie algebra. This concludes the proof.
\end{proof}

\subsubsection{Proof of Corollary~\ref{cor:TIZ-complete}}

The matrix space conjugacy problem corresponds to the orbit problem for $U\otimes U^*\otimes V$. The algebra isomorphism problem corresponds to the orbit problem for $U\otimes U\otimes U^*$. The trilinear form isomorphism problem corresponds to the orbit problem for $U\otimes U\otimes U$. For the Calabi--Yau threefold classification problem, clearly it generalises the trilinear form isomorphism problem. We then reduce it to symmetric trilinear form isomorphism, so to show that it is $\ThreeFTI_\Z$-complete. 

Recall that the Calabi--Yau threefold isomorphism problem considers a pair of a symmetric trilinear form $\phi:\Z^n\times\Z^n\times\dots\times \Z^n\to \N$ and a linear form $f:\Z^n\to \N$. We construct a symmetric trilinear form $\tilde\phi:\Z^{n+1}\times\Z^{n+1}\times\Z^{n+1}$ as follows. First, extend $f:\Z^n\to \N$ to $\tilde f:\Z^{n+1}\to\N$ by $\tilde f(\stdb_i)=f(\stdb_i)$ for $i\in[n]$, and $f(\stdb_{n+1})=0$. We then construct $\tilde\phi:\Z^{n+1}\times\Z^{n+1}\times\Z^{n+1}$ such that (1) when restricting to $\Z^n$ spanned by the first $n$ coordinates, $\tilde\phi|_{\Z_n}=\phi$, (2) the linear form obtained by restricting the first two coordinates to $\stdb_{n+1}$, that is $\tilde\phi(\stdb_{n+1}, \stdb_{n+1}, v)$ for $v\in \Z^{n+1}$, is $\tilde f$, and (3) extend to other unattended entries by symmetry. 

For another pair of $\psi:\Z^n\times\Z^n\times\dots\times \Z^n\to \N$ and $g:\Z^n\to \N$, we apply the above procedure to get $\tilde\psi$. It is easy to see that $(\phi, f)$ and $(\psi, g)$ are isomorphic under $\GL(n, \Z)$ implies that $\tilde\phi$ and $\tilde\psi$ are isomorphic. For the converse direction, we need to ensure that an isomorphism from $\tilde\phi$ to $\tilde\psi$ which is in $\GL(n+1, \Z)$ does not mix up the first $n$ coordinates and the last coordinate. This can be done by adding the same gadget as in the proof of Theorem~\ref{thm:Lambda3U}. Therefore, we blow-up the dimensions even further, but they are polynomially bounded so that would be fine. This concludes the proof of Corollary~\ref{cor:TIZ-complete}.

\subsection{General modules over finite local principal ideal rings}\label{subsec:five-action-PIR}

Over finite local principal ideal rings (PIRs), a suitable notion of degeneracy can be defined, and a reduction from degenerate to non-degenerate cases can be achieved via the GQ gadget. In this part, we present the main ingredients for adapting the GQ gadget to this seting. 

\paragraph{Some key points for adapting GQ gadgets.} The GQ gadget was reviewed in Section~\ref{sec:GQ}. There are three identity matrices in the added slices (cf. $\tilde C_{m+2}$ in Equation~\ref{eq:GQ}): 
\begin{enumerate}
	\item One identity matrix ($\idmat_{2(n+6)}$ in $\tilde C_{m+2}$) responsible for large enough rank so this slice does not add to others,
	\item A second identity matrix ($\idmat_{n}$ in $\tilde C_{m+2}$) responsible for clearing out those parts added to it, 
	\item A third identity matrix ($\idmat_{2}$ in $\tilde C_{m+2}$) for ensuring the transformation matrices to have a zero-block, so it is in a suitable block diagonal structure.
\end{enumerate}
Once the transformation matrices are put in a block diagonal structure, a non-degeneracy condition is used to ensure that the original tensors get transformed by invertible matrices.

All these properties need to be taken care of for $\Z/p^e\Z$ modules, but they are achievable via the following ingredients. Instead of repeating the proofs from \cite{TI4}, we shall only indicate the key points for achieving Theorem~\ref{thm:ptoe-equivalence} as follows. 

Here we give a brief summary. 
\begin{itemize}
	\item First, we need the diagonal-full property, satisfied by local rings as shown in Section~\ref{subsec:diagonal-full}.
	\item In Section~\ref{subsec:nondeg}, we introduce the non-degeneracy definitions suitable for our purpose. We need two conditions for rings: (1) non-zero non-units are zero-divisors, satisfied by finite rings, and (2) the existence of Smith Normal Forms, satisfied by PIRs.
	\item In Section~\ref{subsec:invertible}, we present a result which suggests that, over $\Z/p^e\Z$, if the transformation matrix preserves the gadget slices, then its diagonal blocks are invertible, despite that it may not be in the strict block upper-triangular structure.
	\item After these, we indicate the differences of the construction needed compared with \cite{TI4} in Remark~\ref{rem:TI4}.
\end{itemize}

\subsubsection{The issue of non-degeneracy}\label{subsec:nondeg}

\paragraph{Non-degeneracy of bilinear forms.}
Let $\rng$ be a commutative ring. 
Let $U$ and $V$ be two modules over $\rng$, and let $f:U\times V\to \rng$ be a bilinear form. For $u\in U$, $f_u:V\to \rng$ is an $\rng$-linear form sending $v\in V$ to $f_u(v)=f(u, v)$. Define $\ker_U(f)=\{u\in U\mid f_u=0\}$. We say that $f$ is \emph{degenerate} along $U$, if $\ker_U(f)$ is non-trivial.
\begin{observation}\label{obs:unit-zd}
	Let $\rng$ be a commutative ring such that every non-unit is a zero-divisor. Let $U$ and $V$ be two modules over $\rng$, and let $f, g:U\times V\to \rng$ be bilinear forms. Suppose there exists $L\in\M(m, \rng)$ such that $f(L(u), v)=g(u, v)$. If $g$ is left non-degenerate, then $L\in\GL(m, \rng)$. 
\end{observation}
\begin{proof}
	For the sake of contradiction, suppose $L\not\in\GL(m, \rng)$. So $\det(L)$ is not a unit, so it is a zero-divisor or 0, which implies that there exists a non-zero $u'\in U$, $Lu'=0$. By $f(L(u), v)=g(u, v)$, this implies that $g(u, v)=f(0, v)=0$, contradicting to that $g$ is non-degenerate.
\end{proof}
Note that a finite ring satisfies that every non-zero element is either a unit or a zero-divisor. Also note that in $\Z$ every non-zero non-unit is a non-zero-divisor, so Observation~\ref{obs:unit-zd} would not be useful in that setting.

\paragraph{Reducing to non-degenerate for bilinear forms over PIRs.}
Now set $\rng$ to be a PIR with the maximal ideal $\m$ with nilpotency index $e$. Let $U$ and $V$ be modules over $\rng$. 


Let $f:U\times V\to \rng$ be a bilinear form and assume that $K:=\ker_U(f)$ is non-trivial. 
By Smith Normal Form, there is a basis $\stdb_1,\dotsc,\stdb_s$ for $U$, together with an integer $r\in[s]$, and exponents $0 \leq k_1 \leq k_2 \leq \dots k_r$  such that $t^{k_1} \stdb_1, t^{k_2} \stdb_2, \dotsc, t^{k_r} \stdb_r$ form a basis for $K$. By our assumption, we can compute this SNF in polynomial time, which includes the elements $\stdb_i$ and the exponents $k_i$.

Suppose under this basis, $U\cong \bigoplus_{i\in[s]} \rng/\m^{a_i}$. Suppose that $V\cong \bigoplus_{i\in[t]} \rng/\m^{b_i}$ with $1\leq b_i\leq e$. Then $f$ can be represented as an $s\times t$ matrix $A_f$ over $\rng$, with the first $r$ rows of $A_f$ corresponding to $\lker(f)$ after multiplying suitable powers of $t$. 

As $t^{k_i}\stdb_i\in \lker(f)$ for $i\in[k]$, it follows that the $i$th row of $A_f$ is divisible by $t^{a_i-k_i}$. Set $\tilde U=(\oplus_{i\in[r]}\rng/\langle t^{k_i}\rangle)\oplus (\oplus_{r+1\leq i\leq s}\rng/\langle t^{a_i}\rangle)$, with $k_i=0$ understood as removing that coordinate. Construct the following matrix $\tilde A_f$: starting from $A$, for $i\in[r]$, if $k_i=0$, then delete the $i$th row of $A$. If $k_i>0$, then mod out $t^{e-k_i}$ on the $i$th row of $A_f$. Then $\tilde A_f$ represents a bilinear form $\tilde f:\tilde U\times V\to \rng$. It can be verified easily that $\tilde f$ is non-degenerate. 

Furthermore, let $f, g:U\times V\to \rng$ be two bilinear forms, and let $\tilde f, \tilde g:\tilde U\times V\to\rng$ be the resulting bilinear forms from the above procedure. It can also be verified easily that $f$ and $g$ are left-equivalent (under $\GL(U)$) if and only if $\tilde f$ and $\tilde g$ are left-equivalent (under $\GL(\tilde U)$). 

\paragraph{Non-degeneracy for trilinear forms and bilinear maps, and reductions to non-degenerate case.} Let $\rng$ be a commutative ring. Let $U, V, W$ be three modules over $\rng$, and $f:U\times V\times W\to\rng$ be a trilinear form. We say that $f$ is non-degenerate along $U$, if $f$ viewed as a bilinear form $U\times (V\otimes W)\to \rng$ is non-degenerate along $U$. Non-degeneracy along $V$ or $W$ is defined similarly.

Given a bilinear map $f:U\times U\to V$, and $u\in U$, let $f_u:U\to V$ be the $\rng$-linear map from $U$ to $V$ defined as $f_u(u')=f(u, u')$. Let $f':U\times U\times V^*\to \rng$ be the trilinear form associated with $f$. We can then define left, right, and image non-degeneracies for $f$ by non-degeneracies of $f'$ along the first $U$, the second $U$, and $V^*$. 



The reduction to non-degeneracy in the bilinear form setting applies to trilinear forms and bilinear maps. This is evident because non-degeneracy is defined in accordance with the bilinear form case. 

\subsubsection{The invertible diagonal lemma}\label{subsec:invertible}

\begin{lemma}[{cf. \cite[Lemma~2.2 and Lemma~5.2]{TI4}}]\label{lem:invertible}
	Suppose $P \Gamma P^t = \Gamma$, where
	\[
	\Gamma = \begin{bmatrix}
		A & 0 & 0 \\ 
		0 & p^{e-1} J_r & 0 \\
		0 & 0 & 0 
	\end{bmatrix}
	\]
	where $J_r = \begin{bmatrix} 0 & I_r \\ -I_r & 0 \end{bmatrix}$, and $r > 2n$, and $A$ is some invertible $n \times n$ (alternating) matrix. If we write
	\[
	P = \begin{bmatrix} 
		P_{11} & P_{12} & P_{13} \\
		P_{21} & P_{22} & P_{23} \\
		P_{31} & P_{32} & P_{33}
	\end{bmatrix}
	\]
	with block sizes commensurate with the blocks of $\Gamma$ above, then $P \pmod{p}$ is block-upper-triangular, $P_{11}, P_{22}, P_{33}$ are invertible, and there is an invertible matrix $Q_{11}$ (such that $Q_{11} \equiv P_{11}^{-1} \pmod{p}$) such that $P_{11} A Q_{11}^{-t} = A$.
\end{lemma}

\begin{proof}
	For simplicity of notation, let $Q = P^{-1}$. Then our equation becomes $P \Gamma Q^{-T} = \Gamma$, or equivalently $P \Gamma = \Gamma Q^t$. Write $Q$ as a $4 \times 4$ matrix similar to $P$. Then we have the following equation
	\[
	\begin{bmatrix} 
		P_{11} & P_{12} & P_{13} \\
		P_{21} & P_{22} & P_{23} \\
		P_{31} & P_{32} & P_{33} 
	\end{bmatrix}
	\begin{bmatrix}
		A & 0 & 0 \\
		0 &p^{e-1} J_r & 0 \\
		0 & 0 & 0 
	\end{bmatrix}
	= 
	\begin{bmatrix}
		A & 0 & 0 \\
		0 &p^{e-1} J_r & 0 \\
		0 & 0 & 0 
	\end{bmatrix}
	\begin{bmatrix} 
		Q_{11}^t & Q_{21}^t & Q_{31}^t \\
		Q_{12}^t & Q_{22}^t & Q_{32}^t \\
		Q_{13}^t & Q_{23}^t & Q_{33}^t \\
	\end{bmatrix}
	\]
	Multiplying out, this gives us
	\[
	\begin{bmatrix} 
		P_{11} A & p^{e-1}P_{12} & 0 \\
		P_{21} A & p^{e-1}P_{22} & 0\\
		P_{31} A & p^{e-1}P_{32} & 0
	\end{bmatrix}
	= 
	\begin{bmatrix} 
		A Q_{11}^t & A Q_{21}^t & A Q_{31}^t \\
		p^{e-1}Q_{12}^t & p^{e-1}Q_{22}^t & p^{e-1}Q_{32}^t \\
		0 & 0 & 0
	\end{bmatrix}
	\]
	From these, we immediately get $P_{31} A = 0$ and since $A$ is invertible, we get $P_{31} = 0$. We also get that $p$ divides $P_{32}$, since $p^{e-1} P_{32} = 0$, and we are working modulo $p^e$. We also get that $P_{21} A = p^{e-1} Q_{21}^t$, so $P_{21} A$ is divisible by $p^{e-1}$. Again, since $A$ is invertible, we find $P_{21} = p^{e-1} Q_{12}^t A^{-1}$, so $P_{21}$ is divisible by $p^{e-1}$. Thus $P$ has the form
	\[
	P = \begin{bmatrix}
		P_{11} & * & * \\
		p^{e-1} * & P_{22} & * \\
		0 & p * & P_{33}
	\end{bmatrix}
	\]
	where the $*$'s may be arbitrary. However, a matrix over $\Z/p^e \Z$ is invertible iff it is invertible modulo $p$ \cite{Ranum, HillarRhea}. Since $P$ is invertible by assumption and $P\pmod{p}$ is block-upper-triangular, we have that $P_{ii} \pmod{p}$ is invertible for $i=1,2,3$, and hence $P_{ii}$ itself (not modulo $p$) is invertible for $i=1,2,3$. 
	
	From the (1,1) entry above, note that we have $P_{11} A = A Q_{11}^t$. It remains to show that $Q_{11}$ is invertible, for then we will have $P_{11} A Q_{11}^{-t} = A$, as claimed.
	
	Similar to $P$, we find that $Q_{31} = 0$, $Q_{32}$ must be divisible by $p$, and $Q_{21}$ must be divisible by $p^{e-1}$. Thus $Q$ has the form
	\[
	Q = \begin{bmatrix}
		Q_{11} & * & * \\
		p^{e-1}* & Q_{22} & * \\
		0 & p* & Q_{33}
	\end{bmatrix}.
	\]
	As with $P$, we find that each $Q_{ii}$ is invertible modulo $p$, hence invertible, thus completing the proof.
\end{proof}

\begin{remark}\label{rem:TI4}
	Given the above preparations, in order to accommodate the gadgets of \cite{TI4}, we expand the rows to accommodate two copies of the rows, plus enough copies of $\Z / p \Z$ (n.b.: $p$, not $p^e$!) to make room for the $p^{e-1} J_r$ gadget. This results in the complex parameter increasing by only a polynomial factor, and then with the preceding lemma in hand, as well as the use of inner rank instead of rank, the results of \cite{TI4} go through. 
\end{remark}
%
%

\section{The complexity of $2\times 2\times 2$ tensor isomorphism over $\Z$}

The first two subsections are background on quadratic rings and Bhargava's work relating them to $2 \times 2 \times 2$ integer tensors; we prove Theorem~\ref{thm:tensorZ} in the third subsection.

\subsection{Preliminaries: Quadratic rings}

A \emph{quadratic ring} is a unital commutative ring $\rng$ of free rank $2$ as a $\Z$-module.  Such a ring is equipped with a trace: $\Tr(\alpha)$ for $\alpha \in \rng$ is the trace of the endomorphism of $\rng$ induced by $\alpha$.  The discriminant of the ring is then $\disc(\rng) = \det(\Tr(\alpha_i\alpha_j))$ where $\rng = \alpha_1 \Z + \alpha_2 \Z$.  In particular, if $\rng = \Z + \tau \Z$ where $\tau$ has minimal polynomial $a\tau^2 + b\tau + c = 0$, then $\disc(\rng) = b^2 - 4ac$.  

The discriminant is always $0, 1 \pmod{4}$ and any integer satisfying this congruence is the discriminant of a unique quadratic ring $\rng$, up to isomorphism.  In other words, discriminants parametrize quadratic rings up to isomorphism.
In particular, we can take $\rng = \Z + \tau \Z$ where $\tau$ has minimal polynomial $\tau^2 = \Delta/4$ in the case that $\Delta \equiv 0 \pmod{4}$ and $\tau^2 - \tau - \frac{\Delta-1}{4}$ in the case that $\Delta \equiv 1 \pmod{4}$.  

Quadratic rings can be listed as follows:  the ring $\Z[x]/(x^2)$ having $\disc(\rng)=0$ is the \emph{degenerate ring} (the others being \emph{non-degenerate}); the ring $(1,1)\Z + (\sqrt{\Delta},0)\Z = (1,1)\Z + \sqrt{\Delta}(\Z \oplus \Z) \subseteq \Z \oplus \Z$ of square discriminant $\disc(\rng) = \Delta \ge 1$; and the familiar \emph{number field orders} $\Z \left[ \frac{\Delta+\sqrt{\Delta}}{2} \right]$ of non-square discriminant $\disc(\rng) = \Delta$.  A ring of integers is \emph{imaginary quadratic} if $\Delta < 0$ and \emph{real quadratic} if $\Delta > 0$.

Each quadratic ring $\rng$ of non-square discriminant has an associated quadratic algebra $K = \rng \otimes_\Z \Q$ which is a number field; then $\rng$ is called a number field order.  We consider the collection of fractional ideals of $\rng$, that is, rank two $\rng$-modules in $K$.  Any fractional ideal is of the form $dI$ for $I$ an ideal of $\rng$ and $d\in K$.  The norm $N(I)$ of an ideal of $\rng$ is $|\rng/I|$, 
and we extend this to $N(dI) = d^2 N(I)$ (which is the ratio of the covolume of the fractional ideal to that of $\rng$).  Two fractional ideals $I$ and $I'$ are considered equivalent if $I = k I'$ for some $0 \neq k \in K$.  The collection of equivalence classes of fractional ideals of $\rng$ is denoted $\Cl(\rng)$.

Bhargava states a theorem concerning the action of $\SL(2,\Z)$ instead of $\GL(2,\Z)$.  For this, he adds an orientation to a quadratic number field order $\rng$ of discriminant $\Delta$ by endowing it with a choice of $\sqrt{\Delta}$ (from among the two options).  He also adds an orientation to an ideal $I$ of $\rng$ by specifying the additional data of $\epsilon \in \{ \pm 1 \} \cong \Z/2\Z$ and performing multiplication of oriented ideals $(I,\epsilon)$ componentwise.  Scalar multiplication of oriented ideals is defined as $\alpha (I, \epsilon) = (\alpha I, \operatorname{sign}(N(\alpha))\epsilon)$ for $0 \neq \alpha \in K$.  The norm of an oriented ideal $I$ is $\epsilon N(I)$.

There is also a notion of oriented basis for $\rng$ or an ideal of $\rng$ (equivalently, for $K$).  For this, if $\tau$ is the chosen square root of $\Delta$ in the oriented ring $\rng$, then we say that $1, \tau$ is a positively oriented basis.  Any other basis $\alpha, \beta$ is defined by reference:  it is positively oriented if the change of basis matrix between $\alpha$, $\beta$ and $1, \tau$ has positive determinant; otherwise it is negatively oriented.

\subsection{Preliminaries: Bhargava's Theorem}

\begin{definition}
A triple $(I_1, I_2, I_3)$ of oriented fractional ideals of $\rng$ is considered \emph{balanced} if $$N(I_1)N(I_2)N(I_3) = 1 \text{ and } I_1I_2I_3 \subseteq \rng.$$  
Here, $N$ denotes the norm with respect to $\rng$.  Two balanced triples of ideals $(I_1, I_2, I_3)$ and $(I_1', I_2', I_3')$ for the same ring $\rng$ are considered equivalent if there exist $k_1, k_2, k_3 \in K$ such that $I_i = k_i I_i'$ for $i=1,2,3$, and orientations agree.
\end{definition}

\begin{definition}  Let $\tA \in \Z^2 \otimes \Z^2 \otimes \Z^2$ have slices $A_1$, $A_2$ with respect to one of the indices.  Then $\tA$ has \emph{discriminant} 
\[
\disc(\tA) = \disc(\det( xA_1 - yA_2)),
\]
where the right side is the discriminant of a quadratic form, namely $\disc(ax^2 + bxy + cy^2) = b^2 - 4ac$.
\end{definition}
The discriminant is preserved regardless of the direction in which we slice (the index we choose).  The tensor $\tA$ is considered \emph{non-degenerate} if it has non-zero discriminant.

Suppose $\rng = \Z + \tau \Z$ with $K = \rng \otimes_\Z \Q$ as in the previous section.  Let $\operatorname{FracI}(K)$ represent the fractional ideals of $K$.
Then, there is a map 
\[
	\rho: \Z^2 \otimes \Z^2 \otimes \Z^2 \rightarrow \operatorname{FracI}(K), 
\]
accomplished by the following process.  Take the quadratic form $Q(x,y) := ax^2 + bxy + cy^2 := -\det(xA_1 + yA_2)$.  Let $\rho(\tA)$ be the rank two $\Z$-module $\alpha \Z +  \beta \Z$ with $\rng$-module structure
\[
\tau \cdot \alpha = \frac{b+\eta}{2} \alpha + a \beta, \quad
\tau \cdot \beta = c \alpha + \frac{b-\eta}{2} \beta,
\]
where
\[
\eta = \left\{ \begin{array}{ll}
1 & \Delta \equiv 1 \pmod{4} \\
0 & \Delta \equiv 0 \pmod{4}
\end{array} \right. .
\]
The result $\rho(\tA)$ is a fractional ideal of $K$.

We consider pairs $(\rng, (I_1, I_2, I_3))$ of an oriented quadratic number field order $\rng$ and an equivalence class of balanced triples $(I_1, I_2, I_3)$ of oriented fractional ideals.  Two such pairs are \emph{isomorphic} if the rings are isomorphic and the isomorphism maps the equivalence classes to each other, preserving orientation.  

We can now state Bhargava's theorem.

\begin{theorem}[{\cite[Theorem 11]{Bhargava}}]
	\label{thm:bhargava}
	Let $\Delta$ be a non-zero non-square fundamental discriminant of a quadratic ring (necessarily a number field order).
There is a bijection between the following two sets:
\begin{enumerate}
	\item $\SL(2,\Z) \times \SL(2,\Z) \times \SL(2,\Z)$ orbits of tensors $\tA \in \Z^2 \otimes \Z^2 \otimes \Z^2$ with discriminant $\Delta$,
	\item isomorphism classes of tuples $(\rng, (I_1, I_2, I_3))$ where $\rng$ is a oriented quadratic number field order of discriminant $\Delta$, and $(I_1,I_2,I_3)$ is a balanced triple of oriented ideals.
\end{enumerate}

	The map from tensor $\tA$ to pair $(\rng, (I_1, I_2, I_3))$ is given by taking $\rng$ to be the oriented quadratic number field order with discriminant $\Delta$, taking $I_1 = \rho(\tA)$ and similarly taking $I_2$ and $I_3$ to be the analogous fractional ideals obtained by applying the definition of $\rho$ with respect to slicing along other indices.  In other words, if $\sigma$ is a $3$-cycle rotation of the indices of a tensor, then $I_i = \rho(\tA^{\sigma^i})$.  The orientation on each fractional ideal is given according to whether the basis $\alpha, \beta$ described in the definition of the map $\rho$ is positively or negatively oriented with respect to $\rng$.

	Finally, the map from pair $(\rng, (I_1,I_2,I_3))$ to tensor $\tA$ is given as follows.  Let $I_1 = \alpha_1\Z + \alpha_2\Z$, $I_2 = \beta_1\Z + \beta_2\Z$ and $I_3 = \gamma_1\Z + \gamma_2\Z$; where we have chosen the orientation of these bases to match or be different from that of $\rng = \Z + \tau \Z$ according to whether $\epsilon(I_i) = 1$ or $-1$, respectively.  Write $\alpha_i\beta_j\gamma_k = c_{ijk} + a_{ijk}\tau$, and then set the coordinates of $\tA$ to be $a_{ijk}$.
\end{theorem}

We would prefer to have a version of Bhargava's equivalence for $\GL(2,\Z)\times \GL(2,\Z) \times \GL(2,\Z)$ orbits.  Fortunately, this involves simply forgetting the orientation information. We say two pairs of the form $(\rng, (I_1, I_2, I_3))$ are \emph{unoriented isomorphic} if there is an isomorphism of rings that maps the equivalence classes to each other, but now not necessarily preserving orientation.

\begin{theorem}
	\label{thm:bhargavaGL}
	Let $\Delta$ be a non-zero non-square fundamental discriminant of a quadratic number field order.
There is a bijection between the following two sets:
\begin{enumerate}
	\item $\GL(2,\Z) \times \GL(2,\Z) \times \GL(2,\Z)$ orbits of tensors $\tA \in \Z^2 \otimes \Z^2 \otimes \Z^2$ with discriminant $\Delta$,
	\item isomorphism classes of tuples $(\rng, (I_1, I_2, I_3))$ where $\rng$ is a (unoriented) quadratic number field order of discriminant $\Delta$, and $(I_1,I_2,I_3)$ is a balanced triple of ideals.
\end{enumerate}
	The maps are as for Theorem~\ref{thm:bhargava}.
\end{theorem}

	\begin{proof}
	Observe that $\GL(2,\Z)$ is formed from $\SL(2,\Z)$ by allowing one additional transformation, namely $M = \begin{pmatrix} 0 & 1 \\ 1 & 0 \end{pmatrix}$.  An action by $M$ on the slices $A_1$ and $A_2$ swaps them, which does not change the discriminant.  This results in replacing the oriented ideal $(I_1,\epsilon)$ with $(I_1,-\epsilon)$; and fixing the ideals $(I_i,\epsilon)$ for $i=2,3$.  Thus we obtain a new pair $(\rng, (I_1', I_2', I_3'))$ which differs only in terms of orientations.  Thus the new pair is equivalent when orientations are forgotten.  

		Conversely, suppose two pairs $(R, (I_1, I_2, I_3))$ and $(R', (I_1, I_2, I_3))$ are unoriented isomorphic, i.e. when orientations are forgotten.  Let $\tA$ and $\tA'$ be the associated tensors.  Then $R \cong R'$, so without loss of generality we may assume $R = R'$.  Then by the map in Theorem~\ref{thm:bhargavaGL}, by a suitable application of $M$ in each direction to $\tA$ and $\tA'$, we can arrange that the bases of the $I_i$ are positively oriented in both cases.  So then unoriented isomorphism becomes isomorphism.
	\end{proof}

Suppose that we represent the quadratic number field order $\rng$ by its discriminant $\Delta(\rng)$, implicitly thinking of $\rng$ as having the basis $1, \tau$ as described by the minimal polynomials in the previous section.
Suppose that we then represent an ideal $I$ in terms of a basis $\alpha_1, \alpha_2$, where the $\alpha_i$ are given as $\Q$-linear combinations of $1$ and $\tau$ (i.e., an ordered pair of rationals).  This allows us to represent a tuple $(\rng, (I_1, I_2, I_3))$.
	With this representation, we observe that the maps realizing Bhargava's bijection are polynomially computable.

\subsection{Equivalence of $2 \times 2 \times 2$ TI$_\Z$ and Principality Decision}

\begin{lemma} \label{lem:tensorZ}
Suppose $\tA, \tB \in \Z^2 \otimes \Z^2 \otimes \Z^2$ have square discriminants. Then isomorphism between them (under $\GL(2,\Z) \times \GL(2,\Z) \times \GL(2,\Z)$) can be determined in polynomial time.
\end{lemma}

A 3-hypermatrix over a field is \emph{concise} if, in each of the three directions, its slices are linearly independent; we refer to an integer hypermatrix as concise if it is concise when considered over $\Q$.

\begin{proof}
\textbf{Check the GCD.} First, the GCD of all the entries of $\tA$ is unchanged by the action by $\GL(2,\Z)^3$. If the GCD of all the entries of $\tA$ is not the same as that for $\tB$, then they are not isomorphic.

Otherwise, divide both $\tA$ and $\tB$ by the GCD. Because this division was action by a scalar, it commutes with the action of $\GL(2,\Z)^3$, so the original $\tA$ and $\tB$ were isomorphic iff the rescaled versions are. We may thus now assume that $\tA$ (resp. $\tB$) has the property that the GCD of its entries is 1.

\textbf{The non-concise case.} Next, let $A_1, A_2$ be the frontal slices of $\tA$. If $A_2$ is a $\Q$-scalar multiple of $A_1$, then we can take the GCD, replace $A_1$ by some scalar multiple of itself and $A_2$ by zero. In this case, we check that $\tB$ has the same structure, and thus reduce to the case of the $\GL(2,\Z) \times \GL(2,\Z)$ equivalence of two $2 \times 2$ matrices under the left-right action. This is solved by putting both matrices into Smith Normal Form and checking that the two have equal SNFs.

Thus, we now assume that $A_1$ and $A_2$ are not $\Q$-multiples of one another (and analogously for the slices in the other two directions).

\textbf{The concise case.} Since the discriminant $D$ is a perfect square, the form $\det(x A_1 - y A_2) = ax^2 + bxy + cy^2$ has two integer roots up to scale (which need not be distinct, namely if $D=0$), given by $(x,y) = (-b + \sqrt{D}, 2a)$ and $(x,y) = (-b - \sqrt{D}, 2a)$. Now, let $\varepsilon$ be the GCD of $-b + \sqrt{D}$ and $2a$, $\alpha = (-b + \sqrt{D})/\varepsilon$, $\beta = -2a / \varepsilon$. Since $\alpha$ and $\beta$ are coprime, there exist $\gamma,\delta \in \Z$ such that $\alpha \delta - \beta \gamma = 1$. Define $A_1', A_2'$ by
\[
\begin{pmatrix}
A_1' \\
A_2'
\end{pmatrix}
=
\begin{pmatrix}
\alpha & \beta \\
\gamma & \delta
\end{pmatrix}
\begin{pmatrix}
A_1 \\
A_2
\end{pmatrix}.
\]
Then the tensor $\tA'$ whose frontal slices are $A_1',A_2'$ differs from $\tA$ by an element of $\SL(2,\Z)$. 

Now, by construction we have $\det(A_1') = 0$. Use the action of $\GL(2,\Z) \times \GL(2,\Z)$ by left and right multiplication to put $A_1'$ into Smith Normal Form $A_1''$, while simultaneously transforming $A_2'$ into some matrix $A_2''$. Since $A_1'$ has rank at most 1, its Smith Normal Form is 
\[
A_1'' = \begin{pmatrix} e & 0 \\ 0 & 0 \end{pmatrix}
\]
for some integer $e$.
Let $\tA''$ be the tensor with frontal slices $A_1'', A_2''$, which is again isomorphic to $\tA$. We now have
\[
\det(x A_1'' - y A_2'') = e (A_2'')_{22} xy + \det(A_2'') y^2 = y(e (A_2'')_{22}  x + \det(A_2'') y),
\]
Since the tensor is concise, we have $e \neq 0$, and thus $D = 0 \Leftrightarrow (A_2'')_{22} = 0$.

\textbf{Concise, discriminant zero.} Suppose that $D = 0 = (A_2'')_{22}$.  Then $\det(x A_1'' - y A_2'')$ has a double root at $(1,0)$, so $A_1''$ and its scalar multiples are the only rank-1 matrices in $Span_\Z(A_1'', A_2'')$. Next, the stabilizer of $A_1''$ under the left-right action of $\GL(2,\Z) \times \GL(2,\Z)$ consists of those pairs $(P,Q) \in \GL(2,\Z)^2$ such that $P,Q$ are both upper triangular (where the action is $(P,Q) \cdot A_1'' = P A_1'' Q^T$). Thus we may use this stabilizer to fix $A_1''$ while adding multiples of the (1,2) and (2,1) entries to the (1,1) entry of $A_2''$. We may also add multiples of $A_1''$ to $A_2''$. Thus, we arrive at the form
\[
A_1'' = \begin{pmatrix} e & 0 \\ 0 & 0 \end{pmatrix} \qquad A_2'' = \begin{pmatrix} f & g \\ h & 0 \end{pmatrix},
\]
where $f$ is reduced modulo the GCD of $g$, $h$, and $e$, and at least one of $g$ or $h$ is nonzero (since otherwise $A_1''$ would be a $\Q$-multiple of $A_2''$, which we handled above).

Now, attempt to do the same procedure to $\tB$. And suppose we arrive at 
\[
B_1'' = \begin{pmatrix} e' & 0 \\ 0 & 0 \end{pmatrix} \qquad B_2'' = \begin{pmatrix} f' & g' \\ h' & 0 \end{pmatrix}.
\]
If any part of this procedure does not result in this form, then $\tA$ and $\tB$ are not isomorphic. Since $A_1''$ and its multiples are the only rank-1 matrices in $Span_\Z(A_1'', A_2'')$, and the same is true of $B_1''$, if $e \neq \pm e'$, then $\tA$ and $\tB$ cannot be isomorphic. Otherwise, if needed, multiply by an element of $\GL(2,\Z)$ to fix the sign so that $e = e'$. 

At this point, if $(P,Q,R)$ is an isomorphism from $\tA''$ to $\tB''$, $R$ must fix the first slice (hence be lower triangular), and thus the action of $(P,Q)$ must also fix the first slice, so $P$ and $Q$ must both be upper triangular. Under the action of this subgroup, the only things we can do are to add multiples of $e,g,h$ to $f$. If the pair $(g,h)$ is not equal to one of $(\pm g', \pm h')$, then $\tA$ cannot be isomorphic to $\tB$. Otherwise, multiply the second row and second column by the necessary signs to make $g = g'$ and $h = h'$. Finally, since $f$ and $f'$ are both reduced to their minimal representatives modulo the ideal $\langle e,g,h \rangle = \langle e', g', h' \rangle$ (equivalently, modulo the GCD of $e$, $g$, and $h$), we have that $\tA \cong \tB$ if and only if $f = \pm f'$. If the signs are not equal, note that we can multiply the second slice, the second row, and the second column all by $-1$, and this has the effect of only multiplying the $f$ entry by $-1$. 

\textbf{Concise, non-zero discriminant.} On the other hand, suppose that both $D$ and $(A_2'')_{22}$ are nonzero. Now there are two $\Q$-linearly independent matrices of rank 1 in $Span_\Z(A_1, A_2)$. We may either replace $A_1$ by such a matrix $A_1'$ and get a resulting $A_2'$ as above, or vice versa. We handle the first case here, the other case being the same just swapping the roles of $A_1$ and $A_2$, \emph{mutatis mutandis}. 

Thus, we have arrived at a form
\[
A_1'' = \begin{pmatrix} e & 0 \\ 0 & 0 \end{pmatrix} \qquad A_2'' = \begin{pmatrix} f & g \\ h & k \end{pmatrix},
\]
with $k \neq 0$, and similarly
\[
B_1'' = \begin{pmatrix} e' & 0 \\ 0 & 0 \end{pmatrix} \qquad B_2'' = \begin{pmatrix} f' & g' \\ h' & k' \end{pmatrix}.
\]
(If $\tB$ cannot be put in this form, then the two are not isomorphic.) As above, to preserve the slice $A_1''$, we cannot add any multiple of $A_2''$ to $A_1''$, and we can only apply the left-right action of $(P,Q)$, where both $P$ and $Q$ are upper triangular. Since we cannot alter $A_1''$, if $e \neq \pm e'$, then we try putting $\tB$ into this form using the other rank-1 matrix in the span of its slices. And if neither works, then the two are not isomorphic. 

So we may now assume $e = e'$ (by multiplying by a sign if needed). Then the two are isomorphic, by an isomorphism that sends $A_1''$ to $B_1''$, if and only if there is an integer solution to: 
\[
\begin{pmatrix} 1 & x \\ 0 & 1 \end{pmatrix}
\begin{pmatrix} f + ze & g \\ h & k \end{pmatrix}
=
\begin{pmatrix} f' & g' \\ h' & k' \end{pmatrix}
\begin{pmatrix} 1 & 0 \\ y & 1 \end{pmatrix}.
\]
This is a set of integer linear equations in $x$, $y$, and $z$ and thus can be efficiently solved.

(In more detail, one way to see the solution directly, instead of using a generic algorithm for solving linear Diophantine equations, is as follows. The action cannot change $k$ and $k'$, so we need $k = \pm k'$, and again we may multiply by signs to achieve $k = k'$. Using $x$, we can take $g$ modulo $k$, so we must have $g \equiv g' \pmod{k}$. Using $y$, we can take $h'$ modulo $k$, so we must have $h \equiv h' \pmod{k}$. And after we use $x$ and $y$ in this way, whatever the new values in the positions of $f$ and $f'$ are, they must be equivalent modulo $e$.)
\end{proof}

\begin{proof}[Proof of Theorem~\ref{thm:tensorZ}]
Let $\tA$ and $\tB$ be two elements of $\Z^2 \otimes \Z^2 \otimes \Z^2$.  
Compute $\disc(\tA)$ and $\disc(\tB)$.  If they are not equal, return false.  
If they are equal and square, solve Tensor Isomorphism using Lemma~\ref{lem:tensorZ}.

	It remains to consider the case that $\disc(\tA) = \disc(\tB)$ is non-square.
In this case, we will reduce \algprobm{Tensor Isomorphism} to \algprobm{Principality Decision}.
	Write $\rng$ for the associated quadratic ring of discriminant $\disc(\tA)$.
	We will determine instead if the associated tuples $(\rng, (I_1,I_2,I_3))$ are unoriented isomorphic, which suffices by Theorem~\ref{thm:bhargavaGL}.
Compute $I_1(\tA)$ and $I_1(\tB)$.  These are ideals of $\rng$.  

	Next, we show that to determine whether $I_1(\tA)$ and $I_1(\tB)$ are equivalent ideals of $\rng$ is to check an ideal for principality.
	First, recall that the \emph{order} of an ideal $I$ is the ring $\operatorname{ord}(I) := \{ \alpha \in K : \alpha I \subseteq I \}$.  This can be computed polynomially by a linear algebra algorithm, or by reducing to computing the GCD of the coefficients of the associated quadratic forms \cite[Section 7.A]{Cox}.  If the orders of $I_1(\tA)$ and $I_1(\tB)$ are not the same, the ideals are not equivalent.  If the orders are the same, then call the order $\rng'$ and observe that $\rng \subseteq \rng'$.  In this case the two ideals are invertible in $\rng'$.   
	To determine if they are equal, it suffices to test $I_i(\tA) I_i(\tB)^{-1}$ for principality.

	Assuming we can test for principality, if the forms are not equal, return false.  If they are equal, proceed to test $I_2(\tA)$ and $I_2(\tB)$.  If not equivalent, return false.  If they are equal, proceed to test $I_3(\tA)$ and $I_3(\tB)$.  If equivalent, return true; otherwise, return false.  

	As far as principality testing, if $\disc(\tA) < 0$, then invertible ideals correspond to positive definite primitive integral binary quadratic forms via a polynomially computable isomorphism, and the classical reduction theory of positive definite primitive integral binary quadratic forms can check efficiently whether an ideal is principal \cite[Section 5.3.1]{Cohen}.

	If $\disc(\tA) > 0$, we are in the real quadratic case, and we rely on the Principality Decision Problem to test if $I_i(\tA)$ and $I_i(\tB)$ are equivalent.  
	Since all cases except for this last one can be solved in polynomial time, this gives a polynomial-time reduction to Principality Decision for ideals in real quadratic fields.

Conversely, we reduce \algprobm{Principality Decision} in real quadratic fields to \algprobm{Tensor Isomorphism}.  Let $I$ be an ideal of a real quadratic number field order $\rng$, given by an integral basis.
We will rephrase the principality problem for $I$ as a tensor problem.  First, compute the inverse of $I$, which is $\overline I/N(I)$, which is easily computable from a basis.  Observe that the triple $(\rng, (I, I^{-1},R))$ is balanced, since $I I^{-1} R = R$.  Then we form $\tA$ from the triple $(\rng, (I,I^{-1},R))$ according to Theorem~\ref{thm:bhargavaGL}, which takes polynomial time.  We also form $\tB$ from the trivial ideal $(\rng, (\rng, \rng, \rng))$ (which is also balanced).  Then, $I$ is principal if and only if $\tA$ and $\tB$ are isomorphic.

	Finally, with the above reduction, the statement about $\cc{BQP}$ then follows immediately from \cite[Theorem 4.1]{Hallgren}.
\end{proof}


\section{Faster algorithms for isomorphism of odd order class-2 nilpotent groups}
 

In this section, we prove Theorem~\ref{thm:tensor_isomorphism_main} by giving an algorithm for testing isomorphism of two nilpotent class-2 groups of odd order in time $N^{O((\log N)^{8/9})}$, where $N$ is the order of the two groups.

In Section~\ref{sec:group_iso_to_tensor_iso}, we reduce \algprobm{Group Isomorphism} for class-2 nilpotent groups to \algprobm{Alternating Matrix Space Isometry} over $\Z/p^e \Z$ for some prime $p$ and integer $e$. 
In Section~\ref{sec:tensor_iso_to_matrix_tuple}, we further reduce \algprobm{Alternating Matrix Space Isometry} to \algprobm{Alternating Matrix Tuple Isometry}. 
In Section~\ref{sec:matrix_tuple}, we give a polynomial time algorithm for the \algprobm{Alternating Matrix Tuple Isometry} problem. 

In this section, we say that
an $n \times n \times m$ tensor $\tA = (a_{i, j, k})$ is alternating (between the first and second directions) if, for any $1 \leq i, j \leq n$ and $k \leq m$, $a_{i, j, k} = -a_{j, i, k}$. 
The \algprobm{Alternating Matrix Space Isometry} problem 
is to determine  whether for two $n \times n \times m$ alternating tensors $\tA$ and $\tB$ over $\Z / p^e \Z$, there are two matrices $P \in \GL(n, \Z / p^e \Z)$ and $Q \in \GL(m, \Z / p^e \Z)$ such that $\tB = (P, P, Q) \circ \tA$.
The \algprobm{Alternating Matrix Tuple Isometry} problem 
is to determine whether for two alternating matrix tuples $(A_1, \dots, A_k)$ and $(B_1, \dots, B_k)$ in $\M(n, \Z / p^e \Z)$, there is a matrix $P \in \GL(n, \Z / p^e \Z)$ such that $B_i = P A_i P^t$ for any $1 \leq i \leq k$.

\subsection{Reduction to \algprobm{Alternating Matrix Space Isometry} over $\Z/p^e \Z$}\label{sec:group_iso_to_tensor_iso}

We define some notations for tensors. 
Let $\tA = (a_{i, j, k})$ be an $n \times m \times h$ tensor over $\Z / p^{e} \Z$. For $i\in[n]$, let $\mathfrak{X}_{\tA, i}$ be the $h \times m$ matrix over $\Z / p^{e} \Z$ such that $\mathfrak{X}_{\tA, i}[k, j] = a_{i, j, k}$ for any $1 \leq j \leq m, 1 \leq k \leq h$. For $j\in[m]$, let 
$\mathfrak{Y}_{\tA, j}$ be the $h \times n$ matrix over $\Z / p^{e} \Z$ such that $\mathfrak{Y}_{\tA, j}[k, i] = a_{i, j, k}$ for any $1 \leq i \leq n, 1 \leq k \leq h$. For $k\in[h]$, let $\mathfrak{Z}_{\tA, k}$ be the $n \times m$ matrix over $\Z / p^{e} \Z$ such that $\mathfrak{Z}_{\tA, k}[i, j] = a_{i, j, k}$ for any $1 \leq i \leq n, 1 \leq j \leq m$.
Let $\alpha_{\tA, 1}, \dots, \alpha_{\tA, e}$ be the integers such that the abelian group generated by $\mathfrak{Z}_{\tA, k}$ over $k\in[h]$ is isomorphic to 
\[\bigoplus_{\ell = 1}^{e} \left(\Z / p^{\ell } \Z\right)^{\alpha_{\tA, \ell}},\]
$\beta_{\tA, 1}, \dots, \beta_{\tA, e}$ be the integers such that the abelian group generated by $\mathfrak{X}_{\tA, i}$ over $i\in[n]$ is isomorphic to 
\[\bigoplus_{\ell = 1}^{e} \left(\Z / p^{\ell } \Z\right)^{\beta_{\tA, \ell}},\]
and 
$\gamma_{\tA, 1}, \dots, \gamma_{\tA, e}$ be the integers such that the abelian group generated by $\mathfrak{Y}_{\tA, j}$ over $j\in[m]$ is isomorphic to 
\[\bigoplus_{\ell = 1}^{e} \left(\Z / p^{\ell } \Z\right)^{\gamma_{\tA, \ell}}.\]
We call $\sum_{\ell = 1}^e \ell \cdot (\alpha_{\tA, \ell} + \beta_{\tA, \ell} + \gamma_{\tA, \ell})$ the \emph{complex parameter} for the tensor.

The following theorem reduces the group isomorphism problem for class-2 nilpotent groups to the \algprobm{Alternating Matrix Space Isometry} problem. 

%
%

\begin{theorem} \label{thm:GpI-to-AMSI}
The problem of testing isomorphism of odd order nilpotent groups of class 2 reduces to \algprobm{Alternating Matrix Space Isometry} over the integers modulo prime powers in polynomial time.

Moreover, if the input groups are of order $N$, then for each instance of the \algprobm{Alternating Matrix Space Isometry} problem with tensors over $p^e$ for some prime $p$ and positive integer $e$, 
then the tensors are of dimension $O(\log_p N) \times O(\log_p N) \times O(\log_p N)$ with complex parameter $O(\log_p N)$. 
\end{theorem}

Before coming to the proof, we recall a key component, which is Baer's Correspondence \cite{Bae38} (see, e.g., \cite{khukhro,naik} for expository treatments). Given a finite odd order nilpotent group $G$ of class 2, Baer's Correspondence constructs a finite Lie ring, denoted $\log(G)$, on the same underlying set as $G$, and where the Lie ring operations are defined by
\[
x + y := xy / \sqrt{[x,y]} \qquad -x := x^{-1} \qquad [x,y]_{Lie} := xyx^{-1} y^{-1},
\]
where here the expressions on the LHS are the Lie ring operations and those on the RHS are in the group. We note that $|G|$ is odd, every element has a unique square root, and since the group has class 2, $[x,y]$ is in the center, making the first expression $xy / \sqrt{[x,y]}$ well-defined. In the opposite direction, given a finite nilpotent Lie ring $L$ of odd order and class 2, Baer's Correspondence defines a group, denotes $\exp(L)$, on the same underlying set as $L$, and in which the group operations are defined by:
\[
xy := x + y + \frac{1}{2}[x,y] \qquad x^{-1} := -x,
\]
where here the LHS is in the group and the RHS is in the Lie ring $L$. Baer's Correspondence says that these two constructions are inverses of one another, and that they give an equivalence between the category of these groups and the category of these Lie rings. In particular, two groups $G,H$ are isomorphic iff $\log(G)$ and $\log(H)$ are isomorphic Lie rings. From the formulas above, it is readily seen that, in the Cayley table model, one can construct $\log(G)$ from $G$ in polynomial time, and one can construct $\exp(L)$ from $L$ in polynomial time (in fact, with the additional requirement of exponent $p$, this can also be done efficiently in the setting of matrix groups and matrix Lie rings; see, e.g., \cite[Sec.~6.2]{TI2}). Furthermore, several properties of $G$ are preserved in $\log(G)$ and vice versa:
\begin{description}
\item[Element orders.] For any integer $k$, we have $x + x + \dotsb + x$ ($k$ times) in the Lie ring is the same element as $x^k$ in the group, and thus if $x \in G$ has (multiplicative) order $p^e$, then $x \in \log(G)$ has (additive) order $p^e$. That is, in $\log(G)$, we have $p^e x = 0$ and $p^e$ is the least such integer. Thus the additive group underlying $\log(G)$ contains the information of the orders of elements of $G$.

\item[Nilpotency class.] $\log(G)$ is nilpotent of class 2.

\item[Centers.] The underlying set of $Z(G)$ is the same as the underlying set of $Z(\log(G))$. The group multiplication on $Z(G)$ is identical to the Lie ring addition operation on $Z(\log(G))$---that is, $Z(G)$ and $Z(\log(G))$ are identical as Abelian groups.

\item[Central quotients.] There is a natural isomorphism between $G / Z(G)$ and $\log(G) / Z(\log(G))$, as Abelian groups. That is, in $G/Z(G)$, the group operation is still given by multiplication in $G$, simply working modulo $Z(G)$. Similarly, since addition in $\log(G)$ is given by $x+y = xy / \sqrt{[x,y]}$ and $[x,y] \in Z(G) = Z(\log(G))$, we have that addition in $\log(G)/Z(\log(G))$ is given exactly by group multiplication, modulo $Z(G)$.
\end{description}

Given two finite Lie rings $L,L'$, an \emph{isomorphism} between them is a bijection $\varphi \colon L \to L'$ such that (1) $\varphi$ is linear, that is, $\varphi(x+y) = \varphi(x) + \varphi(y)$, (2) $\varphi([x,y]) = [\varphi(x), \varphi(y)]$ for all $x,y \in L$. As with finite groups, finite Lie rings may be given by their ``Cayley tables'', which actually consists of two tables: one for the addition operation and one for the Lie bracket.

\begin{proof}[Proof of Theorem~\ref{thm:GpI-to-AMSI}]
First, any finite nilpotent group is the direct product of its Sylow subgroups. As factoring a group into its direct products is unique (the Remak--Krull--Schmidt Theorem), and can be done in polynomial time \cite{WilsonProductsI, WilsonProductsII,KN}, for isomorphism testing we thus reduce to the case of testing isomorphism of $p$-groups of class 2, for some odd prime $p$ (which may be given as part of the input).

Next, from Baer's Correspondence, we reduce from testing isomorphism of $p$-groups of class 2 to testing isomorphism $p$-Lie rings of class 2, of the same size as the groups we started with. Let $L,L'$ denote the corresponding Lie rings that we wish to test for isomorphism.

Now, given $L$, consider the following system of linear and bilinear maps between Abelian $p$-groups. Let $U = Z(L)$, $V = L$, $W = L/Z(L)$. Let $\iota \colon U \to V$ be the natural inclusion map of $Z(L)$ as an additive subgroup of $L$, and let $\pi \colon V \to W$ be the natural quotient map $L \to L/Z(L)$ with kernel $Z(L)$. Finally, let $B \colon W \otimes W \to U$ denote the Lie bracket. Let $U',V',W',\iota', \pi', B'$ be the corresponding groups and maps correpsonding to $L'$. 


\begin{lemma} \label{lem:iso-Lie-ring}
With notation as above, $L \cong L'$ if and only if there are Abelian group isomorphisms $\varphi_U \colon U \to U'$, $\varphi_V \colon V \to V'$, $\varphi_W \colon W \to W'$ such that the following conditions hold:
\begin{align} 
\varphi_V \circ \iota & = \iota' \circ \varphi_U \label{eqn:iso-condition-inj} \\
\varphi_W \circ \pi & = \pi' \circ \varphi_V \label{eqn:iso-condition-surj} \\
\varphi_U(B(x,y)) & = B'(\varphi_W(x), \varphi_W(y)) \qquad (\forall x,y \in W) \label{eqn:iso-condition-bracket}
\end{align}
\end{lemma}

This is essentially standard, but we phrase it this way as it will help us below to set up the tensors we are aiming for. Over a field, we do not need such a complicated setup, and instead can just focus on $W=L/Z(L)$ and $U=Z(L)$; by including $L$ itself in the setup, we are automatically including the Abelian group structure on $L$, which need no longer be the direct sum of $Z(L)$ and $L/Z(L)$. Conditions \eqref{eqn:iso-condition-inj} and \eqref{eqn:iso-condition-surj} then essentially say that the isomorphism on the center is just the restriction of the isomorphism on $L$, and the isomorphism on $L/Z(L)$ is the one induced from the isomorphism on $L$.

\begin{proof}[Proof of Lemma~\ref{lem:iso-Lie-ring}]
($\Rightarrow$) Suppose $\varphi \colon L \to L'$ is a Lie ring isomorphism; equivalently, $\varphi$ is an isomorphism of the additive groups such that $\varphi([x,y]) = [\varphi(x), \varphi(y)]$ for all $x,y \in L$. Since the center is a characteristic sub-Lie-ring, $\varphi(Z(L)) = Z(L')$, and therefore $\varphi$ also induces a map $L/Z(L) \to L'/Z(L')$, namely, the map which sends $x + Z(L)$ to $\varphi(x) + Z(L') = \varphi(x) + \varphi(Z(L)) = \varphi(x + Z(L))$. Let $\varphi_U = \varphi|_{Z(L)}$, $\varphi_V = \varphi$, and $\varphi_W$ be the induced map on $L/Z(L)$ just described.

Since $Z(L)$ is characteristic, $\varphi_U$ is an isomorphism of the (Abelian) Lie rings $Z(L) \cong Z(L')$, and therefore in particular an isomorphism of their underlying Abelian groups. By definition $\varphi_V = \varphi$ is an isomorphism of Lie rings, hence of Abelian groups. And then $\varphi_W$ must also be an isomorphism of Abelian groups: it is a homomorphism by construction, surjective because $\varphi$ is surjective, and injective because its kernel is $Z(L)$ by construction. 

We claim that $(\varphi_U, \varphi_V, \varphi_W)$ satisfy \eqref{eqn:iso-condition-inj}--\eqref{eqn:iso-condition-bracket}.
\begin{itemize}
\item[\ref{eqn:iso-condition-inj}: ] Suppose $u \in U = Z(L)$. Then we have $\varphi_V(\iota(u)) = \varphi(u)$ and $\iota'(\varphi_U(u)) = \iota'(\varphi(u)) = \varphi(u)$, where here we have identified elements of $U = Z(L)$ with the corresponding elements of $L$, so that e.g. $\iota(u) = u$ and $\iota'(u') = u'$ for $u \in Z(L), u' \in Z(L')$.

\item[\ref{eqn:iso-condition-surj}: ] Let $x \in L$. Then we have $\pi(x) = x + Z(L)$ (considered as an element of $L/Z(L)$). So we have 
\begin{align*}
\varphi_W(\pi(x)) & = \varphi_W(x + Z(L)) = \varphi(x + Z(L)) = \varphi(x) + \varphi(Z(L)) \\ & = \varphi(x) + Z(L') = \pi'(\varphi(x)) = \pi'(\varphi_V(x)).
\end{align*}

\item[\ref{eqn:iso-condition-bracket}: ] Let $x,y \in L$.  Then we have
\begin{align*}
\varphi_U(B(x + Z(L),y + Z(L))) & = \varphi_U([x,y]) & \text{(since $[Z(L),*]=[*,Z(L)]=0$)} \\
 & = \varphi([x,y]) & \text{(since $[x,y] \in Z(L)$)} \\
 & = [\varphi(x), \varphi(y)] & \text{(since $\varphi$ is a homomorphism)}\\
 & = [\varphi(x) + Z(L'), \varphi(y) + Z(L')] \\
 &  = B'(\varphi_W(x), \varphi_W(y)).
\end{align*}

\end{itemize}

($\Leftarrow$) Conversely, suppose $(\varphi_U, \varphi_V, \varphi_W)$ satisfy \eqref{eqn:iso-condition-inj}--\eqref{eqn:iso-condition-bracket}. Let $\varphi = \varphi_V$; we claim that $\varphi$ is in fact an isomorphism of Lie rings. By assumption, $\varphi_V = \varphi$ is an isomorphism of Abelian groups, so all that remains to show is that $\varphi$ preserves the Lie bracket, i.e., that $\varphi([x,y]) = [\varphi(x), \varphi(y)]$. For all $x,y \in L$, we have:
\begin{align*}
\varphi([x,y]) & = \varphi_V(\iota(B(x + Z(L),y + Z(L)))) & \text{(definition of $B, \varphi, \iota$)}\\ 
 & = \iota'(\varphi_U(B(x + Z(L),y + Z(L)))) & \text{(by \eqref{eqn:iso-condition-inj})} \\
 & = \iota'(B'(\varphi_W(\pi(x)), \varphi_W(\pi(y)))) & \text{(by \eqref{eqn:iso-condition-bracket})} \\
 & = [\varphi_W(\pi(x)), \varphi_W(\pi(y))] & \text{(by definition of $B',\iota'$)} \\
 & = [\pi'(\varphi(x)), \pi'(\varphi(y))] & \text{(by \eqref{eqn:iso-condition-surj} and def'n of $\varphi$)} \\
 & = [\varphi(x) + Z(L'), \varphi(y) + Z(L')] & \text{(by definition of $\pi'$)} \\
 & = [\varphi(x), \varphi(y)] & \text{(since $[Z(L'),*] = [*, Z(L')] = 0$.)}.
\end{align*}

This completes the proof of Lemma~\ref{lem:iso-Lie-ring}.
\end{proof}

\begin{lemma} \label{lem:Lie-ring-to-TI}
There is a polynomial-time algorithm, which, given a nilpotent $p$-Lie ring $L$ of class 2 and exponent $p^e$ (by Cayley tables or by basis and structure coefficients), produces a tensor $T_L$ in $(L \oplus L \oplus L/Z(L) \oplus L/Z(L) \oplus \Z/p^e) \otimes (L \oplus L \oplus L/Z(L) \oplus L/Z(L) \oplus \Z/p^e)  \otimes (\Z/p^e \oplus Z(L) \oplus Z(L))$ such that $L \cong L'$ if and only if there are $X_1 \in \aut(Z(L))$, $X_2 \in \aut(L_+)$ (automorphism of the underlying Abelian group of $L$), and $X_3 \in \aut(L/Z(L))$ such that
\begin{equation} \label{eqn:Lie-ring-to-TI}
(X_2 \oplus X_2^{-1} \oplus X_3^{-1} \oplus X_3 \oplus 1, (X_2 \oplus X_2^{-1} \oplus X_3^{-1} \oplus X_3 \oplus 1)^t, 1 \oplus X_1 \oplus X_1^{-1}) \circ T_L = T_{L'}.
\end{equation}
Furthermore, each frontal slice of the tensors $T_L, T_{L'}$ is an alternating matrix.
\end{lemma}

\begin{proof}
View $T_L$ as a block tensor, as indicated by the shape $(L \oplus L \oplus L/Z(L) \oplus L/Z(L) \oplus \Z/p^e) \otimes (L \oplus L \oplus L/Z(L) \oplus L/Z(L) \oplus \Z/p^e)  \otimes (\Z/p^e \oplus Z(L) \oplus Z(L))$; that is, there are $75$ blocks arranged in a $5 \times 5 \times 3$ fashion, and e.g. the (1,1,1) block is $L \otimes L \otimes \Z/p^e$, and the (3,3,2) block is $L/Z(L) \otimes L/Z(L) \otimes Z(L)$, and the (5,2,3) block is $\Z/p^e \otimes L \otimes Z(L)$. Throughout, choose a basis for $L$, a basis of $Z(L)$, and a basis of $L/Z(L)$. 
\begin{itemize}
\item Let $\tB \colon L / Z(L) \otimes L / Z(L) \to Z(L)$ be the commutator map of $L$; we place $\tB$, written in the chosen bases, in the (3,3,2) block. 

\item Let $M_\pi$ be the matrix corresponding to the projection map $\pi \colon L \to L/Z(L)$ (again, in the chosen bases); we place $M_{\pi}$ in the (4,2,1) block, and $-M_{\pi}^t$ in the (2,4,1) block. 

\item Let $M_{\iota}$ be the matrix corresponding to the inclusion map $\iota \colon Z(L) \to L$; we place $M_{\iota}$ in the (5,1,3) block and $-M_{\iota}^t$ in the (1,5,3) block.
\end{itemize}

Now we show that this tensor $T_L$ has the desired property, namely that $L \cong L'$ if and only if $T_L$ and $T_{L'}$ are isomorphic by a transformation as in the statement of the lemma.

($\Rightarrow$) Suppose $L \cong L'$. By Lemma~\ref{lem:iso-Lie-ring}, there are isomorphisms $\varphi_U \colon Z(L) \to Z(L')$, $\varphi_V \colon L \to L'$, and $\varphi_W \colon L/Z(L) \to L' / Z(L')$ satisfying equations \eqref{eqn:iso-condition-inj}--\eqref{eqn:iso-condition-bracket}. Let $X_1 = \varphi_U$, $X_2 = \varphi_V$, and $X_3 = \varphi_W$. We claim that $X_1, X_2, X_3$ satisfy \eqref{eqn:Lie-ring-to-TI}.
\begin{itemize}
\item Consider the action on $\tB$ in the (3,3,2) block of $T_L$. This is acted on by $\tB \mapsto X_3^{-1} \tB^{X_1} X_3^{-T}$, where here the exponent $\tB^{X_1}$ means that $X_1$ is used to take linear combinations of frontal slices of $\tB$. In terms of the bilinear form, this has the effect of replacing $B(x,y)$ by $X_1 B(X_3^{-1} x, X_3^{-1} y) = \varphi_U(B(\varphi_W^{-1}(x), \varphi_W^{-1}(y))$. By \eqref{eqn:iso-condition-bracket}, this is equal to $B'(x,y)$, which agrees with the (3,3,2) block of $T_{L'}$.

\item Consider the action on $M_{\pi}$ in the (1,4,1) block.  This transforms into $M_\pi \mapsto X_3 M_\pi X_2^{-T}$. But this is the same as $\varphi_W \circ \pi \circ \varphi_V^{-1}$, which by \eqref{eqn:iso-condition-surj} is equal to $\pi'$, hence making the (1,4,1) block into $M_{\pi'}$, the (1,4,1) block of $T_{L'}$. In the (4,1,1) block, the same equation occurs just with everything transposed and negated, resulting in $-M_{\pi'}$, again agreeing with the (4,1,1) block of $T_{L'}$.

\item Similarly, in the (5,1,3) block, we have $M_{\iota} \mapsto X_2 M_{\iota} X_1^{-1}$, corresponding to $\varphi_V \circ \iota \circ \varphi_U^{-1}$, and by \eqref{eqn:iso-condition-inj}, this is equal to $\iota'$, hence the resulting matrix equals $M_{\iota'}$, which is precisely the (5,1,3) block of $T_{L'}$. As in the previous case, the (1,5,3) block is simply the negative transpose in both $T_L$ and $T_{L'}$.
\end{itemize}

($\Leftarrow$) Given invertible $X_1, X_2, X_3$ satisfying \eqref{eqn:Lie-ring-to-TI}, define $\varphi_U = X_1, \varphi_V = X_2, \varphi_W = X_3$. Reading the preceding argument in reverse shows that $\varphi_U, \varphi_V, \varphi_W$ satisfy \eqref{eqn:iso-condition-inj}--\eqref{eqn:iso-condition-bracket}, and therefore by Lemma~\ref{lem:iso-Lie-ring}, we have that $L \cong L'$. This completes the proof of Lemma~\ref{lem:Lie-ring-to-TI}.
\end{proof}

Deciding if two tensors as in Lemma~\ref{lem:Lie-ring-to-TI} are equivalent under the action specified there is an instance of \algprobm{Partitioned \& Linked Alternating Matrix Space Isometry} over $\Z/p^e \Z$, but where the tensors need not be homocyclic (the modules need not be free). The latter problem is the same as \algprobm{Alternating Matrix Space Isometry} except that the matrices doing the isometry must have a block form (specified as part of the input; this is why it is called ``partitioned''), and certain blocks of the matrices doing the isometry are required to be equal, inverses, or transposes of one another (why it is called ``linked''). In \cite{TI4}, over fields gadgets were presented that reduces the latter problem to \algprobm{Alternating Matrix Space Isometry}. We claim that those same gadgets work in our setting, and the only key adaptation we need is Lemma~\ref{lem:invertible} and Remark~\ref{rem:TI4}, which allows us to reduce to (ordinary, non-partitioned, non-linked, but still not necessarily homocyclic)  \algprobm{Alternating Matrix Space Isometry} over $\Z/p^e \Z$. 

Finally, we reduce to \algprobm{Alternating Matrix Space Isometry} over free (equivalently, homocyclic) $\Z/p^e \Z$-modules. 

\begin{definition}\label{def:tensor_from_group}
Let $G$ and $H$ be two Abelian $p$-groups for some odd prime $p$, 
$f: G \times G \rightarrow H$ be a bilinear map, and $\tA$ be the tensor in $G \otimes G\otimes H$ with respect to $f_\tA$. 
Let $e$ be an integer such that the largest order of elements in both $G$ and $H$ is at most $p^e$. 
Suppose $G$ is isomorphic to 
\[\bigoplus_{\ell = 1}^{e} \left(\Z / p^{\ell } \Z\right)^{\alpha_\ell},\]
and $H$ is isomorphic to 
\[\bigoplus_{\ell = 1}^{e} \left(\Z / p^{\ell } \Z\right)^{\beta_\ell}.\]
Let $n = \sum_{\ell = 1}^{e} \alpha_\ell$ and $m = \sum_{\ell = 1}^{e} \beta_\ell $. 
An $n \times n \times m$ tensor $\tB = (b_{i, j, k})$ over $\Z / p^e \Z$ is a tensor derived from $\tA$ if there exist
$c_1, \dots, c_n \in G$ and $d_1, \dots, d_m \in H$ satisfying the following conditions:
\begin{enumerate}
\item $c_1, \dots, c_n$ generate $G$ and for each $1 \leq \ell \leq e$, $c_{j}$ is of order $p^\ell$ for any $(\sum_{i = 1}^{\ell - 1} \alpha_i) + 1 \leq j \leq \sum_{i = 1}^{\ell} \alpha_i$. (Equivalently, the $c_i$ are a basis for $G$ viewed as a $\Z/p^e \Z$-module.)
\item $d_1, \dots, d_m$ generate $H$ and for each $1 \leq \ell \leq e$, $d_{j}$ is of order $p^\ell$ for any $(\sum_{i = 1}^{\ell - 1} \beta_i) + 1 \leq j \leq \sum_{i = 1}^{\ell} \beta_i$. (Equivalently, the $d_i$ are a basis for $H$ viewed as a $\Z/p^e \Z$-module.)
\item For any $1 \leq i, j \leq n, 1 \leq k \leq m$, $b_{i, j, k} = f_\tA(c_i, c_j, d_k)\cdot p^{e - t_k}$, where $t_k$ is the integer such that the order of $d_k$ is $p^{t_k}$.
\end{enumerate}



\end{definition}

\begin{definition}\label{def:matrix_from_group}
Let $G$ and $H$ be two Abelian $p$-groups for some odd prime $p$, 
$f: G \rightarrow H$ be a linear map, and  
$A$ be a matrix in $G \otimes H$ with respect to $f$. 
Let $e$ be an integer such that the largest order of elements in both $G$ and $H$ is at most $p^e$. 
Suppose $G$ is isomorphic to 
\[\bigoplus_{\ell = 1}^{e} \left(\Z / p^{\ell } \Z\right)^{\alpha_\ell},\]
and $H$ is isomorphic to 
\[\bigoplus_{\ell = 1}^{e} \left(\Z / p^{\ell } \Z\right)^{\beta_\ell}.\]
Let $n = \sum_{\ell = 1}^{e} \alpha_\ell$ and $m = \sum_{\ell = 1}^{e} \beta_\ell $. 
An $n \times m$ matrix  $B = (b_{i, j})$ over $\Z / p^e \Z$ is a matrix derived from $A$ if there exist
$c_1, \dots, c_n \in G$ and $d_1, \dots, d_m \in H$ satisfying the following conditions:
\begin{enumerate}
\item $c_1, \dots, c_n$ generate $G$ and for each $1 \leq \ell \leq e$, $c_{j}$ is of order $p^\ell$ for any $(\sum_{i = 1}^{\ell - 1} \alpha_i) + 1 \leq j \leq \sum_{i = 1}^{\ell} \alpha_i$. 
\item $d_1, \dots, d_m$ generate $H$ and for each $1 \leq \ell \leq e$, $d_{j}$ is of order $p^\ell$ for any $(\sum_{i = 1}^{\ell - 1} \beta_i) + 1 \leq j \leq \sum_{i = 1}^{\ell} \beta_i$. 
\item For any $1 \leq i, j \leq n, 1 \leq k \leq m$, $b_{i, j} = f_A(c_i, d_j)\cdot p^{e - t}$, where $t$ is the integer such that the order of $d_j$ is $p^t$. 
\end{enumerate}



\end{definition}

\begin{lemma}\label{lem:reduction_to_homocyclic}
Let $G, H, G', H'$ be four Abelian $p$-groups such that $G \cong G'$ and $H \cong H'$. Then we have 
\begin{enumerate}
\item Let $\tA, \tA'$ be two tensors in $G \otimes G  \otimes H$ and $G' \otimes G' \otimes H'$, and 
$\tB$ and $\tB'$ be two tensors over $\Z / p^e \Z$ for some integer $e$ derived from $\tA$ and $\tA'$ respectively. 
Then there are $X_G \in \aut(G)$ and $Y_H \in \aut(H)$ such that 
$\tA' = (X_G, X_G^t, Y_H) \circ \tA$ 
if and only if there are invertible matrices $X \in \GL(n, \Z / p^e \Z)$ and $Y \in \GL(m, \Z / p^e \Z)$ such that $\tB' = (X, X^t, Y) \circ \tB$, where $n$ is the size of a basis for $G$ and $m$ is the size of a basis for $H$. 
\item Let $A, A'$ be two matrices in $G  \otimes H$ and $G' \otimes H'$, and 
$B$ and $B'$ be two matrices over $\Z / p^e \Z$ for some integer $e$ derived from $B$ and $B'$ respectively. 
Then there are $X_G \in \aut(G)$ and $Y_H \in \aut(H)$ such that 
$A' = X_G A Y_H$ 
if and only if there are invertible matrices $X \in \GL(n, \Z / p^e \Z)$ and $Y \in \GL(m, \Z / p^e \Z)$ such that $B'= X  B Y$, where $n$ is the size of a basis for $G$ and $m$ is the size of a basis for $H$. 

\end{enumerate}
\end{lemma}
\begin{proof}
By Definition~\ref{def:tensor_from_group}, the $\tB$ and $\tB'$ are of dimension $n \times n \times m$. 
Let $c_{\tA, 1}, \dots, c_{\tA, n}, d_{\tA, 1}, \dots, d_{\tA, m}$ and $c_{\tA', 1}, \dots, c_{\tA', n}, d_{\tA', 1}, \dots, d_{\tA', m}$ be the elements used to obtain $\tB$ and $\tB'$  following Definition~\ref{def:tensor_from_group} respectively. 

We first show that if $\tA$ and $\tA'$ are isomorphic, then there are $X$ and $Y$ such that $\tB' = (X, X^t, Y)\circ \tB$. 
Since $\tA$ and $\tA'$ are isomorphic, 
 there are $x_{i, j} \in \Z / p^e \Z$ such that $X_G(c_{\tA, i}) = \sum_{j = 1}^n {x_{i, j}} c_{\tA, j}$ 
 for every $1 \leq i \leq n$. 
 Let $X = (x_{i, j})$. 
 $X$ is in $\GL(n, \Z / p^e \Z)$. 
Similarly, there are $y_{i, j} \in \Z / p^e \Z$ such that $Y_H(d_{\tA, i}) = \sum_{j = 1}^m y_{i, j} d_{\tA, j}$ for every $1 \leq i \leq m$. 
 Let $Y_0 = (y_{i, j})$. 
 $Y_0$ is in $\GL(n, \Z / p^e \Z)$. 
It is easy to verify that $\tB' = (X, X^t, Y_0^{-1}) \circ \tB$. 

Now we show that if there are invertible $X$ and $Y$ such that $\tB' = (X, X^t, Y)\circ \tB$, then $\tA$ and $\tA'$ are isomorphic. 
Denote $X = (x_{i, j})$ and $Y' = Y^{-1} = (y_{i, j}')$. 
Let $X_G$ be the linear function for $G$ such that $X_G(c_{\tA, i}) = \sum_{j = 1}^{n} x_{i, j} c_{\tA, j}$, and 
$Y_H$ be the linear function for $H$ such that $Y_H(d_{\tA, i}) = \sum_{j = 1}^{m} y'_{i, j} d_{\tA, j}$. 
One can verify that both $X_G$ and $Y_H$ are automorphisms for $G$ and $H$ respectively, and $\tA' = (X_G, X_G^t, Y_H)$. 

The proof for the second case of the lemma is similar to the first case. 
This completes the proof of Lemma~\ref{lem:reduction_to_homocyclic}.
\end{proof}

\begin{definition}
Let $G_1, \dots, G_k, H_1, \dots, H_\ell$ be a sequence of Abelian $p$-groups. 
Let $\tA$ be a tensor in 
$(G_1 \oplus \dots \oplus G_k) \otimes (G_1 \oplus \dots \oplus G_k)  \otimes (H_1 \oplus \dots \oplus H_\ell)$ such that each non-zero subtensor on $G_i \otimes G_{j} \otimes H_u$ satisfies one of the following conditions:
\begin{enumerate}
\item $i = i'$ and the subtensor encodes a bilinear map. 
\item One of $G_i, G_j, H_u$ is isomorphic to 1, and the subtensor encodes a linear map. 
\end{enumerate}
The derived tensor over $\Z / p^e \Z$ for $\tA$ is obtained by replacing each non-zero sub-tensor by its derived tensor. 
\end{definition}

\begin{corollary}
Let $G_1, \dots, G_k, H_1, \dots, H_\ell$ and $G_1', \dots, G_k', H_1', \dots, H_\ell'$ be two sequences of Abelian $p$-groups such that $G_i \cong G_i'$ for every $1\leq i\leq k$ and $H_j \cong H_j'$ for every $1 \leq j \leq \ell$. 
Let $\tA$ and $\tA'$ be tensors in $(G_1 \oplus \dots \oplus G_k) \otimes (G_1 \oplus \dots \oplus G_k)  \otimes (H_1 \oplus \dots \oplus H_\ell)$ and  $(G_1' \oplus \dots \oplus G_k') \otimes (G_1' \oplus \dots \oplus G_k')  \otimes (H_1' \oplus \dots \oplus H_\ell')$ such that for each non-zero subtensor on $G_i \otimes G_{j} \otimes H_u$ (or $G_i' \otimes G_{j}' \otimes H_u'$) satisfying one of the following conditions:
\begin{enumerate}
\item $i = i'$ and the subtensor encodes a bilinear map. 
\item One of $G_i, G_j, H_u$ (or $G_i', G_j', H_u'$) is isomorphic to 1, and the subtensor encodes a linear map. 
\end{enumerate}
Then the two tensors are isomorphic if and only if their derived tensors over $\Z / p^e \Z$ are isomorphic.  
\end{corollary}

The dimension of the tensor obtained and the complex parameter is obtained by the construction of the tensor.

This completes the proof of Theorem~\ref{thm:GpI-to-AMSI}.
\end{proof}

\subsection{Reduction to matrix tuple isometry}\label{sec:tensor_iso_to_matrix_tuple}
In this section, we prove the following theorem.

\begin{theorem}\label{thm:tensor_iso_algo}
Let $\tA, \tB$ be two $n \times n \times m$ alternating tensors over $\Z / p^e \Z$ with the same complex parameter $\mu$. 
There is an algorithm to determine if 
there are invertible matrices $P \in \GL(n, \Z / p^e \Z)$ and $Q \in \GL(m, \Z / p^e \Z)$ such that 
$\tB = (P, P, Q) \circ \tA$ in time 
\[p^{O(\mu^{17 / 9} \cdot (\log p)^{8/9})} \cdot t,\]
where $t$ is the running time to determine whether two alternating matrix tuples $(A_1, \dots, A_k)$ and $(B_1, \dots, B_k)$ in $\M(n', \Z / p^e \Z)^{k}$ (for some $n'$ and $k$ that are polynomial in $n$ and $m$) has an invertible matrix $P \in \GL(n', \Z / p^e \Z)$ such that $A_i = P B_i P^t$ for any $1 \leq i \leq k$.
\end{theorem}

\subsubsection{Individualizing tensors over $\Z/p^e \Z$}

In this section, we consider the following problem: Let $\mathfrak{A}$ be a $\Z / p^e \Z$-submodule of $\M(n\times m, \Z / p^e \Z)$, and $r$ be an integer at most $\min\{n, m\}$. We want to choose two matrices $L$ and $R$ such that for any matrix $A \in \mathfrak{A}$ with inner rank at least $r$, $LAR$ is a non-zero matrix. 
Our goal is to choose $L$ and $R$ such that the number of rows of $L$ and the number of columns of $R$ is small. 

\begin{lemma}\label{lem:sampling_p_to_e}
Let $A$ be a matrix in $\M(m\times n,  \Z / p^e \Z)$ with inner rank $r$ and $L$ be a matrix in $M(t\times m,  \Z / p^e \Z)$ such that $LA$ is of inner rank $r'$ for some integer $r' < r$. 
For a randomly sampled row vector $q \in (\Z / p^e \Z)^m$, 
\[\begin{bmatrix}L \\ q\end{bmatrix}A\] is of inner rank $r' + 1$ with probability at least $1 - (1/p)^{r - r'}$. 
\end{lemma}
\begin{proof}

Let $L_{r'}$ be a matrix satisfying the following conditions:
\begin{enumerate}
\item $\rank(L_{r'} A) =\rank(L A) = r'$ and   $\rowspan(L A) \subset \rowspan(L_{r'} A)$, where $\rowspan(A)$ denotes the $\Z / p^e \Z$-span for the rows of matrix $A$.
\item There is no row vector $q \in (\Z / p^e \Z)^m$  such that 
\[\rank\left(\begin{bmatrix} L_{r'} \\ q \end{bmatrix} A\right) = r' \text{ and } 
\rowspan(L_{r'} A) \subsetneq \rowspan\left(\begin{bmatrix} L_{r'} \\ q \end{bmatrix} A\right).\]
\end{enumerate}
For all $r' + 1 \leq r'' \leq r$, let
\[L_{r''}  = \begin{bmatrix}L_{r'' - 1} \\ L_{r''}' \end{bmatrix}\]
where $L_{r''}'$ is a matrix with row vectors in $(\Z / p^e \Z)^m$ 
such that the following conditions hold: 
\begin{enumerate}
\item $\rank(L_{r''} A) = r''$. 
\item There is no $q_{r''}'$ as a row vector in $(\Z / p^e \Z)^m$  such that 
$\rowspan(L_{r''} A) \subsetneq \langle \rowspan(L_{r''}) \cup \{q_{r''}'\} \rangle$.
\item There is no row vector $q \in (\Z / p^e \Z)^m$  such that 
\[\rank\left(\begin{bmatrix} L_{r''} \\ q \end{bmatrix} A\right) = r'' \text{ and } 
\rowspan(L_{r''} A) \subsetneq \rowspan\left(\begin{bmatrix} L_{r''} \\ q \end{bmatrix} A\right).\]

\end{enumerate}
By the fact that  $A$ is of rank $r$, such $L_{r'}, L_{r' + 1}, \dots, L_{r}$ exist. 

Let $S_{r''} = \{q \in (\Z / p^e \Z)^m : q\cdot A \in \rowspan(L_{r''} A) \}$ for all $r' \leq r'' \leq r$.
For any $r' \leq r'' \leq r - 1$, any vector $q \in (\Z / p^e \Z)^m \setminus S_{r''}$ satisfies that the inner rank of 
\[\begin{bmatrix}L_{r''} \\ q\end{bmatrix} A\] is of rank $r'' + 1$,
and $S_{r'} = (\Z / p^e \Z)^m$. 
On the other hand, for any $r'  + 1\leq r'' \leq r$ and any $q \in S_{r''} \setminus S_{r'' - 1}$, the inner rank of 
\[\begin{bmatrix}L_{r'' - 1} \\ i\cdot q + x\end{bmatrix}A\]
is $r''$ for any $1 \leq i \leq p - 1$ and $x \in S_{r'' - 1}$. 
Hence, for a randomly sampled row vector $q \in (\Z / p^e \Z)^m$, 
\[\begin{bmatrix}L \\ q\end{bmatrix}A\] is a matrix of inner rank $r' + 1$ with probability at least $1 - (1 / p)^{r - r'}$. 
\end{proof}

\begin{lemma}\label{lem:individualization_single_matrix}
Let $A$ be a matrix in $\M(m\times n,  \Z / p^e \Z)$ with inner rank $r$. 
Given an integer $r' \leq r / 2$ and a parameter $0 < \delta < 1$ such that $\log(1 / \delta) \geq r$, 
let $Q$ be a random matrix in \[\M\left(\left\lceil \frac{8 r' \log (1/\delta)}{r}\right\rceil\times m,  \Z / p^e \Z\right).\] With probability at least $1 - \delta$, $QA$ is of inner rank at least $r'$. 

\end{lemma}
\begin{proof}

Let $\alpha = \left\lceil \frac{8 r' \log (1/\delta)}{r}\right \rceil$. 
Without loss of generality,
we assume the matrix $Q$ is sampled sequentially by rows $q_1, . . . , q_\alpha$, where $q_i$ is the $i$-th row of $Q$.
Let $Q_\beta$ denote the matrix $Q$ that contains only the first $\beta$ rows for some $1 \leq \beta \leq \alpha$. 

Suppose that after sampling the first $\beta$ rows for some positive integer $\beta$, $Q_\beta A$ is of inner rank $\gamma$. 
By Lemma~\ref{lem:sampling_p_to_e}, the probability of sampling a random row vector $q \in (\Z / p^e \Z)^m$ such that \[\begin{bmatrix}Q_\beta \\ q\end{bmatrix}B\] is a matrix of inner rank $\gamma + 1$ is at least $1 - (1 / p)^{r - \gamma}$.
By sampling $\lceil \log (r' / \delta)  / (r - r')\rceil $ new rows of $Q$, the inner rank of $Q$ increases by at
least one with probability
\[1 - \left(\frac{1}{p} \right)^{(r - \gamma) \lceil \log(r' / \delta) / (r - r')\rceil} \geq 1 - \left(\frac{1}{p} \right)^{\log (r' / \delta)}\geq 1 - \frac{\delta}{r'}. \]
On the other hand, observe that 
\[r' \left\lceil \frac{\log (r' / \delta)}{(r - r')}\right\rceil \leq \frac{2 r' \log (r' / \delta)}{r - r'}\leq \frac{8r' \log (1 / \delta)}{r}. \]
By union bound, with probability at least $1 - \delta$, if $Q$ contains at least $\left\lceil \frac{8 r' \log (1/\delta)}{r}\right\rceil$ rows, the inner rank of $QA$ is a matrix of inner rank at least $r'$. 
\end{proof}

\begin{lemma}\label{lem:individualization}
For any $\Z/p^e \Z$-submodule of $\M(m\times n, \Z / p^e \Z)$ containing $\tau$ different matrices and any integer $r \geq 4$, 
denote 
$t = \left\lceil 32 \max\{\log \tau, r \}/ \sqrt{r} \right\rceil$. 
There is a matrix $L \in \M(t\times m, \Z / p^e \Z)$ and a matrix $R \in \M(n\times t, \Z / p^e \Z)$ such that for any matrix $A$ in the submodule of rank at least $r$, $LAR$ is a non-zero matrix. 
\end{lemma}
\begin{proof}
Let $A$ be an arbitrary matrix of rank at least $r$ in the submodule. By Lemma~\ref{lem:individualization_single_matrix} with $r' = \lfloor \sqrt{r}\rfloor$ and $\delta = \min\{1 / 4\tau, 1 / 2^r \}$, for a random matrix $L \in \M(t, m, \Z/p^e \Z)$, $LA$ is of rank at least $r'$ with probability at least $1 - \delta$. 

If this is the case, then by Lemma~\ref{lem:individualization_single_matrix}, with $r' = 1$ and $\delta = \min\{1 / 4\tau, 1 / 2^r \}$, for a random matrix $R \in \M(n, t, \Z/p^e \Z)$, $LAR$ is of rank at least $1$ with probability at least $1 - \delta$.

By union bound, with constant probability, for random $L$ and $R$, $LAR$ is a non-zero matrix
for all the matrices $A$ in the submodule of rank at least $r$.
\end{proof}

\subsubsection{Low rank matrix submodule characterization over $\Z/p^e \Z$}

We define the rank of a $(\Z / p^e \Z)$-submodule of $(\Z / p^e \Z)^n$ to be the smallest integer $r$ such that 
the submodule has a generating set of size  $r$. 

\begin{lemma}\label{lem:space_dim_prelim}
Let $A_1, A_2, \dots, A_k$ be $k$ matrices in $\M(m\times n, \Z / p^e \Z)$ for some positive integers $k, m, n$. 
If there exist $d$ row vectors
 $x_1, x_2, \dots, x_d \in ( \Z / p^e \Z)^m$ such that for every $1 \leq i < d$, 
the following condition holds
\begin{align}\label{equ:rank_bound}
\rank\big(\left\langle \left\{x_j A_\ell : 1 \leq j \leq i, 1 \leq \ell \leq k\right\}\right\rangle \big) \neq \rank\big(\left\langle \left\{x_j A_\ell : 1 \leq j \leq i + 1, 1 \leq \ell \leq k\right\} \right\rangle\big),
\end{align}
then there is a linear combination of $A_1, A_2, \dots, A_k$ with inner rank at least $(1 - 1 / p)d$.
\end{lemma}
\begin{proof}
Let $X$ be the matrix in $\M(d\times n, \Z / p^e \Z)$ such that $x_i$ is the $i$-th row of $X$.
To prove the lemma, it is sufficient to show that if
$\alpha_1, \dots, \alpha_k $ are uniformly and independently  sampled from $\Z / p^e \Z$, then
we have the following expectation estimation. 
\[\mathbb{E}_{\alpha_1, \dots, \alpha_k}\left[\rank \left({X \left(\sum_{\ell=1}^k \alpha_\ell A_\ell\right) }\right)\right] \geq \left(1 - \frac{1}{p}\right)d.\]
If this is the case, then there exist $\alpha_1, \dots, \alpha_k \in \Z / p \Z$ such that 
\[\rank{ \left(\sum_{\ell=1}^k \alpha_\ell A_\ell \right)} \geq \rank {\left(X \left(\sum_{\ell=1}^k \alpha_\ell A_\ell\right) \right)} \geq \left(1 - \frac{1}{p}\right)d, \]
and then the lemma follows. 

For any $1 \leq i \leq d$, 
let $D_i$ be a $\Z / p^e \Z$-row span constructed as follows:
Let $D_i$ be \[\left\langle \left\{x_j A_\ell : 1 \leq j \leq i, 1 \leq \ell \leq k\right\}\right\rangle\] initially. 
Find a vector $u = x_i (\sum_{j = 1}^k \alpha_j A_j)$ for some $\alpha_1, \dots, \alpha_k \in \Z / p^e \Z$ such that $u$ is not in $D_i$ but $\rank(\langle \{u\}\cup D_i \rangle) = \rank(D_i)$. 
Let $D_i = \langle \{u\}\cup D_i \rangle$. 
Repeat above process until no vector $u$  can be added to $D_i$ to increase the size of $D_i$. 
Let $S_i$ be the set of row vectors
\[\left\{(\alpha_1, \dots, \alpha_k) \in (\Z / p^e \Z)^k : x_i \left(\sum_{\ell = 1}^k \alpha_\ell A_\ell \right) \in D_{i}\right\}.\]
By Inequality~(\ref{equ:rank_bound}), $S_i$ does not equal to $(\Z / p^e \Z)^k$.

Since $D_{i-1}$ is a $\Z / p^e \Z$-submodule of $(\Z / p^e \Z)^n$, 
for any $(\alpha_1, \dots, \alpha_k)$ in $S_i$ and $\mu \in \Z / p^e \Z$, $\mu (\alpha_1, \dots, \alpha_k)$ is in $S_i$, and 
if both $(\alpha_1, \dots, \alpha_k)$ and $(\beta_1, \dots, \beta_k)$ are in $S_i$, then $(\alpha_1 + \beta_1, \dots, \alpha_k + \beta_k)$ is also in $S_i$. 
Hence, the row vectors in $S_i$ form a $\Z / p^e \Z$-submodule of $(\Z / p^e \Z)^k$, and thus 
$|S_i| \leq p^{k e-1}$.

On the other hand, for any $(\alpha_1, \dots, \alpha_k) \notin S_i$, 
since $\rank(\langle \{x_i (\sum_{\ell = 1}^k \alpha_\ell A_\ell)\} \cup D_i\rangle) > \rank(D_i)$, 
we have 
\[\rank\left(\begin{bmatrix} x_1 \\ \vdots \\ x_i \end{bmatrix}\left(\sum_{\ell=1}^k \alpha_\ell A_\ell\right)\right) > \rank\left(\begin{bmatrix} x_1 \\ \vdots \\ x_{i-1} \end{bmatrix}\left(\sum_{\ell=1}^k \alpha_\ell A_\ell\right)\right).\]
Thus,
\[\mathrm{Pr}_{\alpha_1, \dots, \alpha_k}\left[\rank\left(\begin{bmatrix} x_1 \\ \vdots \\ x_i \end{bmatrix}\left(\sum_{\ell=1}^k \alpha_\ell A_\ell\right)\right) > \rank\left(\begin{bmatrix} x_1 \\ \vdots \\ x_{i-1} \end{bmatrix}\left(\sum_{\ell=1}^k \alpha_\ell A_\ell\right)\right)\right] \geq \frac{p^{ek} - |S_i|}{p^{ek}} \geq 1 - \frac{1}{p}.\]

Thus, we have \begin{align*}& \mathbb{E}_{\alpha_1, \dots, \alpha_k}\left[\rank {X \left(\sum_{\ell=1}^k \alpha_\ell A_\ell\right) }\right]  \\ \geq & 
\sum_{i=1}^d \mathrm{Pr}_{\alpha_1, \dots, \alpha_k}\left[\rank\left(\begin{bmatrix} x_1 \\ \vdots \\ x_i \end{bmatrix}\left(\sum_{\ell=1}^k \alpha_\ell A_\ell\right)\right) > \rank\left(\begin{bmatrix} x_1 \\ \vdots \\ x_{i-1} \end{bmatrix}\left(\sum_{\ell=1}^k \alpha_\ell A_\ell\right)\right)\right]  \\
\geq & \left(1 - \frac{1}{p}\right)d.\qedhere\end{align*}

\end{proof}

\begin{lemma}\label{lem:char_main}
Let $\mathfrak{A}$ be a $\Z / p^e \Z$-submodule of $M(m, n, \Z / p^e \Z)$ 
such that for each matrix $A$ in the submodule, the inner rank of $A$ is at most $r$ for some positive integer $r$. 
There exists a matrix $P \in \GL(m, \Z / p^e \Z)$ and an integer $d = O(er)$ such that 
\[\rank\left(\left\langle \bigcup_{i = d+1}^{m} \langle p_i \mathfrak{A}\rangle \right\rangle\right) \leq O(e^2 r^2),\]
where $p_i$ denotes the $i$-th row of $P$.


\end{lemma}

\begin{proof}
Let $A_1, \dots, A_k$ be a generating set of $\mathfrak{A}$.
Let $x_1, x_2, \dots, x_d \in (\Z / p^e \Z )^m$ be $d$ row vectors for some positive integer $d$ 
 such that there is a matrix in $\GL(m, \Z / p^e \Z)$ with $x_i$ being $i$-th row, 
 and for any $1 \leq i \leq d$, there are at least $p^{ek} / 2$ different $\alpha_1, \alpha_2, \dots, \alpha_k \in \Z / p^e \Z$ satisfying
\begin{equation}\label{equ:characterization}\rowspan\left(X_{i-1} \left(\sum_{j =1}^k \alpha_j A_j\right)\right) \subsetneq \rowspan\left(X_i \left(\sum_{j =1}^k \alpha_j A_j\right)\right)\end{equation} where for each $1 \leq i \leq d$,
$X_i$ is the $i \times m$ matrix with $x_j$ as the $j$-th row of $X_i$ for all $1 \leq j \leq i$.
Since every linear combination $\sum_{j =1}^k \alpha_j A_j$ of $A_1, \dots, A_k$ has at most $e \cdot r$ different $i$ such that (\ref{equ:characterization}) holds, 
we have 
\[d \cdot \frac{p^{ek}}{2} \leq e \cdot r \cdot p^{ek},\]
which implies $d \leq 2er$. Suppose there does not exist a row vector $x_{d+1} \in (\Z / p^e \Z)^m$ such that there are at least $p^{ek} / 2$ different $\alpha_1, \alpha_2, \dots, \alpha_k\in\Z / p^e \Z$ satisfying
\[\rowspan\left(X_{d} \left(\sum_{j =1}^k \alpha_j A_j\right)\right) \subsetneq \rowspan\left(X_{d+1} \left(\sum_{j =1}^k \alpha_j A_j\right)\right).\]
Then for each row vector $v \in (\Z / p\Z)^m$ such that $v\notin \langle x_1, \dots, x_d\rangle$, 
there exist $\beta_1, \dots, \beta_d \in \Z / p^e\Z$ such that the following condition holds for at least $\frac{p^{ek}}{ 2p^{ed}}$ different $\alpha_1, \alpha_2, \dots, \alpha_k \in \Z / p^e\Z$
\[v \cdot \left(\sum_{i = 1}^k \alpha_k A_k\right) = \left(\sum_{i=1}^d \beta_i x_i\right)\left(\sum_{i = 1}^k \alpha_k A_k\right).\]
Thus, the rank of $\langle (v -  \sum_{i=1}^d \beta_i x_i) \mathfrak{A}\rangle$ is at most $ed + 1 = O(e^2r)$. 
By Lemma~\ref{lem:space_dim_prelim}, one can construct a matrix $P \in \mathrm{GL}(m, \Z / p^e \Z)$ satisfying the following properties: Let $p_i$ denote the row vector of the $i$-th row of $P$.
\begin{enumerate}
\item $p_i = x_i$  for any $1 \leq i \leq d$.
\item $\langle \cup_{i = d + 1}^m \cup_{j = 1}^k\{ p_i A_j \}\rangle$ is of rank at most $O(e^2r^2)$. 
\end{enumerate} 
Any such matrix $P$ proves the lemma.
\end{proof}

\begin{lemma}\label{lem:char_skew}
Let $r$ be a positive integer and $\mathfrak{A}$ be a $\Z / p^e \Z$-submodule of $\M(n, \Z / p^e \Z)$ 
such that 
for each matrix $A$ in the submodule, $A$ is an alternating matrix and the inner rank of $A$ is at most $r$. 
Then there exists an integer $t = O(e^2 r^2)$ and a matrix $P \in \GL(n, \Z / p^e \Z)$ such that for each matrix $A \in \mathfrak{A}$,
$i$-th row $j$-th column of $P A P^t$ is zero for any $i \geq t, j \geq t$. 
\end{lemma}
\begin{proof}
By Lemma~\ref{lem:char_main}, there exists an integer $d = O(er)$ and a matrix $P_0 \in \GL(n, \Z / p^e \Z)$ such that  
\[\rank\left(\left\langle \bigcup_{i = d+1}^{m} \langle p_{0, i} \mathfrak{A}\rangle \right\rangle\right) \leq O(e^2 r^2),\]
where $p_{0, i}$ denotes the $i$-th row of $P_0$. 
Let $S$ be a smallest generating set of  $\left\langle \bigcup_{i = d+1}^{m} \langle p_{0, i} \mathfrak{A}\rangle \right\rangle$. We have $|S| = O(e^2 r^2)$. 
Denote $S = \{s_1, \dots, s_{|S|}\}$. 

Let $t_0 = d$. 
We perform the following process for $i = 1, \dots, |S|$: 
\begin{enumerate}
\item Find a  $j \in \{t_{i-1} + 1, \dots n\}$ such that $p_{i-1, j} \cdot s_i^t \in \Z / p^e \Z$ (suppose $p_{i - 1, j}$ denote $j$-th row of $P_{i-1}$) is non-zero and generates all the $p_{i-1, k} \cdot s_i^t$ under additions for any $t_{i-1} + 1\leq k \leq n$. 
If there is no such $j$, then let $P_i = P_{i-1}$, $t_i = t_{i - 1}$,  and continue to the process for $i + 1$. 
\item Otherwise, let $t_i = t_{i-1} + 1$ and $P_i'$ be the resulted matrix by switch $j$-th row and $t_i$-th row of $P_{i-1}$.
\item Otherwise, $P_i$ be the matrix such that the first $t_i$ rows are the same as $P_i'$, and for any $t_i + 1 \leq j \leq n$, 
the $j$-th row of $P_i$ is a linear combination of $j$-th row and $t_i$-th row of $P_i'$ such that its inter product with $s_i$ is zero. 
\end{enumerate}
Let $P$ be $P_{|S|}$. 
It is easy to verify that $t_{|S|} = O(e^2 r^2)$, and for each matrix $A \in \mathfrak{A}$,
$i$-th row $j$-th column of $P A P^t$ is zero for any $i \geq t_{|S|}, j \geq t_{|S|}$.
In addition, since $P$ is obtained by applying multiple row switching operations and row addition operations to $P_0$, $P$ is also in $\GL(n, \Z / p^e \Z)$. 
\end{proof}

\subsubsection{Reduction to matrix tuple isometry}
We define some additional notations used in this section. 
Let $\tA$ be an $n \times m \times h$ tensor over $\Z / p^e \Z$. 
Let $X = (x_{i', i})\in \M(n'\times n, \Z/ p^e \Z), Y = (y_{j', j}) \in \M(m'\times m, \Z/ p^e \Z)$, and $Z = (z_{k', k}) \in \M(h'\times h, \Z / p^e \Z)$. 
We define $(X, Y, Z)\circ \tA$ to be the $n' \times m' \times h'$ dimensional tensor $\tB = (b_{i', j' ,k'})$ such that
$b_{i', j', k'} = \sum_{i, j, k} x_{i', i} y_{j', j} z_{k', k} a_{i, j, k}$. 
Here, $X$, $Y$, and $Z$ do not need to be square matrices.

For a row vector $x = (x_1, \dots, x_n) \in (\Z / p^{e} \Z)^n$, we denote $\mathfrak{X}_{\tA, x} = \sum_{\ell = 1}^n x_\ell \cdot \mathfrak{X}_{\tA, \ell}$.
For a row vector $y = (y_1, \dots, y_m) \in (\Z / p^{e} \Z)^m$, we denote $\mathfrak{Y}_{\tA, y} = \sum_{\ell = 1}^m y_\ell \cdot \mathfrak{Y}_{\tA, \ell}$.
For a vector $z = (z_1, \dots, z_h) \in (\Z / p^{e} \Z)^h$, we denote $\mathfrak{Z}_{tA, z} = \sum_{\ell = 1}^h z_\ell \mathfrak{Z}_{\tA, \ell}$.
The order of a matrix $\mathfrak{X}_{\tA, x}$, denoted as $\ord(\mathfrak{X}_{\tA, x})$, is the smallest non-negative integer $t$ such that 
$t \cdot \mathfrak{X}_{\tA, x}$ is a zero matrix. 
The orders of matrices $\mathfrak{Y}_{\tA, y}$ and $\mathfrak{Z}_{\tA, z}$ are defined in the same way. 

Let $O_{\mathfrak{X}, t} = \{x \in (\Z / p^{e} \Z)^n : \ord(\mathfrak{X}_{T, x}) \leq t\}$. Clearly, $\langle O_{\mathfrak{X}, t} \rangle = O_{\mathfrak{X}, t}$ for any $t$.
We also define $O_{\mathfrak{Y}, t}$ and $O_{\mathfrak{Z}, t} $  in the same way.

\begin{lemma}\label{lem:enumeration}
Let $\tA$ be an alternating $n \times n \times m$ tensor over $\Z / p^{e} \Z$, and $e_0$ be an integer such that $1 \leq e_0 \leq e$. 
Suppose the abelian group generated by $\mathfrak{X}_{\tA, x}$ for all the $x \in (\Z / p^{e}\Z)^n$ is isomorphic to 
\[\bigoplus_{\ell = 1}^{e} \left(\Z / p^{\ell } \Z\right)^{\beta_\ell}.\]
For any integer $4 \leq r \leq n$, 
there are matrices $X_0, X_1, X_2, X_3$ and $T$ satisfying the following conditions:
\begin{enumerate}
\item Every row of $X_0, X_1, X_2, X_3$ and $T$ is in $(\Z / p^{e} \Z)^n$. In addition, 
\[\rowspan\left(\begin{bmatrix} X_0\\ X_1 \\ X_2 \\ X_3\end{bmatrix}\right) = (\Z / p^{e} \Z)^n,\]
and 
\[\rowspan\left(\begin{bmatrix} X_1 \\ X_2 \\ X_3\end{bmatrix}\right) = O_{\mathfrak{X}, p^{e_0}}.\]
\item $T$ has $O(e_0^2 r^2)$ rows from $O_{\mathfrak{X}, p^{e_0}}$. 

\item $X_0$ has $\sum_{i = e_0+ 1}^{e} \beta_i$ rows, $X_1$ has 
$O(\max\{\log(p^{\sum_{i = 1}^{e} i \cdot \alpha_i}), r\} /  \sqrt{r} )$ rows, and
$X_2$ has $O(e_0^2 r^2 + \sum_{i = e_0 + 1}^{e} \beta_i)$ rows. 
Rows of $X_3$ correspond to a smallest generating set of 
\[\left\langle \left\{x \in O_{\mathfrak{X}, p^e} : \forall s \in S, s \cdot x^t = 0 \text{ and } 
x \cdot \mathfrak{Z}_{\tA, z} \in S \text{ for all } z \in (\Z / p^e \Z)^m \text{ s.t. } X_1  \mathfrak{Z}_{\tA, z} X_1^t = 0,   \right\}\right\rangle,\]
where \[S = \langle \{s  \in (\Z / p^e \Z)^m : \exists z  \text{ s.t. }X_1 \mathfrak{Z}_{\tA, z} X_1^t = 0, s \in \rowspan(T \mathfrak{Z}_{\tA, z})\}\rangle.\]

\item For any $z \in (\Z / p^e \Z)^m$ such that 
\[\begin{bmatrix} X_0 \\ X_1 \\ X_2 \end{bmatrix} \mathfrak{Z}_{\tA, z} \begin{bmatrix} X_0 \\ X_1  \\ X_2 \end{bmatrix}^t= 0,\]
$X_3 \mathfrak{Z}_{\tA, z}  X_3^t = 0$. 

\end{enumerate} 
\end{lemma}
\begin{proof}
Since the abelian group generated by $\mathfrak{X}_{\tA, x}$ over $x \in (\Z / p^{e}\Z)^n$ is isomorphic to $\bigoplus_{\ell = 1}^{e} \left(\Z / p^{\ell } \Z\right)^{\beta_\ell},$
there is an invertible matrix $X \in \GL(n, \Z / p^{e}\Z)$ such that for any $1 \leq i \leq e$ and any $j\in[\sum_{k = i + 1}^{e} \beta_k + 1, \sum_{k = i}^{e} \beta_k]$,
$\mathfrak{X}_{\tA, j}$ is of order $p^i$, and for $j > \sum_{\ell = 1}^{e} \beta_\ell$,   $\mathfrak{X}_{\tA, j}$ is the zero matrix.
Let $X_0$ be the first $\sum_{\ell = e_0 + 1}^{e} \beta_\ell$ rows of $X$. 

Let $X'$ be a matrix such that the rows of $X'$ correspond to a smallest generating set of $O_{\mathfrak{X}, p^{e_0}}$, and $n'$ denote the number of rows in $X'$. We have $n' \leq n$ and $\langle \rowspan(U_0) \cup \rowspan(X')\rangle = (\Z / p^e \Z)^n$.
For any $x \in \rowspan(X')$, $\mathfrak{X}_{\tA, x}$ is a matrix of order at most $p^{e_0}$. 
$(X', X', I) \circ \tA$ has every entry be a multiple of $p^{e - e_0}$.

Let $\tA'$ be the tensor over $\Z / p^{e_0} \Z$ equivalent to $(X', X', I) \circ \tA$. 
By Lemma~\ref{lem:individualization}, 
there is a matrix 
\[U_{1}' \in \M\left(O\left(\frac{\max\left\{\log(p^{\sum_{i = 1}^{e} i \cdot \alpha_i}), r\right\}}{ \sqrt{r}}\right), n', p^{e_0}\right)\] such that 
for any matrix $\mathfrak{Z}_{\tA', z}$ of inner rank at most $r$, $U_{1}' \mathfrak{Z}_{\tA', z} U_{1}'^t$ is a non-zero matrix. 

Let $\mathfrak{A}$ be the $\Z / p^{e_0}\Z$-submodule of $\M(n', \Z / p^{e_0} \Z)$ spanned by all the $\mathfrak{Z}_{\tA', z}$ such that $U_{1}'\mathfrak{Z}_{\tA’, z} U_{1}'^t = 0$. 
By Lemma~\ref{lem:char_skew}, there exists a matrix $P \in \GL(n, \Z / p^{e_0} \Z)$ such that the $i$-th row $j$-th column of $P A P^t$ is zero for all the $A \in \mathfrak{A}$ and $i, j \geq O(e_0^2r^2)$. 

Hence, let $P'$ be the matrix of $P$ removing the first $O(e_0^2 r^2)$ rows. 
We have \[\left\langle \left\{ s \in (\Z / p^{e_0}\Z)^{n'} : \exists z \in (\Z / p^{e_0} \Z)^m \text{ s.t. } U_{1}' \mathfrak{Z}_{\tA', z} U_{1}'^t = 0 \text{ and } s \in \rowspan(P'  \mathfrak{Z}_{\tA', z}) \right\} \right\rangle\] has a generating set of $O(e_0^2 r^2)$ row vectors such that every element in the generating set multiple $p^e$ equals the zero vector.

Let $U_{1}$ be the matrix over $\Z / p^e \Z$ which has the same dimension as $U_{1}'$ such that every entry of $U_{1}$ equals $U_{1}'$.
By the correspondance between $(X', X', I) \circ \tA$ and $\tA'$,  
there exist matrices $T_0$,
$U_{2}$ and $U_3$ over $\Z / p^e \Z$ all with $n'$ columns satisfying the following conditions:
\begin{enumerate}
\item \[\rowspan\left(\begin{bmatrix} U_{1} \\ U_{2} \\ U_3 \end{bmatrix}\right) = (\Z /  p^e \Z)^{n'}\]
\item $T_0$ contains $O(e_0^2 r^2)$ rows, $U_1$ contains $O(\max\{\log(p^{\sum_{i = 1}^{e} i \cdot \alpha_i}), r\} / \sqrt{r})$ rows, and $U_2$ contains $O(e_0^2 r^2)$ rows. 
\item 
Let 
\[\begin{split}S_0 = \langle \{s \in (\Z / p^{e}\Z)^{n'} : \exists  z \in (\Z / p^{e}\Z)^m \text{ s.t. }& U_1 \mathfrak{Z}_{(X', X', I) \circ \tA, z} U_1^t = 0 \text{ and } \\ & s \in \rowspan(T_0 \mathfrak{Z}_{(X', X', I) \circ \tA, z})\}\rangle.\end{split}\]
$S_0$ has a generating set of size at most $O(e_0^2 r^2)$.
\item 
The rows of $U_3$ is a generating set of 
\[\begin{split} \langle \{u \in (\Z / p^e \Z)^{n'} : 
u \cdot \mathfrak{Z}_{(X', X', I) \circ \tA, z} \in S_0  \text{ for all } 
z \in (\Z / p^e \Z)^m \text{ s.t. }  U_1 \mathfrak{Z}_{(X', X', I) \circ \tA, z} U_1^t = 0 \}\rangle.\end{split}\]
\end{enumerate}

Now we construct $T, X_1, X_2, X_3$ as required by the lemma. 
Let 
$X_1 = U_1 X'$, 
\[S =  \langle \{ s \in (\Z / p^e \Z)^n : \exists z \in (\Z / p^e \Z)^m \text{ s.t. } X_1 \mathfrak{Z}_{\tA, z} X_1^t = 0 \text{ and } s \in \rowspan(T_0 X' \mathfrak{Z}_{\tA, z})\} \rangle. \]
By the relation between $\tA$ and $(X', X', I) \circ \tA$, 
$S_1$ has a generating set of size at most $O(e_0^2 r^2 + \sum_{\ell = e_0 + 1}^e \beta_{\ell})$, and there is a matrix $T$ of $O(e_0^2 r^2 + \sum_{\ell = e_0 + 1}^e \beta_{\ell})$ rows such that 
\[S = \langle \cup_{z \in (\Z / p\Z)^m : X_1 \mathfrak{Z}_{\tA, z}X_1^t = 0}\{\rowspan(T \mathfrak{Z}_{\tA, z})\}\rangle.\]

So let $X_3$ be a smallest generating set of 
\[\langle \{x \in O_{\mathfrak{X}, p^{e_0}} : \forall s \in S, s \cdot x^t = 0 \text{ and } x \cdot \mathfrak{Z}_{ \tA, z} \in S  \text{ for all } 
z \in (\Z / p^e \Z)^m \text{ s.t. }  X_1 \mathfrak{Z}_{\tA, z} X_1^t = 0\}\rangle,\]
and $X_2$ contains all the rows of $U_2 X'$ such that $\rowspan(U_3X')$ is a subset of $\langle \rowspan(X_2) \cup \rowspan(X_3)\rangle$.
There exists such a $X_2$ with $O(e_0^2r^2 + \sum_{\ell = e_0 + 1}^e \beta_\ell)$ rows.


The $X_0, X_1, X_2, X_3$ and $T$ defined above satisfies the requirement of the lemma. 
\end{proof}

\begin{lemma}\label{lem:lexical_order}
Let $G$ be an abelian $p$-group, and let $\rho$ be a complete strict order of all the elements in $G$. 
Suppose comparing the order of two elements under $\rho$ takes constant time, and 
$G$ is isomorphic to 
$\bigoplus_{\ell = 1}^{e} \left(\Z / p^{\ell } \Z\right)^{k_\ell}$.
There is an algorithm with running time polynomial in the order of $G$ to compute $g_1, \dots, g_{\sum_{\ell = 1}^e k_\ell}$ satisfying the following conditions:
\begin{enumerate}
	\item For any $1 \leq \ell \leq e$, $g_{(\sum_{j = 1}^{\ell - 1} k_j) + 1}, \dots, g_{\sum_{j = 1}^{\ell} k_j}$ are of order $p^\ell$. 
	\item $g_1, \dots, g_{\sum_{\ell = 1}^e k_\ell}$ form a generating set of $G$. 
	\item $g_1, \dots, g_{\sum_{\ell = 1}^e k_\ell}$ has the smallest order, i.e., for any $g_1', \dots, g_{\sum_{\ell = 1}^e k_\ell}'$ satisfying the first two conditions but different to $g_1, \dots, g_{\sum_{\ell = 1}^e k_\ell}$, there is an $1 \leq i \leq \sum_{\ell = 1}^e k_\ell$ such that $g_j' = g_j$ for all the $1 \leq j < i$ and $\tau(g_i) < \tau(g_i')$.
\end{enumerate}
\end{lemma}
\begin{proof}
Here is the algorithm:
\begin{enumerate}
\item Let $\ell = 1$, $i = 0$.
\item Repeat until $g_1, \dots, g_i$ generate $G$:
\begin{enumerate}
	\item Find the element $g_i$ satisfying the following conditions:
	\begin{enumerate}
		\item $g_i$ is of order $p^\ell$, not generated by $g_1, \dots, g_{i-1}$, and there does not exist $g' \in G$ such that $g'^t = g_i$ for some positive integer $t$.
		\item For any other $g_i'$ satisfying the first condition, $\rho(g_i) < \rho(g_i')$. 
	\end{enumerate}
	\item If $g_i$ exists, then $i = i + 1$, otherwise $\ell = \ell + 1$. 
\end{enumerate} 
\end{enumerate}
The correctness of the algorithm is obtained by the characterization of the automorphism of abelian groups.
The running time of the algorithm is obtained by the fact that every group has a generating set of size at most logarithmic in the group order.
\end{proof}

\begin{definition}\label{def:semi_canonical}
Given an $n\times n \times m$ alternating tensor $\tA$ over $\Z / p^e \Z$ and matrices $T, X_0, X_1, X_2$ with rows in $(\Z / p^e \Z)^n$ satisfying Lemma~\ref{lem:enumeration}, 
a tensor $\tB$ is a semi-canonical form of $\tA$ if there are $X_3, Z_0, Z_1$ satisfying the following conditions such that the $\tB$ equals 
\[\left(\begin{bmatrix} X_0 \\ X_1 \\ X_2 \\ X_3 \end{bmatrix}, \begin{bmatrix} X_0 \\ X_1 \\ X_2 \\ X_3 \end{bmatrix}, \begin{bmatrix} Z_0 \\ Z_1\end{bmatrix}\right) \circ \tA.\]

\begin{enumerate}
\item $X_3$ is a matrix such that the row span of 
\[\begin{bmatrix} X_0 \\ X_1 \\ X_2 \\ X_3 \end{bmatrix}\]
is $(\Z / p^e \Z)^n$, and the rows of $X_3$ is a smallest generating set of 
\[\left\langle \left\{x \in O_{\mathfrak{X}, p^e} : \forall s \in S, s \cdot x^t = 0 \text{ and }  x \cdot \mathfrak{Z}_{\tA, z} \in S 
\text{ for any } z \in (\Z / p^e \Z)^m \text{ s.t. } X_1  \mathfrak{Z}_{\tA, z} X_1^t = 0 \right\}\right\rangle,\]
where \[S = \langle \{s  \in (\Z / p^e \Z)^m : \exists z  \text{ s.t. }X_1 \mathfrak{Z}_{\tA, z} X_1^t = 0, s \in \rowspan(T \mathfrak{Z}_{\tA, z})\}\rangle.\]

\item Suppose the matrices are ordered by the lexical order. Compute $z_1, \dots, z_d$ as row vectors such that \[\begin{bmatrix} X_0 \\ X_1 \\ X_2\end{bmatrix} \mathfrak{Z}_{T, z_1} \begin{bmatrix} X_0 \\ X_1 \\ X_2 \end{bmatrix}^t, \dots, \begin{bmatrix} X_0 \\ X_1 \\ X_2 \end{bmatrix} \mathfrak{Z}_{T, z_d} \begin{bmatrix} X_0 \\ X_1 \\ X_2 \end{bmatrix}^t\] is the smallest generator of \[\left\langle \left\{\begin{bmatrix} X_0 \\ X_1 \\ X_2 \end{bmatrix} \mathfrak{Z}_{T, z} \begin{bmatrix} X_0 \\ X_1  \\ X_2 \end{bmatrix}^t: z \in (\Z / p^{e_0} \Z)^m\right\}\right\rangle.\]
$Z_0$ is the matrix with $z_i$ as $i$-th row.
\item $Z_1$ is a matrix such that the rows of $Z_1$ is a smallest generating set of \[\left\langle \left\{z : \begin{bmatrix} X_0 \\ X_1  \\ X_2 \end{bmatrix} \mathfrak{Z}_{T, z_i} \begin{bmatrix} X_0 \\ X_1 \\ X_2 \end{bmatrix}^t = 0\right\} \right\rangle. \]
\end{enumerate}
\end{definition}

\begin{lemma}\label{lem:tensor_isomorphism_restricted}
Let $\tA$ and $\tB$ be two $n \times n \times m$ alternating tensors over $\Z / p^e \Z$. 
Suppose $\tA'$ is a semi-canonical form for $\tA$ obtained for $T_{\tA}, X_{\tA, 0}, X_{\tA, 1}, X_{\tA, 2}$,
 and $\tB'$ is a semi-canonical form for $\tB$ obtained for $T_{\tB}, X_{\tB, 0}, X_{\tB, 1}, X_{\tB, 2}$.
If $T_{\tA}, X_{\tA, 0}, X_{\tA, 1}, X_{\tA, 2}$ has the same number of rows to $T_{\tB}, X_{\tB, 0}, X_{\tB, 1}, X_{\tB, 2}$ respectively, then 
\begin{enumerate}
\item If there are invertible matrices $X$ and $Y$ such that 
the following conditions hold:
\begin{enumerate}
\item  $\tB = (X, X, Y) \circ \tA$;
\item $X_{\tB, 0} = X_{\tA, 0} X^{-1} + W_{0}$ for some $W_0$ of the same dimension as $X_{\tA, 0}$;
\item $X_{\tB, 1} = X_{\tA, 1} X^{-1} + W_{1}$ for some $W_1$ of the same dimension as $X_{\tA, 1}$; 
\item $X_{\tB, 2} = X_{\tA, 2} X^{-1} + W_{2}$ for some $W_2$ of the same dimension as $X_{\tA, 2}$; 
\item $T_{\tB} = T_{\tA} X^{-1} + W_{3}$ for some $W_3$ of the same dimension as $T_{\tA}$;
\item  For each $i \in \{0, 1, 2, 3\}$,  $(W_i, I, I) \circ \tB$ is a zero tensor,
\end{enumerate}
then there are invertible matrices
\[P = \begin{bmatrix} I_{n'} & 0 \\ 0 & P_0\end{bmatrix}\]
for some invertible matrix $P_0$
and 
\[Q = \begin{bmatrix} I_{m'} & Q_0 \\ 0 & Q_1\end{bmatrix}\]
for some arbitrary matrix $Q_0$ and invertible matrix $Q_1$
such that $\tB' = (P, P, Q)\circ \tA'$, 
where $n'$ is the sum of the numbers of rows for $X_{\tA, 0}, X_{\tA, 1}$, and $X_{\tA, 2}$,
and $m'$ is the number of rows of $Z_{\tA, 0}$ when constructing the semi-canonical form $\tA'$.


\item If there do not exist invertible matrices $P \in \GL(n, \Z / p^e \Z)$ and $Q \in \GL(m, \Z / p^e \Z)$ such that $\tB = (P, P, Q) \circ \tA$, 
then 
there are no invertible matrices $P'$ and $Q'$ such that $\tB' = (P', P', Q') \circ \tA'$.
\end{enumerate}

\end{lemma}
\begin{proof}
Let $X_{\tA, 3}, Z_{\tA, 0}$, and $Z_{\tA, 1}$ denote the matrices used to construct $\tA$ for $\tA$, and $X_{\tB, 3}, Z_{\tB, 0}, and Z_{\tB, 1}$ denote the matrices used to construct $\tA$ for $\tA$. 

For the first property, by Definition~\ref{def:semi_canonical}, 
we have $X_{\tB, 3} = X_{\tA, 3} X^{-1}$, $Z_{\tB, 1} = Z_{\tA, 1} Y^{-1}$, and 
$Z_{\tB, 0} = Z_{\tA, 0} Y^{-1} + U$ for some $U$ such that the row span of $U$ is a subset of the row span of $Z_{\tB, 1}$. 
Hence, matrices $P$ and $Q$ exist. 

The second property is obtained by the first condition of Definition~\ref{def:semi_canonical} and the fact that 
\[\rowspan\left(\begin{bmatrix} Z_{\tA, 0} \\ Z_{\tA, 1} \end{bmatrix}\right)  = (\Z / p^e \Z)^m \text{ and } \rowspan\left(\begin{bmatrix} Z_{\tB, 0} \\ Z_{\tB, 1} \end{bmatrix}\right)  = (\Z / p^e \Z)^m \] 
by the last condition of Lemma~\ref{lem:enumeration} and the definition of $Z_{\tA, 0}, Z_{\tA, 1}, Z_{\tB, 0}$, and $Z_{\tB, 1}$.  
\end{proof}

\begin{definition}
In the correlated matrix tuple isometry problem, the inputs contains 
two $n \times n$ alternating matrix tuples $(A_1, \dots, A_k)$ and $(A_1', \dots, A_k')$ over $ \Z / p^e \Z$,
two $m \times n$ matrix tuples $(B_1, \dots, B_\ell)$ and $(B_1', \dots, B_\ell')$ over $\Z / p^e \Z$,
and a parameter $t$. 
The problem is to determine if there is an invertible matrix $P \in \GL(n, \Z / p^e \Z)$
and another matrix 
\[Q = \begin{bmatrix} I_{t} & Q_0 \\ 0 & Q_1\end{bmatrix}\] for some $Q_0 \in M(t, m - t, \Z / p^e \Z)$, $Q_1 \in M(m - t, \Z / p^e \Z)$ such that 
$A_i' = P A_i P^t$ for every $1 \leq i \leq k$ and $B_i' = Q B_i P^t$ for every $1 \leq i \leq \ell$.
\end{definition}
\begin{lemma}\label{lem:correlated_matrix_tuple}
The correlated matrix tuple isometry problem reduces to the alternating matrix tuple isometry problem in polynomial time in the following sense: 
For an input instance of the correlated matrix tuple isometry problem containing 
$n \times n$ alternating matrix tuples $(A_1, \dots, A_k)$ and $(A_1', \dots, A_k')$ over $ \Z / p^e \Z$ and 
$m \times n$ matrix tuples $(B_1, \dots, B_\ell)$ and $(B_1', \dots, B_\ell')$ over $\Z / p^e \Z$,
there is an algorithm to construct alternating $g \times g$ matrix tuples 
$(C_1, \dots, C_h)$ and $(C_1, \dots, C_h)$ over $\Z / p^e \Z$ with $g = \mathrm{poly}(n, m)$ and $h =  \mathrm{poly}(k, \ell)$
such that $(C_1, \dots, C_h)$ and $(C_1, \dots, C_h)$ are isometric if and only if 
the input of the correlated matrix tuple isometry  problem 
$(A_1, \dots, A_k)$  and $(B_1, \dots, B_\ell)$ are correlated isometric to $(A_1', \dots, A_k')$ and $(B_1', \dots, B_\ell')$ respectively. 
\end{lemma}

A proof of Lemma~\ref{lem:correlated_matrix_tuple}  for matrix tuples over a finite field is provided in Section III.D of~\cite{IMQSZ24}. The proof for matrix tuples over the ring 
$\mathbb{Z}/p^e\mathbb{Z}$ is identical to the finite field case, so we omit it here.

Finally, we give our algorithm in Figure~\ref{fig:algorithm} and prove Theorem~\ref{thm:tensor_iso_algo}. 

\begin{figure}[th]

\begin{framed}
\noindent \textbf{Alternating Matrix Space Isometry Algorithm}

\noindent \textbf{Input}: Two tensors $\tA$ and $\tA'$ of the same dimension $n \times n \times m$ over $\Z / p^e \Z$.

\noindent \textbf{Output}: Yes or no.

\begin{enumerate}
\item If the parameters $\alpha_1, \dots, \alpha_e, \beta_1, \dots, \beta_e, \gamma_1, \dots, \gamma_e$ of the two tensors are not identical, then return no. 
Otherwise, set  $e_0 = \lceil (\sum_{\ell = 1}^{e} \ell \cdot \beta_\ell)^{1/9} \rceil$, and $r = \lceil ((\sum_{\ell = 1}^{e} \ell \cdot \alpha_\ell) \cdot \log p)^{2/9} \rceil$. 
\item Find arbitrary matrices $X_{\tA, 0}$, $X_{\tA, 1}, X_{\tA, 2}, T_\tA$ satisfying the conditions of Lemma~\ref{lem:enumeration}, and compute the semi-canonical form $\tA'$ for $\tA$ with respect to $X_{\tA, 0}$, $X_{\tA, 1}, X_{\tA, 2}, T_\tA$ using arbitrary matrices $X_{\tA, 3}, Z_{\tA, 0}, Z_{\tA, 1}$ satisfying the conditions of Lemma~\ref{lem:enumeration}. 
\item 
Let $n'$ and $m'$ be the integers such that $\tA'$ is of dimension $n' \times n' \times m'$.  
Let $n''$ be the sum of the numbers of rows for $X_{\tA, 0}$, $X_{\tA, 1}$, and $X_{\tA, 2}$, and $m''$ be the number of rows for $Z_{\tA, 0}$. 
\item For any $X_{\tB, 0}$, $X_{\tB, 1}, X_{\tB, 2}, T_\tB$ for $\tA'$ satisfying the conditions of Lemma~\ref{lem:enumeration} with dimensions the same as $X_{\tA, 0}$, $X_{\tA, 1}, X_{\tA, 2}, T_\tA$ respectively,  
\begin{enumerate}
\item Construct a semi-canonical form $\tB'$ for $\tB$ with respect to $X_{\tB, 0}$, $X_{\tB, 1}, X_{\tB, 2}, T_\tB$. 
\item If $\tA'_{i, j, k} \neq \tB'_{i, j, k}$ for any $1 \leq i \leq n'', 1 \leq j \leq n'', 1 \leq k \leq m''$, then continue. 
\item Let $A_i$ be the induced matrix of $\mathfrak{Z}_{\tA, i}$ on the last $n' - n''$ rows and columns for any $1 \leq i \leq m''$, and $B_i$ be $\mathfrak{X}_{C, i}$ removing the first $n''$ columns for any $1 \leq i \leq n''$. Construct $A_i'$ and $B_i'$ for $\tB'$ in the same way. 
\item If $A_1, \dots, A_{m''}, B_1, \dots, B_{n''}$ and $A_1', \dots, A_{m''}', B_1', \dots, B_{n''}'$ are correlated matrix tuple isometry, then output yes. 
\end{enumerate}
\item Output no.
\end{enumerate}
\end{framed}
\caption{Alternating Matrix Space Isometry Algorithm}\label{fig:algorithm}
\end{figure}

\begin{proof}[Proof of Theorem~\ref{thm:tensor_iso_algo}]
The correctness of the algorithm is obtained by Lemma~\ref{lem:enumeration}, Lemma~\ref{lem:lexical_order}, Definition~\ref{def:semi_canonical},  Lemma~\ref{lem:tensor_isomorphism_restricted}, and Lemma~\ref{lem:correlated_matrix_tuple}.

In the rest of this proof, we bound the running time of the algorithm. When enumerating rows of $X_{\tA, 0}$ and $X_{\tB, 0}$, we only enumerate row vectors $x$ such that $\mathfrak{X}_{T, x}$ are different. 
By Lemma~\ref{lem:enumeration}, the cost of enumerating a single row of $X_0$ is $p^{\sum_{\ell = 1}^{e} \ell \cdot \beta_\ell}$, and the cost of enumerating $X_0$ is 
\[\left(p^{(\sum_{\ell = 1}^{e} \ell \cdot \beta_\ell)} \right)^{(\sum_{\ell = e_0 + 1}^{e} \beta_\ell)}.\]

Similarly, when enumerating 
$X_{\tA, 1}, X_{\tA, 2}, T_{\tA}, X_{\tB, 1}, X_{\tB, 2}$, or $T_{\tB}$, we only enumerate row vectors $x$ such that $\mathfrak{X}_{T, x}$ are different.
By Lemma~\ref{lem:enumeration},
the cost of enumerating a single row of $X_{\tA, 1}, X_{\tA, 2}, T_{\tA}$, $X_{\tB, 1}$, $X_{\tB, 2}$, or $T_{\tB}$ is \[p^{\sum_{\ell = 1}^{e_0} \ell \cdot \beta_\ell} \cdot p^{\sum_{\ell = e_0 + 1}^{e} e_0 \cdot \beta_\ell},\] and thus the cost of enumerating all possible $X_{\tA, 1}, X_{\tA, 2}, T_{\tA}, X_{\tB, 1}, X_{\tB, 2}$, and $T_{\tB}$  is 
\[\left(p^{(\sum_{\ell = 1}^{e_0} \ell \cdot \beta_\ell + \sum_{\ell = e_0 + 1}^{e} e_0 \cdot \beta_\ell)}\right)^{ O(\max\{\log(p^{\sum_{i = 1}^{e} i \cdot \alpha_i}), r\}/ \sqrt{r} + e_0^2 r^2 + \sum_{i = e_0 + 1}^{e} \beta_i)}.\]


By the choices of the parameters, 
the total cost of enumeration is at most 
\[p^{O((\sum_{\ell = 1}^{e} \ell \cdot (\beta_\ell + \alpha_\ell)) \cdot(\sum_{\ell = 1}^{e} \ell \cdot (\beta_\ell + \alpha_\ell)\log p)^{8/9})}.\]
Recall that $\mu = \sum_{\ell = 1}^e (\alpha_\ell + \beta_\ell + \gamma_\ell)$, 
where $\gamma_i=\beta_i$ here as we work with alternating tensors. 
So the total cost of enumeration is at most 
\[(p^\mu)^{O(( \mu \cdot \log p)^{8/9})}. \qedhere\]
\end{proof}

\subsection{Efficient algorithm for isometry of tuples of (anti)symmetric bilinear forms over $\Z/p^e \Z$ 
}\label{subsec:tuple-isometry}\label{sec:matrix_tuple}

Throughout Section~\ref{subsec:tuple-isometry}, $R$ denotes a finite local commutative principal ideal ring (PIR) of odd characteristic (equivalently, $R/\m$ is a field of odd characteristic), and we use the notation and terminology laid out in Section~\ref{sec:PIR}.

Given an $R$-module $M$, integers $1 \leq z_i \leq e$, and $R$-bilinear forms $B_i,C_i \colon M \otimes_R M \to R/\m^{z_i}$, the \algprobm{(Tuple) Isometry Problem} is to decide whether there exists $F \in \aut(M)$ such that $B_i(F(u), F(v)) = C_i(u,v)$ for all $u,v \in M$ and all $i=1,\dotsc,m$, and any such $F$ is called an \emph{isometry} from $\vec{B}$ to $\vec{C}$. If we represent each bilinear form by its corresponding matrix, which we denote by the same letter by abuse of notation, we have $(B_i)_{jk} = B_i(v_j, v_k)$, where the $v_*$ are a basis of $M$, and the congruence problem then asks whether $F^t B_i F = C_i$ for all $i = 1,\dotsc,m$, where now we view $F$ as its matrix according to Theorem~\ref{thm:HillarRheaPIR}. (Note that the matrix $B_i$ can be any matrix with entries in $R/\m^{z_i}$, subject to the constraint that if $z_i \leq \min\{e_j, e_k\}$, then the $(j,k)$ entry must be a multiple of $\m^{z_i - \min\{e_j, e_k\}}$. This follows from bilinearity and the fact that $v_j$ is annihilated by $\m^{e_j}$ and $v_k$ is annihilated by $\m^{e_k}$ in $M$.)

For $\varepsilon \in \{\pm 1\}$, we say a bilinear form $B \colon M \otimes M \to R/\m^z$ is $\varepsilon$-symmetric if $B(u,v) = \varepsilon B(v,u)$ for all $u, v \in P$; thus $\varepsilon=1$ corresponds to symmetric and $\varepsilon=-1$ corresponds to anti-symmetric (equivalently, since $p$ is odd, alternating). 

Our main result in this section is:

\begin{theorem}[{Generalizes \cite[Thm.~1.7(1)]{IQ19} from finite fields to finite local commutative PIRs}]\label{thm:matrix_tuple_isometry}
If $R$ is a finite local commutative PIR with quotient field of odd characteristic $p$, and $M$ is an $n$-generated $R$-module, the \algprobm{Tuple Isometry Problem} for $m$-tuples of $\varepsilon$-symmetric bilinear forms on $M$ can be solved in randomized  time $\poly(n,m,e)$ with operations in $R$ at unit cost, or deterministic time $\poly(n,m,e, p)$ with operations in $R$ at unit cost, where $p$ is the characteristic of $R/\m$.

In particular, when $R = \Z/p^e \Z$, the latter implies a solution in deterministic time $\poly(n,m,p,e)$ on a Turing machine.
\end{theorem}

Our algorithm follows that of Ivanyos and Qiao closely, and uses several of their key results as subroutines. Here we recall the main steps of their algorithm, and give proofs of the key results that need to be generalized from the case of finite fields to local commutative PIRs as above; we maintain the numbering from their paper for ease of comparison. In Appendix~\ref{app:similar}, we include those proofs of the key results here that are identical or nearly identical to the original ones, merely so that it may be checked carefully that the stated proofs do indeed go through in our more general setting.

Throughout the remainder of this section, let $M$ be an $n$-generated $R$-module. Given an $\varepsilon$-symmetric bilinear form $B \colon M \otimes M \to R/\m^z$, we define its kernel as $\ker B = \{m \in M | B(m,m') = 0 \forall m' \in M\}$. (Note that since $B$ is $\varepsilon$-symmetric, its left kernel and right kernel coincide as submodules of $M$.) Given $z_1, \dotsc, z_m$ and $\varepsilon$-symmetric bilinear forms $B_i \colon M \otimes M \to R/\m^{z_i}$, we say the tuple $(B_1, \dotsc, B_m)$ is \emph{non-degenerate} if $\bigcap_{i=1}^m \ker B_i = 0$. 


\paragraph{Steps of the algorithm.}
\begin{enumerate}
\item \textit{Reduce to the non-degenerate case.} 

\begin{lemma}
There is a polynomial-time $R$-algorithm which, given a tuple of $\varepsilon$-symmetric bilinear forms $\vec{B}$ on $M$, produces a quotient module $M'$ of $M$, and a non-degenerate tuple of $\varepsilon$-symmetric bilinear forms $\text{nondeg}(\vec{B})$ on $M'$ such that for any two tuples $\vec{B}, \vec{C}$, we have that $\text{nondeg}(\vec{B})$ and $\text{nondeg}(\vec{C})$ are isometric if and only if $\vec{B}, \vec{C}$ are isometric.
\end{lemma}

Over fields, this is easy by standard linear algebra. Over finite local PIRs, we instead use Smith Normal Form.

Essentially, $\vec{B}$ and $\text{nondeg}(\vec{B})$ are isometric---there is no real loss of information in moving from one to the other---but because they are bilinear forms on different modules $M$ and $M'$, respectively, making this precise takes some work. As we won't need such a statement, we avoid this complication, but we mention it as it may still be helpful for intuition.

\begin{proof}
We show how to compute $\text{nondeg}(\vec{B})$ and the associated quotient $M'$ of $M$; the same procedure is then applied to $\vec{C}$. (If the quotient module produced in computing $\text{nondeg}(\vec{C})$ is not the same as $M'$, then $\vec{B}$ and $\vec{C}$ do not have isomorphic kernels, and therefore the two tuples cannot be isometric.)

Suppose $\vec{B}$ is degenerate, with common kernel $\bigcap_{i=1}^m \ker B_i = K \leq M$. By Smith Normal Form, there is a basis $m_1,\dotsc,m_n$ for $M$, together with exponents $0 \leq k_1 \leq k_2 \leq \dotsb $ and an integer $k$ such that $t^{k_1} m_1, t^{k_2} m_2, \dotsc, t^{k_k} m_k$ is a basis for $K$. By our assumptions on $R,M,\vec{B}$, we can compute this SNF in polynomial time, which includes the elements $m_i$ and the exponents $k_i$. If the $B_i$ were originally written in a different basis, we apply the change of basis matrix (coming from the SNF computation) to put $B_i$ into the $\{m_1, \dotsc, m_n\}$ basis, resulting in an isometric tuple of bilinear forms, which we call $\vec{B}'$.

In this basis, for $j=1,\dotsc,k$, we have that $t^{k_j} R / \m^{e_j}$ (in the $j$-th position) is in the kernel of $\vec{B}'$. Since every such vector is in the kernel of each $B_i'$, it must be the case that every element of the $j$-th row and $j$-th column of $B_i'$ is divisible by $t^{e_j - k_j}$.

Let $M' = M / K$. If $M = \bigoplus_{i=1}^n \langle m_i \rangle = \bigoplus_{i=1}^n R/\langle t^{e_i} \rangle$, then we have $M'$ is the following ``coordinate-wise quotient'': $M' = \left(\bigoplus_{i=1}^k R/\langle t^{k_i}  \rangle\right) \oplus \left(\bigoplus_{i=k+1}^n R/\langle t^{e_i} \rangle \right)$. That is, in the first $k$ coordinates, we mod out by $t^{k_i}$ in the $i$-th coordinate, and the remaining coordinates are untouched. (If some $k_i=0$, we may then omit that coordinate, which corresponds to $M'$ having a smaller basis than $M$.)

In $\text{nondeg}(B)$ we replace each element of the $j$-th row and column of $B'$ with its quotient by $t^{e_j - k_j}$. (Although there are multiple such quotients, they all differ by multiples of $t^{k_j}$, hence they are all equal as elements of $R/\langle t^{k_j} \rangle$, which is how we now view the $j$-th row and column of $\text{nondeg}(B)$.) If we ever have $e_j = k_j$, then that row and column of every $B_i'$ is zero, and we simply remove that row and column from each $B_i'$. For $j=1,\dotsc,k$, since the $j$-th row and column of $\text{nondeg}(B)$ are now viewed as elements of $R/t^{k_j}$, the matrices we constructed in $\text{nondeg}(\vec{B})$ correspond to bilinear forms on $M'$, whose common kernel corresponds to the image of $K$ in $M'$, which is zero. Thus $\text{nondeg}(\vec{B})$ is non-degenerate.

Furthermore, if $\vec{B}$ and $\vec{C}$ are isometric, then so are $\vec{B}'$ and $\vec{C}'$ (since $\vec{B}'$ is isometric to $\vec{B}$ by construction, and similarly for $\vec{C}'$ and $\vec{C}$). If $F$ is an isometry from $\vec{B}'$ to $\vec{C}'$, let $F'$ be the matrix gotten from $F$ by removing the rows and columns corresponding to rows and columns of $\vec{B}'$ that are removed in $\text{nondeg}(\vec{B})$. Then $F'$ is an isometry from $\text{nondeg}(\vec{B})$ to $\text{nondeg}(\vec{C})$. Doing this process in reverse gives the converse.
\end{proof}

From now on, we assume that $\vec{B}$ and $\vec{C}$ are themselves non-degenerate tuples of bilinear forms on $M$.

\item \textit{Solve the twisted equivalence problem.} Test whether $\vec{B}$ and $\vec{C}$ are ``twisted equivalent'', meaning that there are invertible elements $A,D \in \aut(M)$ such that $A^t \vec{B} = \vec{C} D$, or in terms of bilinear forms, that $B_i(Au, v) = C_i(u, Dv)$ for all $u,v \in M$ and all $i=1,\dotsc,m$.

\begin{lemma} \label{lem:twisted}
There is a polynomial-time reduction from \algprobm{Twisted Equivalence} of tuples of $\varepsilon$-symmetric bilinear forms on an $R$-module $M$ to the \algprobm{Module Isomorphism Problem} over $R$. In particular, \algprobm{Twisted Equivalence} for such tuples can be solved in polynomial time.
\end{lemma}

The reduction is the same construction as in Ivanyos \& Qiao \cite[Prop.~3.2]{IQ19}, but the argument of correctness requires some care in the case of a finite local PIR, so we include it in Appendix~\ref{app:similar}. To solve the \algprobm{Module Isomorphism Problem} over $R$, we use the polynomial-time algorithm of Ciocanea-Teodorescu \cite{CT}, which extends that of Brooksbank \& Luks \cite{BL08} from working over fields to working over finite rings.

\item \textit{Reduce to decomposing an element of a $*$-algebra.}

\newcommand{\adj}{\text{Adj}}

\begin{definition}[{cf. \cite[Def.~2.3]{IQ19}}]
Let $R$ be a finite local PIR with quotient field of odd characteristic, and $M$ a finite $R$-module. For $\varepsilon \in \{\pm 1\}$, and $\vec{B} = (B_1, \dotsc, B_m)$ a tuple of $\varepsilon$-symmetric bilinear forms on $M$, the \emph{adjoint algebra} of $\vec{B}$, denoted $\adj(\vec{B})$, is $\{(A,D) : A,D \in \End_R(M), A^t B_i = B_i D \,\forall i\}$. $\adj(\vec{B})$ is a $*$-algebra with $(A,D)^* = (D,A)$.
\end{definition}

By definition we have that $\adj(\vec{B})$ is a sub-algebra of $\End(M)^{op} \oplus \End(M)$, where $\End(M)^{op}$ has the same set of elements of $\End(M)$, but multiplication is reversed (corresponding to the use of the transpose in the above definition). However, when $\vec{B}$ is non-degenerate, we may realize $\adj(\vec{B})$ as isomorphic to a sub-algebra of $\End(M)$ as follows. The idea is that when $\vec{B}$ is non-degenerate, if $(A,D) \in \adj(\vec{B})$, then $A$ uniquely determines $D$ and vice versa. For suppose $(A,D), (A',D)$ are both in $\adj(\vec{B})$. Then we have $A^t B_i = B_i D = (A')^t B_i$ for all $i$. But then we have $(A^t - (A')^t)B_i = 0$ for all $i$, and thus the columns of $A - A'$ are in the common kernel of the $B_i$. But if $\vec{B}$ is non-degenerate, this common kernel is zero by definition, and thus $A - A' = 0$, so $A = A'$. From this it follows that the projection map from $\adj(\vec{B}) \subseteq \End(M)^{op} \oplus \End(M)$ onto the second coordinate $\End(M)$ is injective. Equivalently, when $\vec{B}$ is non-degenerate, we may thus define $\adj(\vec{B})$ as the following sub-algebra of $\End(M)$: $\adj(\vec{B}) \cong \{D \in \End(M) : (\exists A \in \End(M))[A^t \vec{B} = \vec{B} D]\}$, and the $*$ operation is then defined by $D^*$ being the unique $A$ such that $A^t \vec{B} = \vec{B} D$. With this setup, the remainder of \cite[Sec.~3.4]{IQ19} goes through without change. This is because the proofs there are phrased entirely in terms of the $*$ and inverse operations and their basic properties, not relying on anything special to working over fields. In particular, we have the following lemma:

\begin{lemma}[{Cf. \cite[Prop.~3.4]{IQ19}}]
Let $\vec{B}, \vec{C}$ be $\varepsilon$-symmetric bilinear forms on the $R$-module $M$, and suppose that $A,D \in \aut(M)$ are such that $A^t \vec{B} = \vec{C} D$. Let $E = A^{-1} D^{-1}$ and $\mathfrak{A} = \adj(\vec{C})$, viewed as a subalgebra of $\End(M)$ as above. Then $\vec{B}$ and $\vec{C}$ are isometric if and only if there exists $X \in \mathfrak{A}$ such that $X^* X = E$.
\end{lemma}

Given $E$ and $\mathfrak{A}$, finding such an $X$ is the \algprobm{$*$-Symmetric Decomposition Problem}. Note that here we are given $\mathfrak{A}$ by a set of elements of $\End(M)$ (which we may view as matrices over $R$ by Theorem~\ref{thm:HillarRheaPIR}) which form a basis for $\mathfrak{A}$.  

\item \textit{Solve the $*$-symmetric decomposition problem.}

\begin{enumerate}
\item \textit{Compute the algebra structure of $\mathfrak{A}$.} 

\begin{theorem}[{cf. \cite[Thm.~3.6]{IQ19}, which is based on \cite{Ron90,EG00,Iva00}}] \label{thm:structure}
Given an $n$-generated  $R$-module $M$, and a basis of a subalgebra $\mathfrak{A} \leq \End(M)$, there is a Las Vegas $R$-algorithm running in time $\poly(n)$ that computes (1) a basis of $\rad(\mathfrak{A})$, (2) a surjective $R$-algebra homomorphism $\pi \colon \mathfrak{A} \to M(n_1, \F_{q_1}) \oplus \dotsb \oplus M(n_k, \F_{q_k})$ with kernel $\rad(\mathfrak{A})$, where $\F_{q_i}$ is a finite extension field of $R/\m$ that is specified by a basis of $\F_{q_i}$ over $R/\m$. There is also a deterministic $R$-algorithm for the same running in time $\poly(n, p)$, where $p$ is the characteristic of $R/\m$.

Given the output of the aforementioned algorithm, there are also efficient deterministic algorithms to (3) compute $\pi(a)$ given $a \in \mathfrak{A}$, and (4) given $b \in M(n_1, \F_{q_1}) \oplus \dotsb \oplus M(n_k, \F_{q_k})$, compute $a \in \mathfrak{A}$ such that $\pi(a) = b$.
\end{theorem}

A key new ingredient compared to the field case is the following lemma:

\begin{lemma} \label{lem:rad}
Suppose $\mathfrak{A}$ is a finite (or Artinian) $R$-algebra, and let $\pi \colon \mathfrak{A} \to \mathfrak{A} / \m \mathfrak{A}$ be the natural quotient map. Then $\pi(\rad(\mathfrak{A})) = \rad(\mathfrak{A}/\m \mathfrak{A})$.
\end{lemma}

An element $x \in \mathfrak{A}$ is \emph{strongly nilpotent} if $xy$ is nilpotent for all $y$; in a finite (or Artinian) algebra, the Jacobson radical $\rad(\mathfrak{A})$ is equal to the set of all strongly nilpotent elements.

\begin{proof}
By Proposition 4.6 of Lam \cite{Lam}, in any ring $\mathfrak{A}$, if $I$ is an ideal contained in $\rad(\mathfrak{A})$, then $\rad(\mathfrak{A})/I = \rad(\mathfrak{A}/I)$. So it will suffice for us to show that $\m \mathfrak{A}$ is an ideal contained in $\rad(\mathfrak{A})$. If $x \in\m \mathfrak{A}$,
then $x=tx'$ for some $x' \in A$, where $t$ is a generator of $\m$, and thus $xy=tx'y$. As $t$ is central, we have that $(xy)^e$ is divisible by $t^e$, hence zero. Thus every element of $\m \mathfrak{A}$ is strongly nilpotent, so $\m \mathfrak{A} \subseteq \rad(\mathfrak{A})$. Furthermore, since $t$ is central, $\m \mathfrak{A}$ is a (principal, two-sided) ideal. Thus $\m \mathfrak{A}$ is an ideal of $\mathfrak{A}$ contained in $\rad(\mathfrak{A})$, and the result follows.
\end{proof}

\begin{proof}[Proof of Theorem~\ref{thm:structure}] 
(1) To compute $\rad(A)$ we proceed as follows. If $\{a_1, \dotsc, a_d\}$ is a generating set of the additive group of $\mathfrak{A}$, then $\m \mathfrak{A}$ is spanned by the elements $\{ta_i : i = 1, \dotsc, d\}$. Furthermore, we can calculate a basis, exponents, and structure coefficients (as above) of $\mathfrak{A}/\m \mathfrak{A}$ by simply setting each $e_i=1$ and taking each $c_{ij}^k$ to its value modulo $t$. This also specifies for us the quotient map $\pi' \colon \mathfrak{A} \to \mathfrak{A}/\m \mathfrak{A}$ in terms of how it acts on a basis. Then using previous algorithms over finite fields \cite{Ron90,EG00,Iva00} in $\mathfrak{A}/\m \mathfrak{A}$---which is now a finite-dimensional algebra over the finite field $R/\m$---we can compute a basis for $\rad(\mathfrak{A}/\m \mathfrak{A})$ over $R/\m$, and for each basis element of $\rad(\mathfrak{A}/\m \mathfrak{A})$, take any one preimage under $\pi$. Those preimages, together with the aforementioned generating set of $\m \mathfrak{A}$, generate $\rad(\mathfrak{A})$ as an $R$-module, by Lemma~\ref{lem:rad}. Finally, using Smith Normal Form, we can extract from this generating set a basis of $\rad(\mathfrak{A})$.

(2 and 3) In the preceding step, we computed the quotient map $\pi' \colon \mathfrak{A} \to \mathfrak{A} / \m \mathfrak{A}$, the latter of which is an Artinian algebra over the field $R/\m$. Using previous algorithms over finite fields \cite{Ron90,EG00,Iva00}, we can then compute the quotient map $\mathfrak{A} / \m \mathfrak{A} \to \mathfrak{A} / \rad(\mathfrak{A})$ as well as the isomorphism of the latter with $M(n_1, \F_{q_1}) \oplus \dotsb \oplus M(n_k, \F_{q_k})$, where $\F_{q_i}$ is a finite extension field of $R/\m$ that is specified by a basis of $\F_{q_i}$ over $R/\m$.

(4) From the algorithms over finite fields \cite{Ron90,EG00,Iva00}, given $b \in M(n_1, \F_{q_1}) \oplus \dotsb \oplus M(n_k, \F_{q_k})$, we can compute an $a' \in \mathfrak{A} / \m \mathfrak{A}$ that is a preimage of $b$ under the quotient map $\mathfrak{A} \to \mathfrak{A} / \rad(\mathfrak{A})$. We now need to compute an $a \in \mathfrak{A}$ whose image modulo $\mathfrak{m}$ is $a'$. Let $M$ be the inclusion matrix of $\m \mathfrak{A}$ into $\mathfrak{A}$ (say, viewing ring elements here as column vectors). Putting $M$ into Smith Normal Form as in the first step, we get a basis $\{a_1,\dotsc,a_d\}$ of $\mathfrak{A}$ and exponents $f_1, \dotsc, f_d$ such that $t^{f_i} a_i = 0$ and $f_i$ is the least integer with this property, and furthermore we get a $k \leq d$ and integers $0 \leq e_1 \leq e_2 \leq \dotsb \leq e_k$, such that $\{t^{e_1} a_1, t^{e_2} a_2, \dotsb, t^{e_k} a_k\}$ is a basis for $\m \mathfrak{A}$. In this basis on $\mathfrak{A}$, the projection map $\mathfrak{A} \to \mathfrak{A} / \m \mathfrak{A}$ takes the form $\sum_{i=1}^d r_i a_i \mapsto \sum_{i=1}^k (r_i \text{ mod } t^{e_i}) \overline{a}_i$, where $\overline{a}_i$ represent the image of $a_i$ in $\mathfrak{A} / \m \mathfrak{A}$, and each $r_i$ is taken modulo the annihilator of $a_i$ (which is some power of $\m$). Now, say $a' = \sum_{i=1}^k \overline{r}_i \overline{a}_i$ is the element of $\mathfrak{A} / \m \mathfrak{A}$ whose preimage we want to compute. Then it suffices to lift $\overline{r}_i$ from $R/\m^{e_i}$ to $r_i \in R/\Ann_R(a_i)$, and then $\sum_{i=1}^k r_i a_i$ will be a preimage of $a'$.

By induction, it will suffice to be able to lift $\overline{r} \in R/\m^f$ to $r \in R/\m^{f+1}$ such that $r \equiv \overline{r} \pmod{\m^f}$, for arbtirary $0 \leq f \leq e$. However, this follows from our assumption that elements of $R/\m^f$ are represented by elements of $R$; see the beginning of Section~\ref{sec:PIR}.
\end{proof}

\item \textit{Recognize the $*$-algebra structure.} This part of the algorithm only deals with the semisimple quotient of $\mathfrak{A}$, which is necessarily an algebra over a field. Thus the results of \cite[Sec.~3.5.2]{IQ19} apply directly here, since this part of the algorithm is working in the same setting as they were working in, not a generalization thereof.

\item \textit{Reduce to the semi-simple case.} 

\begin{proposition}[{Cf. \cite[Prop.~3.11]{IQ19}}] \label{prop:semisimple}
Let $\mathfrak{A}$ be an $*$-algebra over $R$. Let $E \in \mathfrak{A}$ be an invertible $*$-symmetric element, and suppose there exists $Y \in \mathfrak{A} / \rad(\mathfrak{A})$ such that $Y^* Y \equiv E \pmod \rad(\mathfrak{A})$. Then there exists $X \in \mathfrak{A}$ such that $X^* X = E$, and such an $X$ can be found by a deterministic polynomial-time $R$-algorithm.
\end{proposition}

The one additional ingredient we need in the proof is the following lemma:

\begin{lemma}
Suppose $\mathfrak{A}$ is a finite $R$-algebra. Let $n = \dim_{R/\m} (\mathfrak{A} / \m \mathfrak{A})$. 
Then $(\rad \mathfrak{A})^{2^k} = 0$ for some $k \leq 1 + \lceil \log_2(\max\{n,e\}) \rceil $ (where recall that $e$ is minimal such that $\m^e = 0$).
\end{lemma}

\begin{proof}
By Lemma~\ref{lem:rad}, $\rad \mathfrak{A} \pmod \m = \rad(\mathfrak{A} / \m \mathfrak{A})$. As argued in the proof of \cite[Prop~3.11]{IQ19}, $\rad(\mathfrak{A}/\m\mathfrak{A})^{2^{k'}} = 0$ for some $k' \leq \lceil \log n \rceil$. Lifting to $\mathfrak{A}$, this is equivalent to saying that $\rad(\mathfrak{A})^{2^{k'}}$ is contained in $\m \mathfrak{A}$ for some $k' \leq \lceil \log n \rceil$. But since $\m^e = 0$ and $\m \subseteq R \subseteq \mathfrak{A}$ is central in $\mathfrak{A}$ (by the definition of $R$-algebra), we have that $(\m \mathfrak{A})^e = 0$ as well. Thus $\rad(\mathfrak{A})^{2^{k'} + e} = 0$. Finally, we have $2^{k'} + e \leq 2\cdot \max\{2^{k'}, e\} \leq 2^k$ for some $k \leq  1+ \lceil \log_2(\max\{n,e\}) \rceil$.
 \end{proof}

With this lemma in hand, the proof of Proposition~\ref{prop:semisimple} is nearly identical to that in Ivanyos--Qiao, again because their proof is phrased almost entirely in the language of $*$-algebras without using any special properties of algebras over fields. The bound on $k$ such that $(\rad \mathfrak{A})^{2^k} = 0$ is used in the proof to give an upper bound on the number of iterations of the algorithm, thus the running time of our $R$-algoirthm is still only at most $\poly(\dim_{R/\m} \mathfrak{A}/\m\mathfrak{A},e)$.

\item At this point, we may directly use the algorithm of \cite[Sec.~3.5.4 and 3.5.5]{IQ19} to solve the semi-simple case, since semi-simple algebras are always over fields. Thus the remainder of the algorithm uses their algorithm as a (black box) subroutine.
\end{enumerate}

\end{enumerate}

Finally, Theorem~\ref{thm:tensor_isomorphism_main} is obtained by Theorem~\ref{thm:GpI-to-AMSI}, Theorem~\ref{thm:tensor_iso_algo}, and Theorem~\ref{thm:matrix_tuple_isometry}. 



\appendix


\section{Appendix for proofs that are very similar to prior work} \label{app:similar}
In this appendix, we gather together all the proofs of claimed results where our result generalizes a prior result, but the proof is very nearly the same as the original result, one must merely check carefully that the corresponding statements in our generalized setting remain true. In each proof, while we include much of the proof to check that the result still holds in our generalized setting, we also highlight any differences needed from the original proof in \textbf{bold}.

\begin{proof}[Proof of Theorem~\ref{thm:HillarRheaPIR}]
Following this notation, we can think of elements of $M = \bigoplus_{i=1}^n R/\m^{e_i}$ as column vectors with $n$ entries, where the $i$-th entry is an element of $R/\m^{e_i}$. Let $\iota_i \colon R/\m^{e_i} \to M$ be the natural inclusion map of the $i$-th summand, and let $\pi_j \colon M \to R/\m^{e_j}$ be the natural quotient map with kernel $\bigoplus_{i \neq j} R/\m^{e_i}$, such that $\pi_i \iota_i$ is the identity on $R/\m^{e_i}$ for each $i$. For $m \in M$, let $m_i = \pi_i(m)$, then $m$ can be written as $m = \iota_1(m_1) + \dotsb + \iota_n(m_n)$, or in the more usual vector notation as $(m_1, \dotsc, m_n)^T$ (the transpose just because we will find it convenient to think of these as column vectors). 

For an endomorphism $\varphi \colon M \to M$, let $\varphi_{ij} = \pi_i \circ \varphi \circ \iota_j$, the latter being an element of $\Hom(R/\m^{e_j}, R/\m^{e_i})$. If $\psi \in \End(M)$, let $\sigma = \varphi + \psi$ and $\mu = \psi \varphi$, and define $\psi_{ij}, \sigma_{ij}, \mu_{ij}$ analogously. Then it is routine to verify that $\sigma_{ij} = \varphi_{ij} + \psi_{ij}$ and $\mu_{ij} = \sum_{k=1}^n \psi_{ik} \circ \varphi_{kj}$. Thus this set of matrices forms a ring in the usual manner, and the map taking $\varphi$ to the matrix whose $(i,j)$ entry is $\varphi_{ij}$ for all $i,j \in \{1,\dotsc,n\}$ is a ring homomorphism from $\End(M)$ into this ring of matrices. It is routine to verify that this homomorphism is injective and surjective, and is thus an isomorphism. This completes the proof of the first part of the result.

Next, consider any $R$-module homomorphism $\varphi \colon R/\m^{e_j} \to R/\m^{e_i}$. Because these are cyclic $R$-modules, such a homomorphism is completely determined by $\varphi(1)$, for we have $\varphi(r) = r \cdot \varphi(1)$ for all $r \in R$. Thus $\varphi$ is completely specified by the single element $\varphi(1) \in R/\m^{e_i}$, and the only constraint on the latter is that it must be the case that multiplication by $\m^{e_j}$ annihilates it, since that was true of $1$ in $R/\m^{e_j}$. Now, if $e_i \leq e_j$, this is true of every element of $R/\m^{e_i}$, and there is no constraint on $\varphi(1)$. If $e_i > e_j$, then $\varphi(1)$ must be in $\m^{e_i - e_j}$, or equivalently, must be divisible by $t^{e_i - e_j}$, where $t$ is a generator of $\m$.

Combining our description of elements of $\Hom(R/\m^{e_j}, R/\m^{e_i})$ with our description of endomorphisms of $M$, we see that endomorphisms of $M$ can equivalently be thought of as $n \times n$ matrices $A$ where the elements in the $i$-th row are elements of $R/\m^{e_i}$, and such that $t^{e_i - e_j}$ divides $A_{ij}$ whenever $e_i \geq e_j$. Let $\overline{S}$ denote the preceding set of matrices, which we have shown is isomorphic to $\End(M)$.

Finally, it remains to characterize which matrices in $\overline{S}$ are invertible. Suppose $A,B \in \overline{S}$ satisfy $AB = I_n$, the $n \times n$ identity matrix. Taking the latter equation modulo $\m$, we find that the matrix $\overline{A}$ gotten by taking every entry of $A$ modulo $\m$ must be invertible. All the remains is to show the converse.

So suppose $A$ is an element of $\overline{S}$, such that $A \pmod{\m}$ is invertible as a matrix over the field $\F = R/\m$. By Corollary 11 of Hungerford \cite{hungerford} (\textbf{the use of this corollary is essentially the only new ingredient compared to the proof of Hillar and Rhea}), $R$ is the quotient of a PID which we call $Z$ (reminiscent of the role the integers $\Z$ play in the proof of Hillar and Rhea), and let $Q$ be the field of fractions of $Z$. (If $R$ itself is a domain, we may take $Z=R$.) Let $\hat{A} \in M_n(Z)$ be a lift of $A$, in the sense that when $\hat{A}_{ij}$ is taken modulo $\m^{e_i}$ for all $i$, we get back precisely $A_{ij}$. Let $S$ be the subset of $M_n(Z)$ whose image under the preceding map lies in $\overline{S}$; thus $\hat{A}$ is in $S$.

\begin{lemma}[{Cf. \cite[Lemma~3.2]{HillarRhea}}]
$S$ forms a ring under matrix multiplication. In particular, if we let $T$ be the diagonal matrix whose diagonal entries are $t^{e_1}, t^{e_2}, \dotsb, t^{e_n}$, then $A \in M_n(Z)$ is in $S$ if and only if there exists a matrix $A' \in M_n(Z)$ such that $A = TAT^{-1}$.
\end{lemma}

The proof of this lemma is almost immediate, and is identical, \emph{mutatis mutandis}, to that in Hillar \& Rhea.

By usual properties of the adjugate matrix (defined in terms of minors), there is a unique matrix $\hat{B} \in M_n(Z)$ such that $\hat{A}\hat{B} = \hat{B} \hat{A} = \det(\hat{A}) I_n$. We will show that $\hat{B}$ is in $S$, and thus its image in $\overline{S}$ is an inverse of $A$ in $\overline{S}$. 

By the preceding lemma, write $\hat{A} = TA' T^{-1}$, where $T$ is as in the lemma and $A' \in M_n(Z)$. Let $B'$ be the adjugate of $A'$. Let $B'' = TB'T^{-1}$. Then we have $\hat{A} B'' = T A' B' T^{-1} = \det(\hat{A}) I_n = T B' A' T^{-1} = B'' \hat{A}$. By the uniqueness of the adjugate, we have $B'' = \hat{B}$. Since $\hat{B}$ can thus be expressed as $T B' T^{-1}$ for some $B' \in M_n(Z)$, by the lemma we have that $\hat{B}$ is in $S$ as desired, and thus its image in $\overline{S}$ is an inverse of $A$, so $A$ is invertible.
\end{proof}

\begin{proof}[Proof of Lemma~\ref{lem:twisted}]
Given a tuple $\vec{B} = (B_1, \dotsc, B_m)$ of $\varepsilon$-symmetric bilinear forms over $M$, let $B_1, \dotsc, B_m$ also represent the corresponding matrices relative to some basis of $M$. If $M$ has a basis of size $n$, then let $M' \cong M \oplus M$, which has basis size $2n$. In a basis of $M'$ consisting of a basis of the first copy of $M$, followed by a basis of the second copy of $M$, let $\vec{B}' = (B_0', B_1', \dotsc, B_m')$ be the following tuple of $R$-module endomorphisms $M' \to M'$, viewing elements of $M'$ as column vectors of length $2n$:
\[
B_0' = \begin{pmatrix} \text{id}_M & 0 \\ 0 & 0 \end{pmatrix}
\qquad
B_i' = \begin{pmatrix} 0 & B_i \\ 0 & 0 \end{pmatrix} \quad (i=1,\dotsc,m).
\]
Similarly build a tuple $\vec{C}'$ from $\vec{C}$.

We claim that there exists $A,D \in \aut(M)$ such that $A \vec{B} = \vec{C}D$ if and only if there is $E \in \aut(M')$ such that $E \vec{B}' E^{-1} = \vec{C}'$, or equivalently $E \vec{B} = \vec{C} E$, or equivalently, the module structures on $M'$ given by $\vec{B}'$ and $\vec{C}'$ are isomorphic. To see this, write $E = \begin{pmatrix} A & G \\ H & D \end{pmatrix}$, where each of $A,G,H,D$ is an $R$-module homomorphism $M \to M$. From $E B_0' = C_0' E$, we get 
\[
\begin{pmatrix} A & 0 \\ H & 0 \end{pmatrix} = \begin{pmatrix} A & G \\ 0 & 0 \end{pmatrix},
\]
and therefore we have $G = H = 0$. Since $E$ is an invertible block-diagonal matrix, each diagonal block must be invertible, hence we have $A,D \in \aut(M)$. Furthermore, from $E B_i' = C_i' E$, we immediately get $A B_i = C_i D$. Thus $\vec{B}$ and $\vec{C}$ are twisted equivalent via the pair $(A^t, D)$. 

Conversely, if $A \vec{B} = \vec{C} D$ with $A,D$ invertible, then it is readily verified that taking $E = \begin{pmatrix} A & 0 \\ 0 & D \end{pmatrix}$ results in $A \vec{B}' = \vec{C}'$, and that $E$ is invertible. This completes the proof of the reduction.

The module isomorphism problem is then solved in polynomial time over the finite ring $R$ by the algorithm of Ciocanea-Teodorescu \cite{CT}.
\end{proof}


\bibliographystyle{alphaurl}
\bibliography{references}

\end{document}